\documentclass[final,5p,times]{elsarticle}

\usepackage{amsmath,amssymb,amsfonts}
\usepackage{algorithmic}
\usepackage{graphicx}
\usepackage{textcomp}
\usepackage{xcolor}
\usepackage{booktabs}
\usepackage[hidelinks]{hyperref}

\usepackage{import}
\usepackage{tabularx}
\usepackage{colortbl}
\usepackage{multirow}
\usepackage{color}
\usepackage{multicol}
\usepackage{array}
\newcolumntype{C}[1]{>{\centering\arraybackslash}m{#1}}
\newcolumntype{L}[1]{>{\raggedright\arraybackslash}m{#1}}
\usepackage{makecell}
\usepackage{caption}
\usepackage{subcaption}
\usepackage{xurl}
\usepackage{xspace}
\usepackage{comment}
\usepackage{enumitem}
\usepackage{xfrac}
\usepackage{adjustbox}
\usepackage[most]{tcolorbox}
\usepackage{float}

\usepackage{pifont}% http://ctan.org/pkg/pifont

\usepackage{lscape}
\usepackage{longtable}
\usepackage{tikz}
\usepackage{bbding}
\usepackage{cancel}
\newcommand*\halfcirc[1][0.55ex]{%
  \begin{tikzpicture}
    \draw[fill] (0,0)-- (90:#1) arc (90:270:#1) -- cycle;
    \draw (0,0) circle (#1);
  \end{tikzpicture}}

\newcolumntype{N}{>{\centering\arraybackslash}m{.02in}}
\newcolumntype{G}{>{\centering\arraybackslash}m{0.5in}}

\newcommand{\cmark}{\ding{51}}
\newcommand{\xmark}{\ding{55}}

\begin{document}

\begin{frontmatter}

\title{Rethinking Satellite Cybersecurity: A System-Level Taxonomy and Longitudinal Analysis}

\author[aff1]{Roy Peled\corref{cor1}}
\author[aff1]{Eran Aizikovich}
\author[aff1]{Edan Habler}
\author[aff1]{Yuval Elovici}
\author[aff1]{Asaf Shabtai}

\cortext[cor1]{Corresponding author}

\address[aff1]{Ben-Gurion University of the Negev, Beer-Sheva, Israel 8410501 }

% --------------------------------------------------
% Abstract
% --------------------------------------------------

\begin{abstract}
Satellite systems are increasingly targeted by cyber and electronic-warfare adversaries as their roles in communication, navigation, Earth observation, and defense continue to expand.
Although prior work has surveyed attacks on space systems, adversarial behavior across the full attack lifecycle has not been comprehensively characterized, and emerging satellite attack surfaces, such as adversarial machine learning (AML), have not been adequately considered.
This paper presents a comprehensive, satellite-specific taxonomy of tactics, techniques, and procedures (TTPs), developed primarily for low Earth orbit (LEO) systems and informed by empirical evidence from satellite systems operating across multiple orbital regimes, including LEO, MEO, and GEO.
The proposed taxonomy provides a structured, system-level framework for analyzing adversarial behavior, supporting threat modeling, defensive planning, and the design of detection and mitigation strategies for satellite systems.
To build the taxonomy, we first analyze the satellite ecosystem across the space, ground, communication, and user segments to identify architectural exposures and operational attack surfaces.
We then compile a dataset of more than 200 publicly reported satellite incidents spanning 1962 to 2026, including over 80 recent incidents not covered in prior work.
The dataset supports a longitudinal empirical analysis that reveals structural shifts in satellite threats, including the growing prevalence of ground-segment compromise, GNSS interference, broadcast and communication disruption, proximity-based counterspace activity, and deception-oriented attacks.
Building on this analysis, we propose a MITRE ATT\&CK-inspired satellite attack lifecycle taxonomy that is adapted to satellite-specific operational contexts.
The taxonomy integrates multiple threat modalities, including subsystem-level exploitation, radio-frequency (RF) interference, on-orbit operations, adversarial machine learning (AML), and deception techniques that are not consistently represented in existing taxonomies.
We further demonstrate the practical utility of the taxonomy through case studies of the 2022 Viasat KA-SAT cyberattack and a simulation-based ICARUS constellation-scale denial-of-service scenario designed to evaluate large-scale disruption dynamics.
The resulting framework combines longitudinal evidence, real-world incidents, and emerging attack modalities to support structured analysis of satellite cyber threats for both research and defense.
\end{abstract}

% --------------------------------------------------
% Keywords
% --------------------------------------------------


\end{frontmatter}

% --------------------------------------------------
% Main Sections
% --------------------------------------------------

\section{\label{sec:intro}Introduction}

\subsection{Satellite Security}

Since the launch of Sputnik in 1957, space technology has played a central role in shaping modern information infrastructure and continues to heavily influence contemporary society.
Today, satellites are indispensable for a wide range of civilian, commercial, and military services, including meteorology~\cite{kim2020new,menzel1994introducing,goodman2019goes}, navigation~\cite{parkinson1995history,morton2021position,morales2019survey}, communication (SATCOM)~\cite{darwish2022leo,kodheli2017integration,di2019ultra,zhen2020energy,kodheli2020satellite}, and intelligence gathering~\cite{early2021spying,shaofei2018analysis}.
Advances in miniaturization, launch services, and commercial off-the-shelf (COTS) technologies~\cite{falco2019cybersecurity,rawlins2022death} have reduced mission costs significantly~\cite{jones2018recent}, accelerating satellite deployment in all orbital regimes.
As a result, the number of active satellites has grown rapidly in recent years, increasing from approximately 4,867 in 2019 to over 12,000 by 2025~\cite{yale2025satellites}, with long-term projections suggesting continued near-exponential growth~\cite{miraux2022environmental}.

This rapid expansion, together with society's increasing reliance on space-based infrastructure, has amplified the consequences of disruptions and strengthened adversaries' incentives to target satellite systems.
In recognition of these risks, industry stakeholders have established initiatives such as Space ISAC~\cite{s_isac} to coordinate threat intelligence and defensive practices, and organizations such as the Aerospace Corporation publish threat intelligence assessments that highlight the evolving cyber and electronic-warfare landscape~\cite{harrison2022space}.
Despite these developments and the continued expansion of satellite attack surfaces across the space, ground, communication, and user segments, satellite cybersecurity has received comparatively less research attention than other established security domains, such as enterprise IT and network security.

Recent studies reported a sustained increase in publicly documented satellite cyber incidents.
Manulis et al.~\cite{manulis2021cyber} reported that the number of documented attacks between 2009--2018 was approximately five times higher than that recorded between 2000--2008.
Other work highlighted persistent regulatory gaps in the space domain~\cite{falco2021cubesat,de2019degree,falco2019cybersecurity,falco2018vacuum} and demonstrated practical exploitation paths that affect satellite subsystems and ground infrastructure~\cite{willbold2023space}.
However, while prior work provides important foundations for understanding satellite and space-system security, existing studies typically focus on specific parts of the threat landscape, such as communication links, ground infrastructure, software exploitation, or high-level incident chronologies.
As a result, there remains a need for a reproducible, incident-driven analysis that connects historical satellite incidents to system segments, spacecraft subsystems, adversary capabilities, lifecycle tactics, and operational impacts.

\subsection{Scope and Purpose}

Several knowledge bases provide structured frameworks for analyzing adversarial activity.
In this work, we build on the lifecycle-oriented structure of MITRE ATT\&CK~\cite{strom2018mitre} to develop a MITRE-inspired satellite attack lifecycle taxonomy adapted to satellite-specific operational contexts.
While ATT\&CK provides a general model of adversarial behavior, it does not capture key satellite-specific characteristics such as RF-layer threats, orbital dynamics, spacecraft subsystem interactions, and cross-segment dependencies.
We also consider SPARTA~\cite{sparta2023}, a space-focused TTP framework.
Our goal is not to replace SPARTA, but to complement it with an incident-grounded and operationally oriented taxonomy that connects adversarial behavior to satellite segments, communication links, spacecraft subsystems, mission functions, and operational impacts.

The taxonomy is developed through a structured process.
First, we examine the satellite ecosystem across the space, ground, communication, and user segments to identify architectural exposures and attack surfaces (Section~\ref{sec:space_env}).
Second, we characterize spacecraft subsystems, vulnerabilities, and adversary capability tiers (Sections~\ref{sec:subsystem}--\ref{sec:AdversariesThreat}).
Third, we compile and code a longitudinal dataset of publicly reported satellite cyber and electronic-warfare incidents to capture evolving threat patterns (Section~\ref{sec:Satellite_Attacks}).
Finally, we use this coded corpus to derive and evaluate a satellite-specific MITRE-inspired lifecycle taxonomy, while selectively aligning with SPARTA concepts where applicable (Section~\ref{sec:MITRE}).
The taxonomy focuses primarily on low Earth orbit (LEO) systems, while remaining applicable to medium Earth orbit (MEO) and geostationary Earth orbit (GEO) systems.
We demonstrate its utility through case studies of the 2022 Viasat KA-SAT cyberattack and a simulation-based ICARUS constellation-scale denial-of-service scenario (Section~\ref{sec:demonstration}).

\begin{table*}[ht!]
%\tiny 
\centering
\scriptsize
\setlength{\tabcolsep}{7pt}
\renewcommand{\arraystretch}{1.2}

\caption{Categorization of related work. \emph{Legend:} \cmark{} = fully covered or explained in depth; \bcancel{\Checkmark} = partially addressed/mentioned at a high level (e.g., only part of a taxonomy, no deep dive); \xmark{} = not covered.}
\label{tab:related_work}

\begin{adjustbox}{max width=\textwidth}
\begin{tabular}{l|ccc|cccc|ccccccc|c|c|c|c}
\hline
%\multicolumn{18}{c}{\textbf{Related Work}} & 
%\\
   &
    \multicolumn{3}{c|}{\textbf{GS}} &
    \multicolumn{4}{c|}{\textbf{Communication}} &
    \multicolumn{7}{c|}{\textbf{Subsystems}} &
    \multicolumn{1}{c|}{\textbf{User}} &
    \multicolumn{1}{c|}{\textbf{History}} &
    \multicolumn{1}{c|}{\textbf{AML}} &
    \multicolumn{1}{c}{\textbf{Attack Tax.}} \\ 
    % \multicolumn{2}{c|}{\textbf{Insider}} \\ 

    & \rotatebox{90}{Cloud}
    & \rotatebox{90}{Standard}
    & \rotatebox{90}{Custom}
    & \rotatebox{90}{ISL}
    & \rotatebox{90}{S2G}
    & \rotatebox{90}{G2S}
    & \rotatebox{90}{S2U}
    & \rotatebox{90}{EPS}
    & \rotatebox{90}{ADCS}
    & \rotatebox{90}{COMMS}
    & \rotatebox{90}{TT\&C}
    & \rotatebox{90}{TCS}
    & \rotatebox{90}{PCS}
    & \rotatebox{90}{C\&DHS}
    % & \rotatebox{90}{SDR}
    & \rotatebox{90}{}
    & \rotatebox{90}{}
    & \rotatebox{90}{}
    % & \rotatebox{90}{AC19}
    % & \rotatebox{90}{AC20}
    \\
% \textbf{Table Head} & Table column subhead & Subhead & Subhead \\
 \hline

Our paper & 
    \cmark & % Cloud
    \cmark & % Standard
    \cmark & % Custom
    \cmark & % ISL
    \cmark & % S2G
    \cmark & % G2S
    \cmark & % S2U
    \cmark & % EPS
    \cmark & % ADCS
    \cmark & % COMMS
    \cmark & % TT&C
    \cmark & % TCS
    \cmark & % PCS
    \cmark & % C&DHS
    % \cmark & % SDR
    \cmark & % USER
    \cmark & % HISTORY
    \cmark & % AML
    \cmark   % TAXONOMY
    \\ 
\hline

% lessons from 60 years of spaceflight
\cite{pavur2022building} & 
    \xmark & % Cloud
    \cmark & % Standard
    \xmark & % Custom
    \xmark & % ISL
    \cmark & % S2G
    \cmark & % G2S
    \xmark & % S2U
    \xmark & % EPS
    \xmark & % ADCS
    \xmark & % COMMS
    \xmark & % TT&C
    \xmark & % TCS
    \xmark & % PCS
    \xmark & % C&DHS
    % \xmark & % SDR
    \xmark & % USER
    \cmark & % HISTORY
    \xmark & % AML
    \xmark   % TAXONOMY
\\ 

% SOK: Building a Launchpad for Impactful Satellite Cyber-Security Research
\cite{pavur2020sok} & 
    \xmark & % Cloud
    \cmark & % Standard
    \xmark & % Custom
    \xmark & % ISL
    \cmark & % S2G
    \cmark & % G2S
    \xmark & % S2U
    \bcancel{\Checkmark} & % EPS
    \bcancel{\Checkmark} & % ADCS
    \xmark & % COMMS
    \bcancel{\Checkmark} & % TT&C
    \bcancel{\Checkmark} & % TCS
    \bcancel{\Checkmark} & % PCS
    \bcancel{\Checkmark} & % C&DHS
    % \xmark & % SDR
    \xmark & % USER
    \cmark & % HISTORY
    \xmark & % AML
    \bcancel{\Checkmark}   % TAXONOMY
\\ 

% Cyber security in New Space
\cite{manulis2021cyber} & 
    \cmark & % Cloud
    \cmark & % Standard
    \cmark & % Custom
    \cmark & % ISL
    \cmark & % S2G
    \cmark & % G2S
    \cmark & % S2U
    \xmark & % EPS
    \bcancel{\Checkmark} & % ADCS
    \cmark & % COMMS
    \cmark & % TT&C
    \xmark & % TCS
    \xmark & % PCS
    \bcancel{\Checkmark} & % C&DHS
    % \cmark & % SDR
    \cmark & % USER
    \cmark & % HISTORY
    \xmark & % AML
    \xmark   % TAXONOMY
\\ 

% Satellite-based communications security: A survey of threats, solutions, and research challenges.
\cite{tedeschi2022satellite} & 
    \xmark & % Cloud
    \cmark & % Standard
    \xmark & % Custom
    \cmark & % ISL
    \cmark & % S2G
    \cmark & % G2S
    \cmark & % S2U
    \xmark & % EPS
    \xmark & % ADCS
    \xmark & % COMMS
    \xmark & % TT&C
    \xmark & % TCS
    \xmark & % PCS
    \xmark & % C&DHS
    % \bcancel{\Checkmark} & % SDR
    \cmark & % USER
    \cmark & % HISTORY
    \xmark & % AML
    \bcancel{\Checkmark}   % TAXONOMY
\\ 

% Survey on security issues of routing and anomaly detection for space information networks
\cite{zhuo2021survey} & 
    \xmark & % Cloud
    \cmark & % Standard
    \xmark & % Custom
    \cmark & % ISL
    \cmark & % S2G
    \cmark & % G2S
    \cmark & % S2U
    \xmark & % EPS
    \xmark & % ADCS
    \xmark & % COMMS
    \xmark & % TT&C
    \xmark & % TCS
    \xmark & % PCS
    \xmark & % C&DHS
    % \xmark & % SDR
    \cmark & % USER
    \cmark & % HISTORY
    \xmark & % AML
    \bcancel{\Checkmark}   % TAXONOMY
\\ 

% Space Odyssey: An Experimental Software Security Analysis of Satellites.
\cite{willbold2023space} & 
    \cmark & % Cloud
    \cmark & % Standard
    \cmark & % Custom
    \cmark & % ISL
    \cmark & % S2G
    \cmark & % G2S
    \cmark & % S2U
    \cmark & % EPS
    \cmark & % ADCS
    \cmark & % COMMS
    \cmark & % TT&C
    \xmark & % TCS
    \xmark & % PCS
    \cmark & % C&DHS
    % \cmark & % SDR
    \cmark & % USER
    \xmark & % HISTORY
    \xmark & % AML
    \bcancel{\Checkmark}  % TAXONOMY
\\

% Characterizing Cyber Attacks against Space Systems/Infrastructures with Missing Data
\cite{ear2023characterizing,ear2025characterizing} &
    \xmark & % Cloud
    \bcancel{\Checkmark} & % Standard
    \xmark & % Custom
    \bcancel{\Checkmark} & % ISL
    \bcancel{\Checkmark} & % S2G
    \bcancel{\Checkmark} & % G2S
    \bcancel{\Checkmark} & % S2U
    \bcancel{\Checkmark} & % EPS
    \bcancel{\Checkmark} & % ADCS
    \bcancel{\Checkmark} & % COMMS
    \bcancel{\Checkmark} & % TT&C
    \bcancel{\Checkmark} & % TCS
    \bcancel{\Checkmark} & % PCS
    \bcancel{\Checkmark} & % C&DHS
    % \xmark & % SDR
    \bcancel{\Checkmark} & % USER
    \cmark & % HISTORY
    \xmark & % AML
    \bcancel{\Checkmark} % TAXONOMY
\\

% SoK: Space Infrastructures: Vulnerabilities, Attacks, and Defenses
\cite{remy2025sok} &
    \xmark & % Cloud
    \cmark & % Standard
    \xmark & % Custom
    \bcancel{\Checkmark} & % ISL
    \bcancel{\Checkmark} & % S2G
    \bcancel{\Checkmark} & % G2S
    \bcancel{\Checkmark} & % S2U
    \bcancel{\Checkmark} & % EPS
    \bcancel{\Checkmark} & % ADCS
    \bcancel{\Checkmark} & % COMMS
    \bcancel{\Checkmark} & % TT&C
    \bcancel{\Checkmark} & % TCS
    \bcancel{\Checkmark} & % PCS
    \bcancel{\Checkmark} & % C&DHS
    % \xmark & % SDR
    \cmark & % USER
    \xmark & % HISTORY
    \xmark & % AML
    \bcancel{\Checkmark} % TAXONOMY
\\

% A Survey on Satellite Communication System Security
\cite{kang2024survey} &
    \bcancel{\Checkmark} & % Cloud
    \cmark & % Standard
    \xmark & % Custom
    \bcancel{\Checkmark} & % ISL
    \cmark & % S2G
    \cmark & % G2S
    \cmark & % S2U
    \bcancel{\Checkmark} & % EPS
    \bcancel{\Checkmark} & % ADCS
    \cmark & % COMMS
    \bcancel{\Checkmark} & % TT&C
    \xmark & % TCS
    \xmark & % PCS
    \bcancel{\Checkmark} & % C&DHS
    % \xmark & % SDR
    \cmark & % USER
    \bcancel{\Checkmark} & % HISTORY
    \xmark & % AML
    \bcancel{\Checkmark} % TAXONOMY
\\
\cite{planta2026satbleed} & 
    \xmark{}  & % Cloud
    \cmark{} & % Standard
    \bcancel{\Checkmark} & % Custom
    \bcancel{\Checkmark} & % ISL
    \cmark{} & % S2G
    \cmark{} & % G2S
    \xmark{} & % S2U
    \bcancel{\Checkmark} & % EPS
    \bcancel{\Checkmark} & % ADCS
    \cmark{} & % COMMS
    \cmark{} & % TT&C
    \xmark{} & % TCS
    \xmark{} & % PCS
    \bcancel{\Checkmark} & % C&DHS
    \xmark{} & % USER
    \xmark{} & % HISTORY
    \xmark{} & % AML
    \bcancel{\Checkmark} % TAXONOMY
\\
\hline
\end{tabular}
\end{adjustbox}
\end{table*}

As shown in Table~\ref{tab:related_work}, prior work addresses important parts of the satellite and space-cybersecurity landscape, but usually focuses on specific attack surfaces, system layers, or analytical goals.
Pavur and Martinovic provide a valuable chronology of satellite cyber incidents and a cross-domain threat-modeling baseline~\cite{pavur2022building,pavur2020sok}.
Manulis et al.~\cite{manulis2021cyber} examine threats and enabling technologies in New Space, while Tedeschi et al.~\cite{tedeschi2022satellite}, Zhuo et al.~\cite{zhuo2021survey}, and Kang et al.~\cite{kang2024survey} focus primarily on satellite communication, networking, and communication-system security.
Experimental studies such as Willbold et al.~\cite{willbold2023space} and Planta et al.~\cite{planta2026satbleed} demonstrate practical exploitation paths affecting real satellite platforms and COTS communication modules, but do not aim to provide an incident-driven, system-level taxonomy spanning historical incidents, segments, link directions, subsystem exposure, adversary capabilities, and operational impacts.

Recent incident-driven and system-level studies provide additional foundations.
Ear et al.~\cite{ear2023characterizing,ear2025characterizing} analyze real-world space-cyber incidents under missing information and use ATT\&CK/SPARTA-based reasoning to reconstruct plausible attack-chain variants.
A direct row-level comparison with Ear et al.~\cite{ear2025characterizing} was not possible because their complete incident-level corpus was not publicly available at the time of our analysis.
While their work extrapolates missing attack-chain details into plausible USCKCs, our corpus preserves source-level uncertainty by distinguishing reported evidence, supported high-level inference, and unknown information.
Remy et al.~\cite{remy2025sok} provide a broad systematization of space-cybersecurity literature and model spacecraft functions through a general infrastructure/module-flow abstraction.
In contrast, our taxonomy adopts the conventional satellite-engineering subsystem framing and explicitly connects incidents and adversarial techniques to EPS, ADCS, COMMS, TT\&C, TCS, PCS, C\&DHS, payload functions, and mission-level impacts.

Building on these foundations, this paper addresses the lack of an incident-grounded framework that connects satellite attacks across three levels of analysis: the technical system level, including segments, communication links, and spacecraft subsystems; the adversarial lifecycle level, including tactics, techniques, and capabilities; and the empirical level, including historical incidents and longitudinal trends.
The proposed taxonomy integrates RF, cyber, physical, deception-oriented, and adversarial machine learning threats within a single lifecycle-oriented framework.
AML-related attack surfaces are treated as emerging and prospective threats, grounded in recent research and the increasing use of ML-enabled satellite functions, while remaining distinct from historically confirmed incident categories.
\subsection{Contributions}

The main contributions of this paper are as follows:

\textit{(i)} We present a reproducible methodology for constructing a satellite cybersecurity incident corpus and deriving a satellite-specific threat taxonomy. 
The methodology defines source selection, inclusion and exclusion criteria, deduplication rules, metadata fields, uncertainty handling, and the mapping process used to connect incidents to segments, links, adversary capabilities, lifecycle phases, and operational impacts.

\textit{(ii)} We compile and release a source-backed dataset of more than 200 publicly reported satellite cyber and electronic-warfare incidents spanning 1962--2026, including more than 80 incidents from 2019--2026.
The dataset extends prior open incident collections~\cite{spacesecurity2020,pavur2022building,pavur2020sok} and complements recent missing-data work~\cite{ear2025characterizing} by providing structured metadata, source references, uncertainty indicators, taxonomy mappings, and subsystem-level interpretation where supported by public evidence.

\textit{(iii)} We introduce a satellite-specific taxonomy inspired by MITRE ATT\&CK and selectively aligned with SPARTA where applicable.
The taxonomy integrates ground-segment intrusions, RF interference, GNSS disruption, on-orbit operations, deception, and AML-related attack surfaces within a single lifecycle-oriented framework.

\textit{(iv)} We use the dataset to perform a longitudinal empirical analysis of satellite threats across RF, cyber, physical, and deception-oriented domains.
The analysis reveals major shifts in adversary behavior, including the growing operationalization of GNSS interference, increasing reliance on ground-segment compromise, and the emergence of deception-driven counterspace strategies.

\textit{(v)} We demonstrate the practical utility of the taxonomy through case studies of the 2022 Viasat KA-SAT cyberattack and a simulation-based ICARUS constellation-scale denial-of-service scenario.
These case studies show how the framework connects observed or modeled adversarial behavior to tactics, techniques, segments, links, and mission-level impacts.

Together, these contributions extend prior survey-based and incident-driven work by providing three key elements.
First, we introduce a reproducible, incident-driven methodology for constructing a satellite-specific threat taxonomy without inferring unsupported missing details.
Second, we construct and release an extended longitudinal dataset of satellite cyber and electronic-warfare incidents spanning multiple decades, enabling empirical analysis of threat evolution and allowing future work to perform direct incident-level comparisons, validations, and extensions.
Third, we derive a subsystem-aware operational framework that connects adversarial behavior across system segments, communication links, lifecycle tactics, and mission-level impacts.

The remainder of the paper is organized as follows.
Section~\ref{sec:methodology} describes the methodology used to construct the incident corpus, including source selection, inclusion and exclusion criteria, deduplication, uncertainty handling, and coding.
Section~\ref{app:threat_analysis} introduces the threat-analysis framework and terminology used throughout the paper.
Sections~\ref{sec:space_env}--\ref{sec:subsystem} define the satellite ecosystem, system assets, trust boundaries, and spacecraft subsystems.
Section~\ref{sec:satellite_vuln} identifies the vulnerabilities that expose these assets.
Section~\ref{sec:AdversariesThreat} defines the adversary categories and capabilities needed to exploit them.
Section~\ref{sec:Satellite_Attacks} uses the incident corpus to analyze how these threats appear historically.
Section~\ref{sec:MITRE} derives a satellite-specific lifecycle taxonomy from these observations.
Finally, Section~\ref{sec:demonstration} demonstrates how the taxonomy supports attack modeling and mitigation analysis through case studies.
\section{\label{sec:methodology}Methodology}

We constructed the incident corpus using a source-driven, iterative methodology over a 4-year period.
Candidate incidents were identified from established sources in space security and satellite cybersecurity, including annual \emph{Space Threat Assessment} reports, prior academic studies, historical incident collections, government publications, company reports, news articles, and public online reporting.
The corpus was then expanded as new satellite-related incidents were publicly reported or identified through additional sources.
Whenever possible, candidate incidents were cross-checked against additional references to verify the event description, affected target, attack vector, operational effect, and relevant context.

\begin{figure}[!ht]
\centering
\includegraphics[width=\columnwidth]{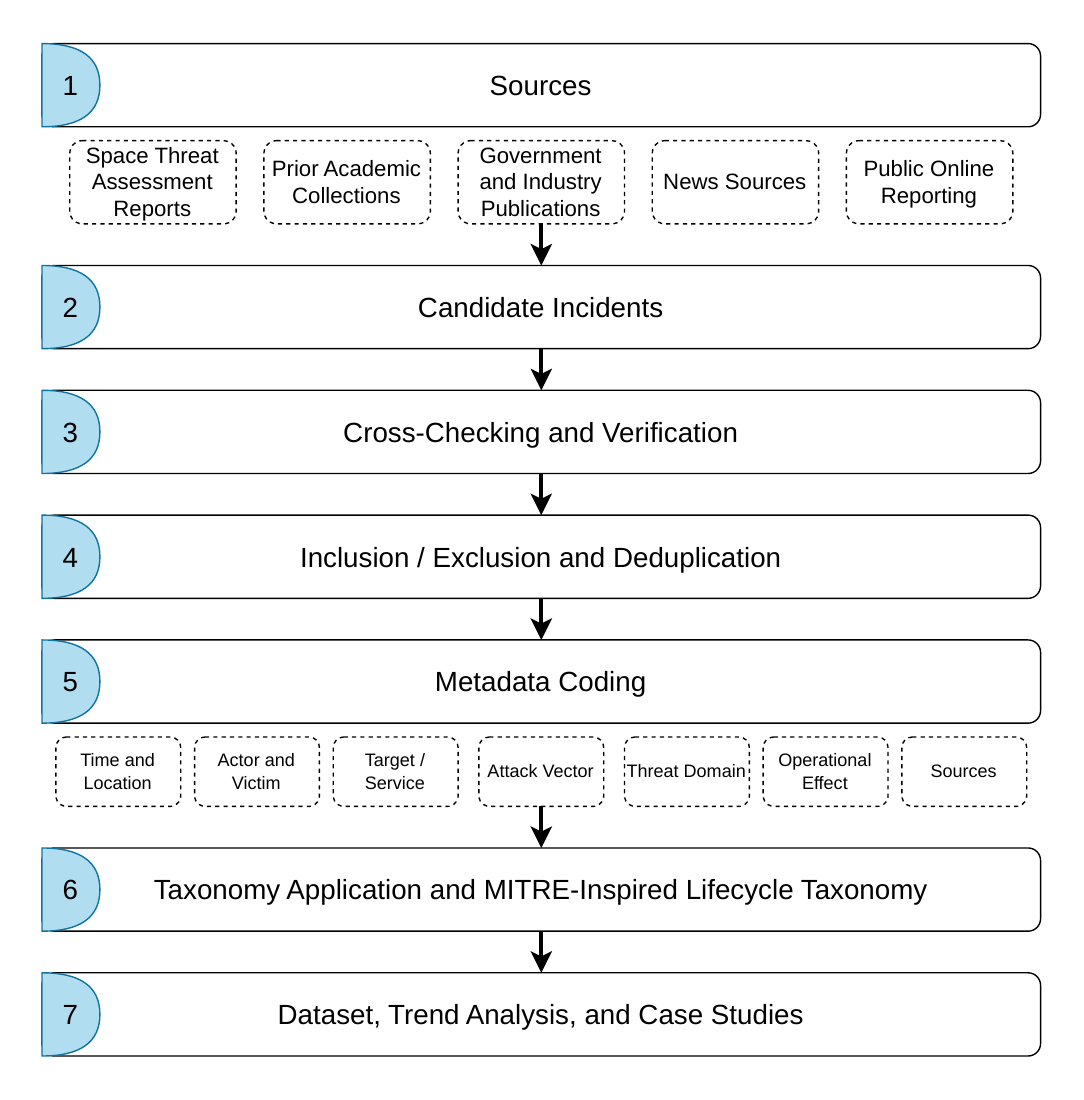}
\caption{Overview of the methodology used to construct the incident corpus and derive the satellite-specific taxonomy. Candidate incidents were extracted from established sources, cross-checked, filtered, deduplicated, coded with structured metadata, and used to support taxonomy application, trend analysis, and case studies.}
\label{fig:methodology_overview}
\end{figure}

Figure~\ref{fig:methodology_overview} summarizes the workflow.
The methodology consisted of four main stages.
First, we extracted candidate incidents from threat-assessment reports and prior research.
Second, we expanded and verified these incidents using external sources, including government or industry publications, company statements, technical reports, news articles, and public online reporting.
Third, we applied inclusion, exclusion, and deduplication rules to determine whether each candidate represented a distinct incident.
Finally, we coded each included incident using structured metadata and applied the proposed satellite-specific taxonomy, including system segment, communication link, adversary capability, threat modality, operational impact, and MITRE-inspired lifecycle interpretation.

\subsection{\label{sec:relation_existing_datasets}Relation to Existing Incident Collections}

Several prior resources provide foundations for documenting space- and satellite-related security incidents.
The Space Attacks Open Database provides an open historical collection of space-related attacks~\cite{spacesecurity2020}.
Fritz provides an early collection of satellite-hacking cases~\cite{fritz2013satellite}, while Manulis et al.~\cite{manulis2021cyber} and Pavur and Martinovic~\cite{pavur2020sok,pavur2022building} analyze satellite cybersecurity threats and historical attack trends.
These works provide valuable baselines, but their coverage focuses primarily on earlier incidents and does not provide a substantially updated post-2019 corpus of satellite cyber and electronic-warfare incidents.

Recent works address related but different objectives.
Ear et al.~\cite{ear2025characterizing} study missing information in space-cyber incidents and extrapolate plausible attack-chain variants using ATT\&CK and SPARTA; however, their complete incident-level corpus was not publicly available at the time of our analysis, preventing direct row-level comparison.
Other resources, such as the JHU space-security risk database and the ENISA Space Threat Landscape dataset, focus primarily on risks, controls, threat categories, or lifecycle mappings rather than historical attack incidents~\cite{falco2021security,enisa2025space}.

Our corpus differs from these resources in four main ways.
First, it substantially extends coverage of publicly reported satellite cyber and electronic-warfare incidents after 2019.
Second, it provides an inspectable incident-level dataset with structured metadata, source references, uncertainty indicators, attack categories, affected segments, communication links, and taxonomy mappings.
Third, each incident is mapped to a MITRE-inspired lifecycle interpretation rather than forced into an exhaustive official ATT\&CK or SPARTA mapping.
Fourth, it preserves uncertainty in suspected or contested cases rather than reconstructing unsupported attack-chain details as confirmed historical evidence.

\subsection{\label{sec:incident_collection}Incident Corpus Construction}

The annual \emph{Space Threat Assessment} reports published between 2020 and 2025 were used as primary sources for recent satellite-related cyber, electronic-warfare, RF interference, GNSS disruption, counterspace, and ground-infrastructure incidents~\cite{harrison2020space,harrison2021space,harrison2022space,bingen2023space,swope2024space,swope2025space}.
Prior academic studies and historical incident collections were used as seed sources for earlier and previously documented incidents~\cite{pavur2020sok,pavur2022building,manulis2021cyber,ear2025characterizing}.

Incidents reported in prior work were not copied directly into the corpus.
Each candidate incident was reviewed and, where possible, cross-checked against additional sources to verify the reported time period, affected country or operator, target system or service, attack vector, and operational effect.
When public reports described similar events, we compared the reported time, location, target, affected service, attack vector, suspected actor, and operational impact to determine whether they referred to the same incident or to separate events.

\subsection{\label{sec:inclusion_exclusion}Inclusion and Exclusion Criteria}

The corpus includes publicly reported events that affected, targeted, or plausibly threatened satellite systems, satellite-enabled services, or their supporting ground, communication, space, or user segments.
We included confirmed attacks as well as suspected or ambiguous incidents when public reporting described a clear satellite connection and documented operational or security relevance.

The corpus covers cyber operations, RF interference, GNSS jamming and spoofing, satellite communication disruption, signal hijacking, unauthorized access, malware, data exfiltration, terminal exploitation, supply-chain compromise, physical or kinetic counterspace activity, and other operational attacks affecting satellite services.
Electronic-warfare and counterspace events were included when they affected satellite communication, navigation, sensing, control, or service availability, even when they did not involve conventional computer network intrusion.

The main corpus focuses on real-world publicly reported incidents.
Research demonstrations and experimental attacks were included only when conducted against a real satellite, an operational satellite service, real ground infrastructure, or an operationally realistic setting relevant to satellite security risk.
Purely theoretical attacks, simulations without operational grounding, ordinary technical failures, space-weather effects, and accidental outages were excluded unless public reporting described an adversarial, suspicious, or security-relevant component.

\subsection{\label{sec:deduplication}Unit of Analysis and Deduplication}

The unit of analysis is a distinct satellite-related security incident or incident episode.
Repeated disruptions were not counted separately when public reporting indicated that they were part of the same campaign, affected the same area or service, occurred within a close time window, and followed a similar attack pattern.
For example, a surge of GNSS spoofing affecting many flights over the same region during the same period was coded as a single incident episode rather than as separate incidents for each affected flight.

Reports describing the same activity were merged.
Events were retained as separate incidents when they differed meaningfully in target, affected service, time window, attack vector, country or region, or operational effect.
This conservative approach may undercount repeated attacks, but it reduces the risk of inflating the corpus by counting multiple reports of the same activity as separate incidents.

\subsection{\label{sec:metadata_coding}Metadata Fields and Coding Process}

Each included incident was coded using a structured metadata schema.
The schema captures the incident year or time period, affected country or region, suspected or reported attacker, victim or affected operator, target type, affected satellite service, attack vector, threat domain, operational effect, and supporting references.
Where public evidence supported additional detail, incidents were also mapped to affected segment, communication link direction, adversary capability tier, lifecycle phase, satellite-specific technique, and operational impact.

These lifecycle fields follow a MITRE-inspired structure, but they should not be interpreted as an exhaustive row-level mapping to official MITRE ATT\&CK or SPARTA technique identifiers.
Fields were assigned only when supported by cited sources or clearly marked high-level inference.
When the available information supported only a broader category, the incident was coded at that level rather than assigned an unsupported, more specific label.

Spacecraft subsystems such as COMMS, TT\&C, ADCS, EPS, C\&DHS, TCS, PCS, and payload are used as part of the satellite-system taxonomy and operational interpretation, rather than as mandatory incident-level metadata fields.
Because public incident reports rarely provide enough detail to determine the affected internal spacecraft subsystem with confidence, subsystem-level interpretation was applied only where supported by evidence.

\subsection{\label{sec:missing_data}Treatment of Missing and Uncertain Information}

Because the corpus is based on public reporting, not all incidents contain the same level of technical detail.
We therefore adopted a conservative coding strategy that distinguishes between reported evidence, supported inference, and unknown information.

Our objective was not to reconstruct complete attack chains from incomplete reports.
Instead, we coded information at the highest level of specificity supported by the available evidence.
When a field was directly described by a source, it was coded as reported evidence.
When a high-level lifecycle step was strongly implied by the nature of a confirmed or well-supported attack, it could be coded as a supported inference and marked explicitly using the ``Inferred:'' qualifier in the supplementary dataset.
However, we did not infer detailed techniques, tools, infrastructure, attacker identity, attribution, or subsystem effects unless supported by the incident description, cited sources, or clear operational reasoning.

For suspected or contested incidents, disputed attack causality, attribution, and intermediate steps were treated as uncertain rather than confirmed historical evidence.
When sources disagreed, we retained the conflict in the source notes and coded the field according to the most conservative interpretation supported by the evidence.
This differs from prior work that reconstructs hypothetical but plausible attack chains from incomplete reports using frameworks such as ATT\&CK and SPARTA~\cite{ear2025characterizing}; our goal is to preserve the distinction between reported evidence, supported high-level inference, and unknown information.

\subsection{\label{sec:mapping_process}Taxonomy Application and Consistency Checking}

After coding the incident metadata, each incident was used to populate and evaluate the proposed satellite-specific taxonomy.
The taxonomy application considered communication link, adversary capability type, attacker type, victim type, affected country or region, source evidence, operational description, and lifecycle interpretation.
The taxonomy follows the general logic of adversarial lifecycle frameworks such as MITRE ATT\&CK, but is adapted to satellite-specific attack patterns.
SPARTA was reviewed and selectively incorporated where its space-specific concepts aligned with the evidence, but the supplementary dataset does not claim exhaustive row-level SPARTA mapping.

The mapping followed five rules.
First, lifecycle fields were assigned only when supported by the incident description, cited sources, or clearly marked high-level inference.
Second, when the evidence supported only a broad adversarial phase, the incident was mapped at that broader level rather than forced into a specific unsupported technique.
Third, inferred fields were marked using the ``Inferred:'' qualifier and were used only for broad lifecycle phases or prerequisites, not for detailed tools, attacker identity, attribution, or subsystem effects.
Fourth, satellite-specific behaviors such as RF interference, GNSS disruption, broadcast hijacking, proximity operations, subsystem exposure, and constellation-scale effects were represented using the proposed satellite-specific taxonomy.
Fifth, SPARTA was considered where a space-specific technique, asset, or operational behavior aligned with the evidence, but the mapping was not forced when the public record did not support it.

To improve consistency, the final dataset was reviewed for duplicate entries, inconsistent terminology, unsupported classifications, missing references, and inconsistent use of inferred fields.
Cases involving incomplete or ambiguous information were retained only when they satisfied the inclusion criteria and were marked or described accordingly.
\section{Threat Analysis} \label{app:threat_analysis}

\begin{figure}[!ht]
\centering
  \includegraphics[width=0.47\textwidth]{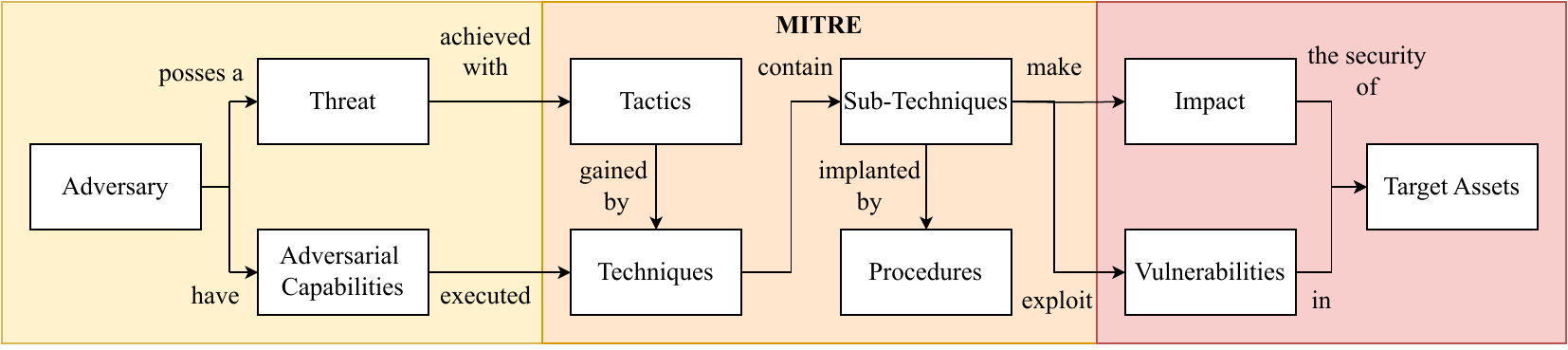}
  \caption{Threat analysis framework based on the NIST Ontology for Modeling Enterprise Security.}
  \label{fig:analysis_graph}
\end{figure}

Our security analysis is informed by the NIST Ontology for Modeling Enterprise Security, which provides a structured framework for reasoning about adversaries, vulnerabilities, assets, and attack processes. 
Figure~\ref{fig:analysis_graph} illustrates the core entities and relationships that underpin this framework and guide the development of the proposed taxonomy.

We adopt the following terminology throughout the paper:

\noindent\textit{\textbf{Adversary}}: An individual, organization, or state actor capable of intentionally compromising the security or safety of a satellite system. Adversaries differ in resources, objectives, and operational capabilities.

\noindent\textit{\textbf{Threat}}: A potential or realized action by an adversary that may compromise system security properties, including confidentiality, integrity, and availability.

\noindent\textit{\textbf{Adversarial Capabilities}}: The resources and competencies available to threat actors. In this work, we distinguish among access-based, positional, knowledge-based, and material capabilities.

\noindent\textit{\textbf{Tactics}}: The adversary's operational objectives, reflecting the intent behind an action within the attack lifecycle.

\noindent\textit{\textbf{Techniques}}: The methods adversaries use to achieve tactical goals.

\noindent\textit{\textbf{Sub-Techniques}}: More specific forms of adversarial behavior used to realize techniques; a technique may include multiple sub-techniques~\cite{strom2019}.

\noindent\textit{\textbf{Procedures}}: Concrete implementations of techniques or sub-techniques, describing how attacks are executed in practice across academic and operational contexts.

\noindent\textit{\textbf{Vulnerability}}: A weakness in an asset, protocol, or operational process that can be exploited by an adversary.

\noindent\textit{\textbf{Impact}}: The operational or security consequences resulting from a successful attack.

\noindent\textit{\textbf{Target Assets}}: The systems, components, data, services, and operational processes within satellite ecosystems that require protection.

Building on this framework, we develop a satellite-specific threat model and taxonomy by following these steps. 
First, we examine the satellite ecosystem across the space, ground, communication, and user segments to identify architectural exposures and targetable assets (Sections~\ref{sec:space_env} and~\ref{sec:subsystem}). 
Second, we characterize vulnerabilities associated with these assets and evaluate how adversaries may exploit them (Section~\ref{sec:satellite_vuln}). 
Third, we categorize adversaries by capabilities and operational constraints, organizing them into six tiers based on skill level, potential impact, and available resources (Section~\ref{sec:AdversariesThreat}). 

We then provide a concise historical overview of satellite attacks to incorporate empirical evidence into the analysis (Section~\ref{sec:Satellite_Attacks}). 
Finally, by mapping adversaries, capabilities, target assets, vulnerabilities, and documented incidents from both academic and operational sources, we construct a satellite-specific lifecycle taxonomy inspired by MITRE ATT\&CK and selectively aligned with SPARTA where applicable. This taxonomy is presented in Section~\ref{sec:MITRE} and evaluated through detailed case studies in Section~\ref{sec:demonstration}.
\section{Satellite Ecosystem} \label{sec:space_env}

This section presents an operational view of the satellite ecosystem across the ground, space, communication, and user segments, with emphasis on architectural exposures and attack surfaces relevant to adversarial activity. 
Rather than providing a purely descriptive overview, we focus on structural dependencies and interfaces that influence how threats propagate across satellite systems and inform the taxonomy developed later in the paper.

\subsection{Ground Segment}

The ground segment comprises terrestrial infrastructure responsible for spacecraft operations and data exchange between space and Earth. 
It typically includes ground stations, mission control centers, data-processing facilities, and remote terminals~\cite{guest2017telemetry,manulis2021cyber,perez2019signal}. 
Because these components are often connected to enterprise IT networks and external service providers, they represent one of the most exposed and frequently targeted attack surfaces in modern satellite operations.

\textbf{Ground stations.}  
Ground stations (GS) provide the radio-frequency (RF) interface to the space segment for telemetry, tracking, and command (TT\&C), as well as payload data, using antennas, feeds/waveguides, transmitters/receivers, and baseband systems~\cite{perez2019signal,guest2017telemetry,hewitson2020}. 
Mission-specific requirements determine frequency selection, tracking capabilities, and supporting infrastructure such as radios, modems, signal processing chains, and control systems~\cite{SOBOLEWSKI2003277}. 
Common GS models include:

\begin{itemize}
  \item \textbf{Traditional GS:} Dedicated physical installations operated directly by mission stakeholders~\cite{hewitson2020}.
  \item \textbf{Cloud-based GS (GSaaS):} Commercial services (e.g., AWS Ground Station, Azure Orbital) providing on-demand global coverage integrated with cloud workflows~\cite{baker2019amazon,werner2022}.
  \item \textbf{Custom GS:} Systems assembled from COTS components tailored to mission needs. Experimental studies demonstrate that such setups can interact with satellite links and may be exploited if misconfigured or operated maliciously~\cite{willbold2023space,SOBOLEWSKI2003277}.
\end{itemize}

\textbf{Mission control.}  
Mission operations centers coordinate planning and execution, uplink telecommands, and monitor telemetry, orbit state, and spacecraft health~\cite{guest2017telemetry,manulis2021cyber}. 
Because all command and telemetry traffic passes through ground infrastructure, compromise at this layer may enable full-system takeover or stealthy operational manipulation.

\textbf{Data processing.}  
Payload products and housekeeping telemetry are received, processed, calibrated, and distributed to users and archival systems. 
These workflows often rely on heterogeneous software stacks and external infrastructure, expanding the attack surface through supply-chain dependencies~\cite{SOBOLEWSKI2003277,perez2019signal,manulis2021cyber}.

\textbf{Remote terminals.}  
Distributed antennas and teleports extend coverage and operational flexibility but introduce additional exposure points through geographically dispersed infrastructure and network dependencies~\cite{andres2022,guest2017telemetry}.

\subsection{Space Segment}\label{sec:payloads}

The space segment consists of satellites operating across major orbital regimes:

\begin{itemize}
  \item \textbf{Geostationary Earth Orbit (GEO)} at approximately 36,000 km~\cite{li2014geostationary}.
  \item \textbf{Medium Earth Orbit (MEO)} between roughly 2,000 and 36,000 km.
  \item \textbf{Low Earth Orbit (LEO)} between approximately 160 and 2,000 km~\cite{vatalaro1995analysis,hiriart2009comparative}.
\end{itemize}

Satellite constellations—particularly in LEO—enable global services including communications, Internet connectivity, and Earth observation~\cite{curzi2020large}. 
This work focuses primarily on LEO systems due to their rapid growth, reliance on commercial off-the-shelf (COTS) hardware, and distributed architectures, which expand the attack surface compared to traditional single-satellite missions~\cite{karvinen2015using}.

A typical satellite comprises multiple tightly integrated subsystems, including electrical power (EPS), attitude determination and control (ADCS), communications (COMMS), telemetry tracking and command (TT\&C), thermal control (TCS), propulsion (PCS), and command \& data handling (C\&DHS). 
These subsystems support spacecraft operation while presenting distinct attack surfaces due to differing trust boundaries, communication pathways, and dependencies.

In addition to platform subsystems, satellites carry mission-specific payloads supporting applications such as communications, Earth observation, weather monitoring, navigation, and scientific research. 
Because payload architectures often integrate third-party software, onboard processing pipelines, and external data feeds, they introduce additional risks not present in core platform systems.

A detailed subsystem-level analysis is provided in Section~\ref{sec:subsystem}. 
These components communicate over internal spacecraft buses implemented in linear or star architectures~\cite{addaim2010design}, which may serve as lateral-movement pathways under adversarial control.

\subsection{Communication Segment}\label{subsec:signal}

\begin{figure*}[t]
\centering
\includegraphics[width=0.87\textwidth]{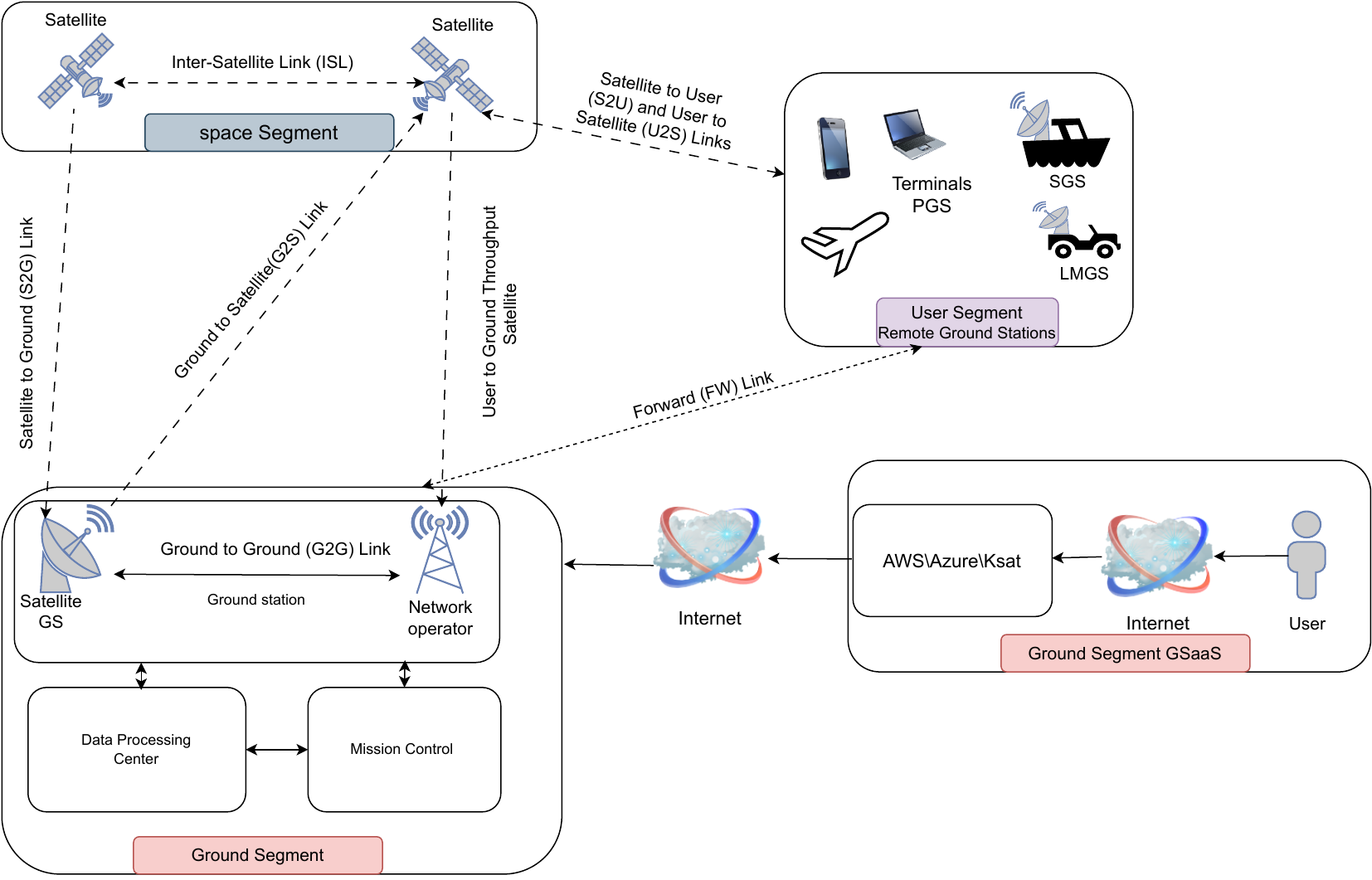}
\caption{Overview of satellite communication (SATCOM) segments and user interfaces.}
\label{fig:comm_fig}
\end{figure*}

The communication segment interconnects space, ground, and user components through diverse RF and networked links (Figure~\ref{fig:comm_fig}). 
These interfaces represent critical operational dependencies and common targets for RF-based and cyber-enabled attacks.

\emph{Space segment communication} includes inter-satellite links (ISL) and satellite-to-ground (S2G) links.

\emph{Ground segment communication} supports ground-to-satellite (G2S), ground-to-ground (G2G), satellite-to-ground (S2G), and satellite-to-user (S2U) communication flows.

\emph{User segment communication} includes terminals aboard aircraft, ships, and handheld devices enabling user-to-ground (U2G) and user-to-satellite (U2S) connectivity~\cite{tedeschi2022satellite,leyva2020leo}.

Modern constellations increasingly support direct-to-device architectures in which user equipment communicates directly with satellites without intermediary infrastructure~\cite{rainbow2023}. 
The adoption of 3GPP non-terrestrial network standards is further reshaping SATCOM architectures and expanding the protocol-level attack surface~\cite {jaffar2022ntn,lin20215g}.

\emph{Security considerations:}  
The diversity of communication pathways introduces vulnerabilities across physical, protocol, and application layers. 
RF interference, spoofing, protocol manipulation, and network-layer compromise may disrupt operations or enable adversarial control.

\subsection{User Segment}\label{subsec:user}

The user segment encompasses end-user systems interacting with satellite services, including GNSS receivers, satellite terminals, airborne and maritime platforms, and cloud-based service consumers. 
These systems represent indirect yet high-impact attack surfaces due to their scale, heterogeneity, and integration with terrestrial infrastructure.

Users may access satellite services through dedicated receivers (e.g., GNSS devices or satellite handsets) or through operator-managed terrestrial distribution networks. 
Many Earth-observation and intelligence services are delivered through APIs, cloud platforms, and web-based tools, extending exposure beyond traditional RF interfaces.

The user segment also includes distributed ground systems such as maritime ship ground stations (SGS), land mobile ground stations (LMGS), and portable personal ground stations (PGS)~\cite{ilvcev2017users}. 
These systems rely on diverse antenna technologies and configurations, which influence operational reliability and susceptibility to interference or misconfiguration.

Major providers such as Inmarsat, Iridium, and O3B deliver global voice, data, and broadband services through integrated space--ground--user ecosystems~\cite{ses-o3b,dvb2023}. 
As satellite services increasingly integrate with terrestrial networks and cloud infrastructure, the user segment plays a growing role in the overall threat landscape.

These segment-level dependencies define the external attack surfaces of satellite systems; the next section examines the internal spacecraft subsystems through which such attacks may propagate.
\section{\label{sec:subsystem}Satellite Subsystems}

Satellite architectures are typically divided into two functional domains: the \emph{payload} and the \emph{bus}.  
The bus comprises the core infrastructure and subsystems that sustain spacecraft operation, while the payload delivers mission-specific functionality~\cite{hassanien2020machine,nguyen2020communication}.  
From a security perspective, these components represent distinct trust boundaries and attack surfaces, each with different operational dependencies and adversarial implications.

The payload consists of onboard instruments, sensors, transponders, or processing systems designed to support the spacecraft's mission objectives, such as communications, Earth observation, navigation, weather monitoring, or scientific research.
Typical examples include communication transponders, software-defined radio (SDR) payloads, imaging sensors, GNSS receivers, synthetic-aperture radar (SAR) payloads, and specialized scientific instruments.
Unlike platform subsystems, which primarily sustain spacecraft operation, payloads often generate mission-specific data products and may require dedicated processing, storage, compression, scheduling, and high-rate downlink capabilities.

Modern payloads frequently incorporate onboard processing, third-party software, reconfigurable RF components, high-rate storage, and external data interfaces, which may introduce additional risks beyond those present in core spacecraft infrastructure~\cite{willbold2023space,nguyen2020communication}.
For example, imaging and SAR payloads may require onboard digitization, buffering, compression, and mission-data downlink, while SDR-based payloads may introduce software-defined waveform processing and reconfigurable communication functions.
As a result, compromise of the payload may affect not only the mission product itself, but also onboard processing, data storage, downlink scheduling, RF behavior, and interfaces between the payload and the spacecraft bus.

In some satellite architectures, the payload may include a dedicated payload data-handling unit or payload-side control interface that is separated from the main spacecraft bus.
This creates additional trust boundaries between payload operations and platform subsystems such as C\&DHS, COMMS, EPS, and ADCS.
From a security perspective, these interfaces are important because an adversary that compromises a payload-facing interface, mission-data path, or reconfigurable payload component may attempt to influence platform-level functions or use the payload as an entry point for lateral movement.

\begin{figure}[!ht]
\centering
  \includegraphics[width=0.7\columnwidth]{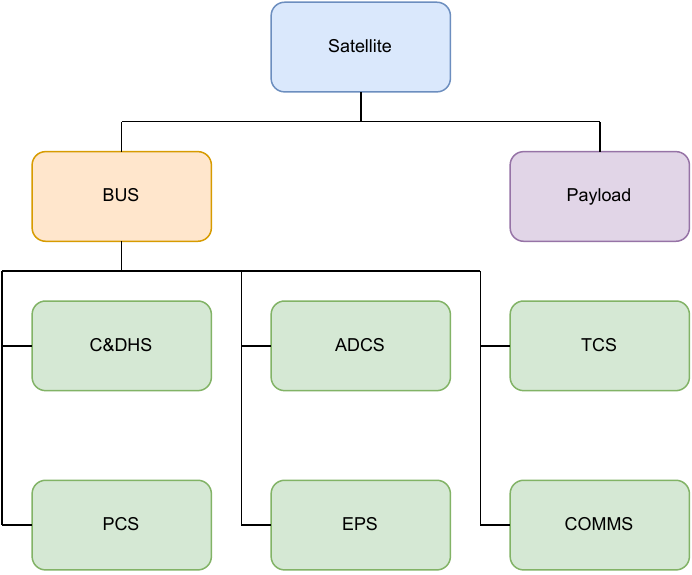}
  \caption{Satellite subsystems and their functional relationships.}
  \label{fig:subsys_fig}
\end{figure}

The spacecraft bus typically includes several tightly integrated subsystems—electrical power (EPS), attitude determination and control (ADCS), communications (COMMS), telemetry tracking and command (TT\&C), thermal control (TCS), propulsion (PCS), and command \& data handling (C\&DHS)—as illustrated in Figure~\ref{fig:subsys_fig}.  
Together, these subsystems support spacecraft operation while creating multiple potential entry points and lateral-movement pathways under adversarial control.

\subsection{\label{subsec:EPS}Electrical Power Subsystem (EPS)}

The EPS supplies energy to all spacecraft components by converting solar energy into electrical power and storing it in onboard batteries~\cite{nguyen2020communication}.  
It regulates power distribution and maintains operational continuity during eclipse phases.  
Because all subsystems depend on EPS availability, disruption or manipulation of power flows may propagate across the spacecraft, leading to degraded functionality or complete mission failure.

\subsection{Attitude Determination and Control Subsystem (ADCS)}

The ADCS maintains spacecraft orientation and stability using sensors such as star trackers, magnetometers, GNSS receivers, and gyroscopes, together with actuators including reaction wheels and magnetic torquers~\cite{hassanien2020machine}.  
Loss or manipulation of ADCS functionality may degrade pointing accuracy, disrupt communications or payload operations, and, in extreme cases, result in loss of mission control.

\subsection{Communication Subsystem (COMMS)}

The COMMS subsystem provides the spacecraft's RF communication capability, supporting downlink transmission and uplink reception between the satellite and external ground or space-based infrastructure~\cite{nguyen2020communication,hassanien2020machine}.
Depending on the mission architecture, COMMS may carry both platform-related traffic, such as command and housekeeping telemetry, and payload-related traffic, such as mission data or payload products.
It typically includes transmitters, receivers, antennas, modulation and coding equipment, and beacon systems used for spacecraft identification, link acquisition, and tracking.
Because COMMS interfaces directly with external infrastructure, it represents a primary exposure point for RF interference, spoofing, jamming, signal hijacking, and protocol-level attacks.

\subsection{Telemetry, Tracking, and Command (TT\&C)}

TT\&C refers to the spacecraft control and monitoring functions carried over the communication link.
While it is closely related to COMMS and may share the same RF hardware, TT\&C is functionally focused on platform control, spacecraft health monitoring, tracking, and command execution rather than on general mission data delivery~\cite{nguyen2020communication}.
It supports uplink reception of telecommands, downlink transmission of housekeeping telemetry, and tracking or ranging functions used to determine and maintain spacecraft operations.
The TT\&C data path is primarily associated with the spacecraft bus and platform subsystems, including C\&DHS, EPS, ADCS, COMMS, TCS, and PCS, although payload commanding or payload housekeeping may also be routed through TT\&C depending on the mission design.
Compromise of TT\&C may enable adversaries to disrupt operations, manipulate telemetry, interfere with tracking, or issue unauthorized commands, making it one of the most security-critical spacecraft functions.

\subsection{Thermal Control Subsystem (TCS)}

The TCS maintains spacecraft components within allowable operational and survival temperature ranges by dissipating, redistributing, or retaining heat generated by onboard equipment and the external space environment~\cite{guerra2018thermalanalysise,delcastillo2024s}.  
Thermal control failures or malfunctions may degrade subsystem performance, accelerate component degradation, or cause irreversible hardware damage, particularly on long-duration missions or in spacecraft that rely on temperature-sensitive payloads and commercial-off-the-shelf components.

\subsection{Propulsion Control Subsystem (PCS)}

The PCS manages propellant storage, distribution, valves, and thruster operations used for orbit maintenance, maneuvering, station keeping, deorbiting, and, where applicable, collision-avoidance maneuvers~\cite{Leomanni_2020}.  
Because propulsion directly affects spacecraft trajectory and orbital lifetime, compromise or disruption of PCS functions may result in trajectory deviation, loss of station-keeping capability, mission degradation, premature deorbiting, or increased collision risk in densely populated orbital regimes.

\subsection{Command \& Data Handling Subsystem (C\&DHS)}

The C\&DHS is the central command-processing and data-management function of the spacecraft and is often implemented by, or tightly coupled with, the onboard computer (OBC)~\cite{eger2008orion,nguyen2020communication}.  
It manages telemetry processing, telecommand execution, onboard scheduling, data storage, and internal communication across spacecraft subsystems.  
Because C\&DHS coordinates interactions among COMMS, TT\&C, EPS, ADCS, TCS, PCS, and payload systems, compromise of this subsystem may enable persistent access, unauthorized command execution, subsystem manipulation, or broader spacecraft control~\cite{willbold2023space}.

Subsystem interconnections are governed primarily by two architectural paradigms: star and linear bus configurations.

\subsection{\label{subsec:starc}Star Architecture}

\begin{figure}[!ht]
\centering
  \includegraphics[width=0.7\linewidth]{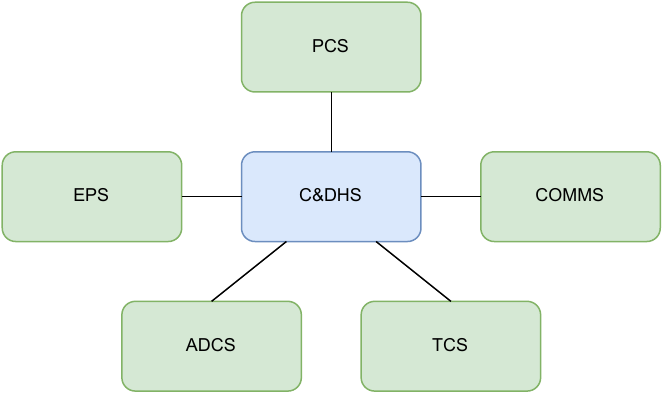}
  \caption{Star bus architecture.}
  \label{fig:starc_fig}
\end{figure}

In star architectures~\cite{Addaim10} (Figure~\ref{fig:starc_fig}), the C\&DHS maintains dedicated communication links with each subsystem.  
While this design enables isolation and direct control, it increases wiring complexity and introduces multiple interface points that may be targeted for disruption or exploitation.

\subsection{\label{subsec:bus}Linear Bus Architecture}

\begin{figure}[!ht]
\centering
  \includegraphics[width=0.8\linewidth]{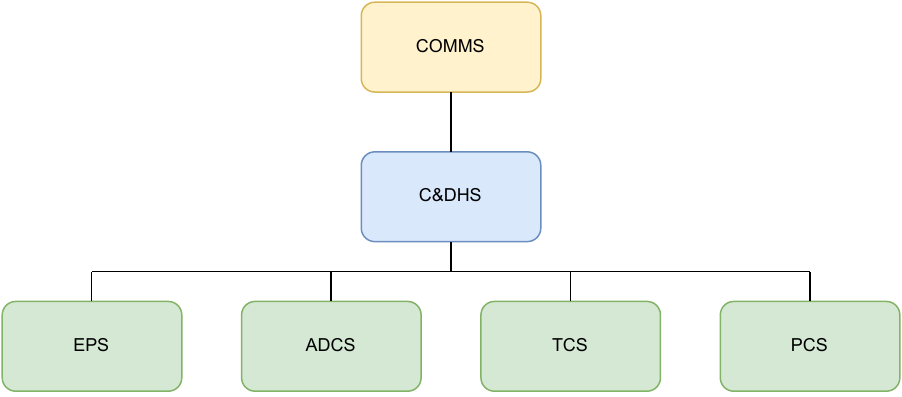}
  \caption{Linear bus architecture.}
  \label{fig:linkarc_fig}
\end{figure}

In linear bus architectures~\cite{Addaim10} (Figure~\ref{fig:linkarc_fig}), subsystems share a common communication backbone similar to an internal network.  
Protocols such as CAN, SPI, and I\textsuperscript{2}C are widely used for subsystem coordination.  
While this approach simplifies integration and reduces wiring complexity, shared-bus designs may weaken subsystem isolation and facilitate lateral movement once an adversary gains initial access through COMMS, TT\&C, a payload interface, or another bus-connected component~\cite{willbold2023space,planta2026satbleed}.

\noindent\textbf{Common onboard bus and interface types.}
Satellite subsystems commonly communicate through I\textsuperscript{2}C, SPI, UART, RS-485, CAN, SpaceWire, Ethernet, MIL-STD-1553, or PC/104-based interconnects.
These interfaces differ in bandwidth, reliability, addressing, and isolation properties, but many provide limited native authentication or privilege separation.
As a result, the selected bus architecture can influence whether compromise of COMMS, TT\&C, C\&DHS, or a payload interface remains localized or propagates across other spacecraft subsystems.

Having defined the spacecraft subsystems and internal communication architectures, we next analyze the vulnerabilities that expose these components to adversarial activity.
\section{Satellite Vulnerabilities} \label{sec:satellite_vuln}

This section analyzes the principal vulnerabilities affecting the satellite ecosystem, as introduced in Section~\ref{sec:space_env}, and the spacecraft subsystems described in Section~\ref{sec:subsystem}. 
Rather than treating vulnerabilities as isolated technical weaknesses, we consider them as operational exposures that adversaries may exploit across the ground, space, communication, user, and subsystem layers of satellite systems.

\subsection{Ground Segment Vulnerabilities}

The ground segment is often the most accessible and operationally exposed part of a satellite system, making it a frequent target for adversaries seeking remote influence over spacecraft behavior~\cite{wainscott-sargent2022,pavur2022building}. 
Its attack surface spans physical facilities, enterprise networks, cloud services, supply chains, and human operators.

\noindent \textbf{1) Physical security.}  
Ground stations and mission facilities remain vulnerable to unauthorized physical access, theft, tampering, and misuse of removable media. 
Such access may expose sensitive software, credentials, or operational data and can enable manipulation of mission systems. 
For example, NASA reported the theft of an encrypted laptop containing command-and-control algorithms related to the International Space Station~\cite{martin2012nasa}. 
Even low-complexity physical vectors, such as unauthorized USB usage, may introduce malware or corrupt mission-critical systems~\cite{gedeon2023}.

\noindent \textbf{2) Network exposure.}  
Ground infrastructure is typically connected to conventional IP networks and is therefore susceptible to phishing, credential theft, exploitation of vulnerable services, and lateral movement through enterprise environments. 
Because mission control and telemetry-processing systems are often integrated with broader IT infrastructure, compromise at this layer may enable data theft, command manipulation, or persistent operational access.

\noindent \textbf{3) Cloud and virtualized infrastructure.}  
The growing adoption of cloud-based ground services expands operational flexibility but also introduces risks associated with remote administration, API exposure, misconfiguration, shared infrastructure, and identity compromise. 
If attackers gain access to these environments, they may inject malicious data, disrupt payload operations, or interfere with command workflows.

\noindent \textbf{4) Ground-segment supply chain.}  
Ground systems frequently depend on third-party hardware, proprietary software, publicly available documentation, and externally managed services. 
This creates opportunities for adversaries to exploit leaked tools, exposed datasheets, or vulnerable dependencies and to abuse trust relationships across the supply chain~\cite{vanlyssel2025spychain}.

\noindent \textbf{5) Third-party services and COTS dependencies.}  
Ground systems increasingly rely on commercial-off-the-shelf components and outsourced services that may be unpatched, outdated, or contain known vulnerabilities~\cite{mitre2023}. 
These dependencies can create indirect compromise paths, as illustrated by the exploitation of outdated remote-access infrastructure in the Viasat incident discussed in Section~\ref{sec:Satellite_Attacks}.

\subsection{Space Segment Vulnerabilities}

Although many publicly documented attacks target terrestrial infrastructure or communication links, spacecraft themselves also exhibit significant vulnerabilities. 
These arise from the growing use of software-defined functionality, COTS hardware, remote reconfiguration mechanisms, and globally distributed supply chains.

\noindent \textbf{1) COTS components.}  
The use of commercial off-the-shelf components lowers costs and accelerates deployment, particularly in LEO systems, but it also introduces substantial security risks~\cite{falco2018job}. 
Widely available hardware and software can be acquired and analyzed by adversaries, patches may be delayed or infeasible after launch, and open development ecosystems may increase the risk of embedded vulnerabilities or backdoors~\cite{nussbaum2020cybersecurity}.

\noindent \textbf{2) Supply-chain dependence.}  
Space systems rely on highly specialized, multi-tier supply chains in which vendors may control critical hardware, firmware, or software modules. 
A defect or malicious modification in even a minor component can propagate into mission-critical failures. 
Because spacecraft are difficult to service after deployment, supply-chain weaknesses may have long-lived and mission-wide consequences~\cite{falco2018vacuum,vanlyssel2025spychain}.

\noindent \textbf{3) Limited standards and regulatory enforcement.}  
Unlike more mature safety-critical sectors, space cybersecurity remains governed by fragmented standards and relatively weak enforcement mechanisms. 
Although organizations such as the ITU regulate spectrum use and orbital coordination, there is no universally enforced cybersecurity baseline for satellite systems. 
This regulatory gap increases the likelihood that insecure designs, weak update practices, or insufficient protections persist in deployed missions~\cite{falco2018vacuum}.

\subsection{Communication Segment Vulnerabilities}

The communication segment is a primary exposure point because it mediates interactions across space, ground, and user components. 
Its vulnerabilities arise from insecure protocols, limited authentication, legacy waveform assumptions, and the physical openness of RF communication.

\noindent \textbf{1) Weak or absent encryption.}  
Some satellite systems do not adequately protect telemetry or telecommand channels with strong encryption, authentication, replay protection, or privilege separation~\cite{willbold2023space,planta2026satbleed}. 
This can enable eavesdropping, replay, spoofing, jamming-assisted manipulation, and signal hijacking~\cite{manulis2021cyber,pavur2022building}. 
Because these links often carry operationally critical information, compromise may affect both confidentiality and control integrity.

\noindent \textbf{2) Incomplete cybersecurity protections in deployed protocols.}  
Analyses of operational satellite communication standards have revealed missing or weak cybersecurity safeguards in parts of the protocol stack, including legacy emergency and beaconing systems~\cite{costin2023cybersecurity}. 
Such gaps create opportunities for signal abuse, impersonation, and service disruption.

\noindent \textbf{3) COM/FEP and TT\&C gateway exposure.}
Recent experimental work shows that satellite communication modules and corresponding ground-station front-end processors may introduce vulnerabilities beyond weak encryption alone.
COTS COM modules can expose flaws in authentication, replay protection, routing, debug interfaces, frame processing, and telemetry/telecommand handling, allowing adversaries to move from RF-layer access toward command-path compromise or ground-side cryptographic exposure~\cite{planta2026satbleed}.
Because COM modules often act as the gateway between external RF links and internally trusted spacecraft buses, compromise at this layer may affect both communication availability and spacecraft control integrity~\cite{willbold2023space,planta2026satbleed}.
\noindent \textbf{4) ISL and constellation-scale link congestion.}
Modern LEO constellations rely on ISLs, satellite-to-ground links, and dynamic routing to provide continuous service.
These links may be vulnerable to congestion or link-flooding attacks when adversaries exploit predictable orbital motion, limited routing diversity, or bottleneck ground paths.
Unlike RF jamming, such attacks can use legitimate-looking network traffic to degrade availability across selected links, gateways, or geographic regions~\cite{giuliari2021icarus}.

\subsection{User Segment Vulnerabilities}

The user segment represents a broad and heterogeneous attack surface that includes end-user terminals, service-provider infrastructure, mobile access technologies, and downstream service ecosystems~\cite{steinberger2008survey}. 
Because these systems bridge space services and terrestrial connectivity, compromise in this segment may have disproportionate operational effects.

\noindent \textbf{1) Satellite service providers (SSPs).}  
Service-provider infrastructures may be targeted through Internet-facing services, VSAT hubs, and associated network services. 
Relevant threats include interception, eavesdropping, VoIP abuse, malware, spoofing, VPN compromise, routing abuse, and rogue device insertion into VSAT environments~\cite{mario2008,mcgann2005analysis,wu2018approach}. 
In centralized topologies, the compromise of a core hub may cause large-scale disruption or denial-of-service attacks.

\noindent \textbf{2) Points of presence (PoPs) and access networks.}  
PoPs and access-layer infrastructure may be vulnerable to router compromise, Internet/Intranet intrusion, abuse of cellular networks, and exploitation of protocols such as BGP, OSPF, RIP, PPP, Ethernet, and SNMP~\cite{center2002cert,steinberger2008survey}. 
These attack paths may allow adversaries to degrade service availability, intercept traffic, or manipulate routing behavior.

\noindent \textbf{3) User terminals and receivers.}
End-user terminals and receivers, including VSAT modems, GNSS receivers, maritime and aviation SATCOM terminals, and direct-to-device equipment, may expose vulnerabilities through firmware flaws, weak credentials, insecure management interfaces, or poor protection against spoofing and jamming.
Because these devices operate at scale outside controlled mission environments, compromise or disruption can cause widespread service impact, as illustrated by the Viasat KA-SAT incident~\cite{viasat2022,viasatt2022,sentinelone2022acidrain}.

\subsection{Subsystem Vulnerabilities}

Beyond segment-level exposures, spacecraft subsystems exhibit vulnerabilities that vary by function, environmental dependencies, and control interfaces.

\noindent \textbf{1) Radiation and solar activity.}  
Subsystems including EPS, ADCS, COMMS, TCS, PCS, and C\&DHS are exposed to radiation effects and solar disturbances that may degrade solar cells, corrupt memory, disrupt control logic, or reduce communication reliability~\cite{falco2020satellites,hosel2014failure}.

\noindent \textbf{2) Space debris and micrometeoroids.}  
Physical damage from debris may impair power generation, attitude control, communications, and structural integrity, with cascading effects across multiple subsystems~\cite{hosel2014failure}.

\noindent \textbf{3) Equipment malfunction and aging.}  
Hardware degradation, latent defects, and subsystem faults may lead to loss of power, reduced pointing accuracy, communication failures, and degraded thermal or propulsion performance~\cite{mayo1963command,kim2011satellite}.

\noindent \textbf{4) Cyber compromise of control-critical subsystems.}  
Subsystems such as ADCS, COMMS, and C\&DHS are particularly sensitive to cyber compromise because they directly affect command execution, data integrity, and operational control~\cite{redteam2023}. 
Successful exploitation may enable stealthy persistence, degraded mission performance, or total spacecraft takeover.

\noindent \textbf{5) Inadequate design and engineering assumptions.}  
Weak architectural isolation, insufficient fault handling, poor interface design, or insecure command logic may create exploitable weaknesses in subsystems such as TCS, PCS, and C\&DHS~\cite{fayyaz2014fault}.

\noindent \textbf{6) Human error.}  
Operational mistakes in software configuration, command sequencing, or maintenance procedures may introduce vulnerabilities even when subsystem designs are otherwise sound.

\noindent \textbf{7) Propulsion-system anomalies.}  
Fuel leakage or propulsion mismanagement can cause orbit deviation, reduced maneuverability, or collision risk, making PCS a safety-critical subsystem with both accidental and adversarial exposure.

\noindent \textbf{8) Payload-specific vulnerabilities.} 
Payloads introduce mission-dependent risks. 
Communication payloads may be disrupted at the RF layer, software-defined radios may expose firmware and control-plane attack surfaces, and imaging payloads may be affected by both hardware degradation and manipulated inputs. 
In networked constellations, the compromise of a single payload or crosslink interface may threaten broader intersatellite operations.

\noindent \textbf{9) Weak interface isolation and internal trust assumptions.}
Small satellites often rely on tightly integrated subsystems connected through shared buses, payload interfaces, or COM--C\&DHS command paths.
Experimental studies show that insufficient privilege separation, dangerous telecommands, insecure update mechanisms, and weak bus or payload-interface isolation can allow malicious inputs accepted by one component to influence other spacecraft functions~\cite{willbold2023space,planta2026satbleed}.
These weaknesses make subsystem boundaries security-critical, particularly between COMMS, TT\&C, C\&DHS, payload interfaces, ADCS, and EPS.

\noindent \textbf{10) Onboard bus communication vulnerabilities.}
Spacecraft subsystems often communicate through internal buses, and interfaces such as CAN, I\textsuperscript{2}C, SPI, UART/RS-485, SpaceWire, Ethernet, MIL-STD-1553, or PC/104-based interconnects.
These buses are usually designed for reliability, determinism, low complexity, or fault tolerance rather than adversarial security.
As a result, many provide limited native support for authentication, encryption, message provenance, or privilege separation.
Prior work on MIL-STD-1553 has shown that legacy aerospace buses can be vulnerable to spoofing, unauthorized-device access, and other cyber-physical attack scenarios~\cite{stan2017protecting,levy2024anomili}.
Similarly, fault-injection work on nanosatellite I\textsuperscript{2}C subsystem integration demonstrates that bus-level traffic can be intercepted or modified in ways that affect interactions between onboard components~\cite{batista2021using}.
When such buses connect COMMS, C\&DHS, EPS, ADCS, payload, or TT\&C-related components, a compromised node or interface may inject commands, replay messages, suppress traffic, corrupt telemetry, or support lateral movement across spacecraft subsystems.

\subsection{Vulnerabilities of Machine Learning in Satellite Systems}

Machine learning (ML) is increasingly integrated into satellite missions for communications, control, perception, scheduling, and anomaly detection. 
Its use introduces additional cyber-physical vulnerabilities rooted in sensor dependence, data integrity, constrained hardware, and cross-segment operational dependencies.

Recent work has begun to highlight the emerging risks of adversarial machine learning (AML) in satellite systems. 
For example, Thummala et al.~\cite{thummala2024adversarial} analyze the attack surface of ML-enabled space systems and map potential adversarial vulnerabilities across different components and operational contexts. 
Complementary to this, Shigol et al.~\cite{shigolrisk} focus on the risk perspective and propose a framework for assessing and quantifying the impact of AML threats on satellite systems.

\noindent\textbf{1) Spoofable and noisy sensor inputs.}  
ML models often rely on star trackers, magnetometers, GNSS receivers, thermal sensors, and imagers. 
These inputs may be manipulated through RF interference, GNSS spoofing, or physical perturbations that induce adversarial features~\cite{tang2023adversarial,czaja2018adversarial}. 
Such corruption can propagate into ADCS, autonomy logic, and anomaly detectors, producing mispointing, unnecessary maneuvers, or incorrect resource allocation.

\noindent\textbf{2) Manipulation of telemetry and communication pipelines.}  
Telemetry, command streams, and spectrum measurements may feed ML components used for anti-jamming~\cite{morales2019jammer,han2024deep,ding2023few}, beam hopping~\cite{xie2025multi,martinez5214437hybrid}, and anomaly detection~\cite{herrmann2024unmasking,rath2020security,sitouah2022deep}. 
Jamming, spoofing, replay, or man-in-the-middle attacks can inject stale, misleading, or adversarial data into these pipelines, while limited bandwidth and compressed telemetry make such corruption difficult to detect.

\noindent\textbf{3) Onboard hardware and execution constraints.}  
Spaceborne ML runs under strict power, memory, radiation, and compute constraints. 
Quantized or pruned models may be more sensitive to bit flips, timing faults, and radiation-induced perturbations~\cite{fourati2021artificial}. 
Supply-chain compromise or side-channel leakage may further expose model parameters or affect inference accelerators.

\noindent\textbf{4) Poisonable training and update processes.}  
Architectures that support adaptive retraining or onboard learning, for example, RL-based control~\cite{harris2023autonomous,li2025mission}, anti-jamming agents~\cite{han2020spatial,li2025achieving}, and telemetry anomaly detection~\cite{hundman2018detecting,thangavel2024artificial}---are vulnerable to poisoning and backdoor insertion. 
Even limited corruption of trusted telemetry or environmental data may degrade policies, distort decision boundaries, or destabilize control behavior.

\noindent\textbf{5) Operational impact of ML misclassification.}  
Misclassification in anomaly detection may trigger unnecessary safe-mode transitions or cause real failures to be missed~\cite{herrmann2024unmasking}. 
Adversarial perturbations in vision models may distort payload products~\cite{tang2023adversarial}, while errors in ML-driven scheduling or traffic management may degrade service quality and network efficiency~\cite{feng2022improved,woo2022detecting}.

\noindent\textbf{6) Ground-segment and cross-segment ML exposure.}  
Many ML-enabled functions, including mission planning~\cite{bourriez2023spacecraft,pinto2020towards}, collision avoidance~\cite{yao2025meta}, network management~\cite{3gpp_ntn_overview}, and cyber monitoring~\cite{rath2020security}, depend on ground systems and cloud-based data pipelines. 
Compromise of these external infrastructures may indirectly corrupt models, training data, or decision support systems used on board~\cite {nasa_oig_ig_19_022}.

\noindent\textbf{7) Absence of ML-specific protections.}  
Existing space standards primarily address deterministic software and traditional fault tolerance, but provide limited guidance on ML robustness, adversarial-input detection, or post-deployment model integrity. 
As a result, ML introduces a cross-cutting vulnerability layer that may affect TT\&C, ADCS, C\&DHS, EPS, and payload subsystems.

To date, we are not aware of any publicly confirmed satellite incidents in which AML techniques
were explicitly identified as the root cause of failure or compromise. Instead, this subsection
characterizes AML-related vulnerabilities as an emerging class of \emph{prospective} threats,
grounded in existing ML deployments in satellite systems and analogies to terrestrial AML attacks.

These vulnerabilities define what can be exploited; the next section characterizes the adversaries and capabilities required to exploit them in practice.
\section{\label{sec:AdversariesThreat}Adversaries \& Capabilities}

This section characterizes the principal adversaries relevant to satellite systems and the capabilities they may leverage to carry out attacks. 
We distinguish between \emph{who} the adversary is and \emph{how} the adversary operates, following space-cyber guidance that separates actor categories from capabilities, tactics, and techniques~\cite{sparta-overview-2023,ccsds-350-1-g3-2022,nist-ir-8270-2023,enisa-2025-space-threat}. 
This distinction is later used to map real-world incidents and taxonomy entries to specific actor tiers and feasible capability profiles.

\subsection{Threat Actors}

We adopt a six-tier model of adversaries with distinct resources, intent, and tradecraft maturity. 
The tiers are organized according to expected capabilities, potential impact, and operational resources, as summarized in Table~\ref{table:threat-actors}.

\begin{enumerate}
    \item \textbf{Tier 1: Opportunistic individuals and small groups} — 
    lone hackers and small activist collectives with widely varying skills, 
    Typically operating with commodity hardware, publicly available tools, and limited funding. 
    Motivations include notoriety, experimentation, or issue-driven disruption.

    \item \textbf{Tier 2: Competitors} — commercially motivated organizations seeking strategic advantage through technology theft, service degradation, or fraud, typically with sustained funding but constrained by legal and reputational risk.

    \item \textbf{Tier 3: Terrorists} — ideologically motivated groups aiming to coerce, intimidate, or disrupt, willing to tolerate collateral damage and legal risk, but usually limited in technical depth and supply-chain reach.
    
    \item \textbf{Tier 4: Insiders (operator or supply-chain)} — authorized personnel, contractors, or suppliers who may abuse privileged access to operational networks, ground systems, or manufacturing pipelines to corrupt data, manipulate systems, steal technology, or sabotage operations.
    
    \item \textbf{Tier 5: Organized criminals} — well-resourced groups focused on monetization (fraud, extortion, ransomware, resale of access), often able to acquire advanced infrastructure and tooling and collaborate with insiders.
    
    \item \textbf{Tier 6: State actors} — government-sponsored teams pursuing strategic, intelligence, or military objectives, with potential access to classified tooling, intelligence, supply-chain compromise, and non-kinetic counterspace capabilities.
\end{enumerate}

These tiers reflect both established threat-intelligence categorizations and the actors observed in our incident corpus (Section~\ref{sec:Satellite_Attacks}), where many documented events can be mapped to Tier~1, Tier~4, and Tier~6 adversaries.

\begin{table}[!ht]
\centering
\caption{Primary adversaries and illustrative goals (non-exhaustive). The model aligns with SPARTA's focus on tactics and techniques and with ENISA/CCSDS/NIST guidance on separating actor types from capability acquisition and attack execution~\cite{sparta-overview-2023,enisa-2025-space-threat,ccsds-350-1-g3-2022,nist-ir-8270-2023}.}
\scriptsize
\setlength{\tabcolsep}{3pt}
\begin{tabular}{|p{0.35\columnwidth}|p{0.60\columnwidth}|}
\hline
\textbf{Threat actor} & \textbf{Typical goals / threats} \\
\hline\hline
Individual hackers & Fame / notoriety; opportunistic exploitation \\
\hline
Activists & Tampering with network communication; message disruption / propaganda \\
\hline
Competitors & Sabotage / destruction; technology theft; financial gain \\
\hline
Terrorists & Tampering with network communication; propaganda; sabotage / destruction; financial gain \\
\hline
Insiders (operator or supply-chain) & Data corruption; technology theft; gaining control; financial gain; sabotage / destruction \\
\hline
Organized criminals & Tampering / eavesdropping; sabotage / destruction; denial of service; extortion; financial gain \\
\hline
State actors & Tampering; technology theft; eavesdropping; financial theft; gaining control; sabotage / destruction; attacks on ground endpoints; denial of service \\
\hline
\end{tabular}
\label{table:threat-actors}
\end{table}

\subsection{Capability Categories}

\begin{figure}[!ht]
  \includegraphics[trim={0 2.5cm 2cm 2.5cm},clip,width=\linewidth]{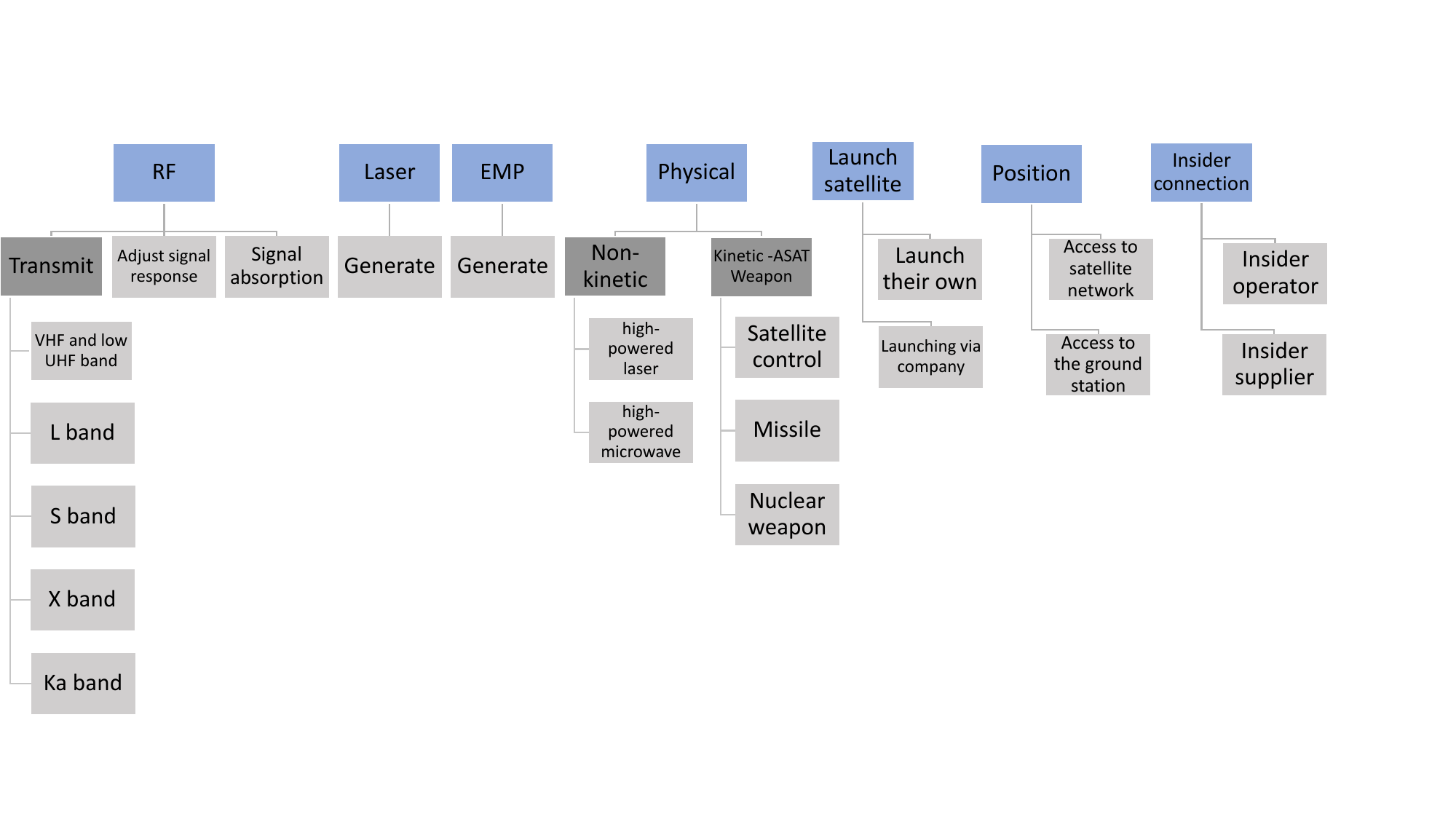}
  \caption{Adversary capability categories relevant to satellite attacks.}
  \label{fig:Adversarycap_fig}
\end{figure}

To complement the actor model, we identify the operational capabilities required to execute different classes of satellite attacks. 
These capabilities capture the practical means available to adversaries, ranging from RF transmission and signal processing to launch access, physical interference, and insider support. 
Figure~\ref{fig:Adversarycap_fig} organizes these capabilities into major categories, while Table~\ref{tab:capabilities} details the individual capabilities considered in this work.

\noindent\textbf{Radio-frequency (RF) signaling capabilities.}  
Satellite communications span a wide range of frequencies, and many attacks depend on the ability to transmit, receive, or manipulate signals within these bands. 
For CubeSats, VHF and UHF are common due to relatively simple ground equipment and favorable propagation characteristics~\cite{nassa2022,You2021}. 
However, modern small satellites increasingly rely on S-, X-, and Ka-band communications, expanding both operational performance and the technical requirements for adversarial interference.

\noindent\textbf{Laser communication capabilities.}  
Optical communication uses laser beams rather than RF signals to transfer data~\cite{majumdar2010free}. 
While it offers advantages such as low link loss and abundant spectrum~\cite{hemmati2020near}, it also introduces specialized requirements, including precise alignment and an unobstructed line of sight. 
Consequently, adversarial interaction with optical inter-satellite links generally requires significantly greater sophistication and, in some cases, access to spaceborne optical infrastructure.

\noindent\textbf{Electromagnetic pulse (EMP) capabilities.}  
EMP effects may arise from natural or artificial sources and can disrupt telecommunication systems, electronic components, and power infrastructure~\cite{kopp1996electromagnetic,longmire1978electromagnetic,dhs2022,wang2019power}. 
In the satellite context, such capabilities are relevant both as direct disruption mechanisms and as part of broader high-impact counterspace scenarios.

\noindent\textbf{Physical attack capabilities.}  
Physical threats include both kinetic and non-kinetic means of degrading or damaging spacecraft and supporting infrastructure. 
Kinetic attacks involve direct physical interaction, whereas non-kinetic approaches may impair systems without contact, for example through directed energy or environmental disruption. 
Even when such attacks do not cause total mission loss, they may reduce performance, interrupt service, or shorten spacecraft lifespan.

\noindent\textbf{Launch capability.}\label{launch_sat}  
Some adversarial scenarios assume the ability to place assets into orbit, either directly or through access to launch providers. 
Given the increasing availability of commercial launch services, the barrier to deploying spaceborne infrastructure is lower than in earlier eras, which has implications for co-orbital interference and satellite-to-satellite threat models~\cite{satellitetoday2021_companies,bbc2021_space_companies}.

\noindent\textbf{Positional access.}  
Some attacks require access to a specific operational environment, such as a ground station, a satellite network, or a constellation operated by a commercial provider. 
Once such access is obtained, the adversary may exploit this position to expand control or stage additional attacks.

\noindent\textbf{Insider access and relationships.}  
Adversaries may rely on insider operators or suppliers to gain privileged access to systems, sensitive information, or supply-chain components. 
Such access can substantially reduce attack complexity and increase the feasibility of stealthy compromise.

Table~\ref{tab:adversary_capabilities} compares the six actor tiers according to the feasibility of acquiring the capabilities required to implement different classes of attacks. 
Each cell reflects the relative likelihood that a given actor type can obtain or sustain the corresponding capability.

%\onecolumn
\begin{table*}[!ht]%{|p{0.025\textwidth}|p{0.15\textwidth}|p{0.75\textwidth}|} 
\scriptsize
%\centering
\caption{Adversaries’ capabilities.}
\label{tab:capabilities}
\begin{tabular}{|p{0.025\textwidth}|p{0.15\textwidth}|p{0.75\textwidth}|} 
\hline
\textbf{} & \textbf{Name} & \textbf{Description} \\ \hline 
%\endhead
\rowcolor{lightgray}\multicolumn{3}{|c|}{Radio frequency (RF) - Signaling Capabilities}\\ \hline 
AC1 & Transmit VHF and low UHF signals & In the past, the VHF band, ranging from 136-138 MHz, was heavily used, but now most activity is restricted to 137-138 MHz.
Meteorological satellites mainly use this band to transmit data and low-resolution images and for low data rate mobile satellite downlinks. 
The UHF band, with frequencies from 399.9-403 MHz, is used for navigation, positioning, time and frequency standards, mobile communication, and meteorological satellites. 
Satellites transmitting on 150 MHz use the companion band around 400 MHz. The 432-438 
MHz range includes a popular amateur satellite band and a few Earth resources satellites. \\ \hline 

AC2 & Transmit L-band signals & The L-band refers to a section of radio spectrum that covers frequencies ranging from 1 to 2 GHz. 
This band has numerous applications in satellite technology. 
For example, it is used to transmit signals for crucial systems like the GPS and to enable satellite mobile phone communication like Iridium. 
In addition, Inmarsat leverages the L-band to establish communication across sea, land, and air. 
WorldSpace satellite radio also broadcasts within the L-band~\cite{strozzi2005survey,antesky2023,esa2023}. \\ \hline

AC3 & Transmit S-band signals & The S-band refers to frequencies ranging from 2-4 GHz in the radio spectrum. 
It has various applications in satellite technology, such as weather radar, surface ship radar, and some communications satellites. 
NASA utilizes the S-band to communicate with the International Space Station and Space Shuttle. 
In May 2009, the European Commission awarded Inmarsat and Solaris Mobile a 2×15 MHz portion of each S-band~\cite{esa2023}. \\ \hline

AC4 & Transmit X-band signals & The X-band refers to a specific range of frequencies in the radio spectrum, those falling between 8-12 GHz. 
This frequency range is mainly used by military personnel for radar applications, such as continuous-wave, pulsed, single-polarization, dual-polarization, synthetic aperture radar, and phased arrays. 
Civil, military, and government organizations also use the X-band radar frequency sub-bands for various purposes, including weather monitoring, air traffic control, maritime vessel traffic control, defense tracking, and vehicle speed detection for law enforcement~\cite{esa2023}. \\ \hline

AC5 & Transmit Ka-band signals & The Ka-band refers to frequencies ranging from 26-40 GHz in the radio spectrum. 
It has multiple applications, including satellite communications, with uplink capability in the 27.5 GHz and 31 GHz bands. 
It is also used for close-range targeting radars on military aircraft, providing high-resolution capabilities~\cite{esa2023}. \\ \hline

AC6 & Adjust signal response & To accurately determine the location of a transmitter based on signal arrival times, it is crucial to be able to coordinate responses during interrogations and adjust transmission rates as needed. \\ \hline

AC7 & Signal absorption and processing & An adversary can use receivers, processors, and parsers to pick up, listen to, and decode signals.\\ \hline

\rowcolor{lightgray}\multicolumn{3}{|c|}{Laser Communication}\\ \hline

AC8 & Emit laser & Eavesdropping or jamming can be performed by using a laser~\cite{yahia2022optical}, and in those cases, the following components are needed, in addition to the laser generator itself: a powerful electrical power supply, optical components, a control system, and a cooling system. 
These components work together to create and manage the laser beam, aiming it at the intended target.\\ \hline

\rowcolor{lightgray}\multicolumn{3}{|c|}{Electromagnetic Pulse (EMP)}\\ \hline 
AC9 & Generate EMP & To create an EMP, several components are required, including a high-energy source, power supply, pulse generator, antenna, enclosure, and trigger mechanism. 
These components work together to produce a powerful electromagnetic field that can disrupt or cause damage to electronic devices. 
An adversary can use an EMP weapon on the ground or in space, via a spacecraft~\cite{falco2021security,falco2020satellites}.\\ \hline

\rowcolor{lightgray}\multicolumn{3}{|c|}{Physical Attack}\\ \hline 

AC10 & Emit high-powered laser & A high-powered laser beam can quickly disable or destroy a target by accurately aiming the shaft for a specific amount of time. 
Some countries have already developed this capability, and ongoing efforts are underway to mount such weapons on orbiting satellites for faster deployment~\cite{harrison2020space}. 
These attacks are difficult to trace and can be launched from any location on the planet. \\ \hline 

AC11 & Emit high-powered microwave &
An attack using a high-powered microwave involves directing a significant amount of energy toward a specific target. 
The consequences of such an attack can vary from the satellite's destruction to the temporary disablement of its subsystems~\cite{kai2020handbook}.\\ \hline 

AC12 & Satellite control & By taking control of a satellite or spacecraft or creating their own, adversaries could attack other satellites by causing collisions or damaging their sensors.
They could equip the spacecraft with weapons to increase their impact. \\ \hline 

AC13 & Missile & Direct-ascent Anti-satellite weapons (ASAT) involves missiles that are launched from the ground or air, and ascend in order to reach a target in orbit. \\ \hline 

AC14 & Nuclear weapon & Since the 1963 Treaty Banning Nuclear Weapon Tests (PTBT) in the atmosphere, outer space, and underwater, no further nuclear weapon tests have been conducted in the upper atmosphere~\cite{schwelb1964nuclear}.
Nevertheless, several nations possess the capability to do so today. 
A high-altitude nuclear explosion (HANE) has three main consequences for space infrastructure~\cite{colgate1965phenomenology}. 
The first one is kinetic, which causes destructive effects within the explosion's radius. 
The last two consequences are related to electromagnetic pulses that HANE weapons produce, which can cause temporary or irreversible damage to satellites and ground stations~\cite{falco2021security}. \\ \hline

\rowcolor{lightgray}\multicolumn{3}{|c|}{Launch Satellite}\\ \hline 

AC15 & Launch on their own & Most launch companies do not permit individuals to launch their customized satellites into space, so an adversary interested in deploying a customized satellite with weapons would likely have to have the ability to launch the satellite him/herself. \\ \hline

AC16 & Connection to launching company & For a similar reason as~{\it (AC15)}, the adversary may need a connection to a launching company in order to launch their weaponized customized satellite into space.\\ \hline

\rowcolor{lightgray}\multicolumn{3}{|c|}{Position Access.}\\ \hline 

AC17 & Access to satellite network/constellation & Adversaries can access satellite networks/constellations by compromising GSs, exploiting communication vulnerabilities, or employing social engineering tactics. 
Alternatively, adversaries can gain access by paying for satellite services. \\ \hline 

AC18 & Access to the ground station & Adversaries can access ground stations through physical infiltration, remote attacks exploiting network vulnerabilities, or social engineering or own a GS or pay to utilize a cloud service's GS.\\ \hline 

\rowcolor{lightgray}\multicolumn{3}{|c|}{Insider Connection.}\\ \hline 

AC19 & Insider operator & An adversary can use an insider operator to access sensitive information, systems, or satellite components. 
Insiders may possess knowledge of the satellite's vulnerabilities, security procedures, and operational protocols, making them valuable assets for attackers.
With insider assistance, attackers can bypass security controls, inject malicious code, or manipulate satellite operations to achieve their objectives, such as disrupting communications, stealing sensitive information, or sabotaging the satellite.
Additionally, insiders may provide physical access to the satellite or its components, allowing the attacker to perform a physical attack. 
Connecting with an insider operator can significantly enhance an adversary's chances of success in a satellite attack. \\ \hline

AC20 & Insider supplier & Through insider suppliers, an adversary can obtain malicious hardware or software components to compromise the satellite's security. 
Insiders with access to the supply chain can introduce counterfeit or modified components that bypass security measures or allow unauthorized access to the satellite's systems or data.
By compromising the supply chain, an attacker can simultaneously compromise multiple satellites and magnify their attack's impact. \\ \hline

\end{tabular}

\end{table*}
%}
%\twocolumn

\begin{table*}[ht]
\scriptsize 
\caption{Adversary Capabilities.}\label{tab:adversary_capabilities}
\begin{center}
\begin{tabular}{c|ccccccc|c|c|ccccc|cc|cc|cc}
\hline
 & \multicolumn{20}{c}{\textbf{Adversary Capabilities}} 
\\
    &
   
    \multicolumn{7}{c|}{\textbf{RF}} &
    \multicolumn{1}{c|}{\textbf{Laser}} &
    \multicolumn{1}{c|}{\textbf{EMP}} &
    \multicolumn{5}{c|}{\textbf{Physical}} &
    \multicolumn{2}{c|}{\textbf{Launch}} &
    \multicolumn{2}{c|}{\textbf{Position}} &
    \multicolumn{2}{c}{\textbf{Insider}} \\ 

    & \rotatebox{90}{AC1}
    & \rotatebox{90}{AC2}
    & \rotatebox{90}{AC3}
    & \rotatebox{90}{AC4}
    & \rotatebox{90}{AC5}
    & \rotatebox{90}{AC6}
    & \rotatebox{90}{AC7}
    & \rotatebox{90}{AC8}
    & \rotatebox{90}{AC9}
    & \rotatebox{90}{AC10}
    & \rotatebox{90}{AC11}
    & \rotatebox{90}{AC12}
    & \rotatebox{90}{AC13}
    & \rotatebox{90}{AC14}
    & \rotatebox{90}{AC15}
    & \rotatebox{90}{AC16}
    & \rotatebox{90}{AC17}
    & \rotatebox{90}{AC18}
    & \rotatebox{90}{AC19}
    & \rotatebox{90}{AC20}
    \\
% \textbf{Table Head} & Table column subhead & Subhead & Subhead \\
\hline

\scriptsize{Tier 1 (Individual hackers \& activists)}
    & $\bullet$ %AC1 
    & $\bullet$ %AC2   
    & $\bullet$ %AC3
    & $\bullet$ %AC4 
    & $\bullet$ %AC5   
    & $\bullet$ %AC6
    & $\bullet$ %AC7 
    & $\circ$ %AC8   
    & $\circ$ %AC9
    & $\circ$ %AC10
    & $\circ$ %AC11   
    & $\circ$ %AC12
    & $\circ$ %AC13
    & $\circ$ %AC14   
    & $\circ$ %AC15
    & $\halfcirc$ %AC16
    & $\bullet$ %AC17
    & $\halfcirc$ %AC18
    & $\halfcirc$ %AC19
    & $\halfcirc$ %AC10
    \\
\scriptsize{Tier 2 (Competitors)}
    & $\bullet$ %AC1 
    & $\bullet$ %AC2   
    & $\bullet$ %AC3
    & $\bullet$ %AC4 
    & $\bullet$ %AC5   
    & $\bullet$ %AC6
    & $\bullet$ %AC7 
    & $\halfcirc$ %AC8   
    & $\halfcirc$ %AC9
    & $\halfcirc$ %AC10
    & $\halfcirc$ %AC11   
    & $\bullet$ %AC12
    & $\circ$ %AC13
    & $\circ$ %AC14   
    & $\halfcirc$ %AC15
    & $\bullet$ %AC16
    & $\bullet$ %AC17
    & $\halfcirc$ %AC18
    & $\bullet$ %AC19
    & $\bullet$ %AC10
    \\
\scriptsize{Tier 3 (Terrorists)}
    & $\bullet$ %AC1 
    & $\bullet$ %AC2   
    & $\bullet$ %AC3
    & $\bullet$ %AC4 
    & $\bullet$ %AC5   
    & $\bullet$ %AC6
    & $\bullet$ %AC7 
    & $\halfcirc$ %AC8   
    & $\halfcirc$ %AC9
    & $\circ$ %AC10
    & $\halfcirc$ %AC11   
    & $\bullet$ %AC12
    & $\halfcirc$ %AC13
    & $\circ$ %AC14   
    & $\circ$ %AC15
    & $\halfcirc$ %AC16
    & $\circ$ %AC17
    & $\circ$ %AC18
    & $\halfcirc$ %AC19
    & $\halfcirc$ %AC10
    \\
\scriptsize{Tier 4 (Insider
operator)}
    & $\circ$ %AC1 
    & $\circ$ %AC2   
    & $\circ$ %AC3
    & $\circ$ %AC4 
    & $\circ$ %AC5   
    & $\circ$ %AC6
    & $\circ$ %AC7 
    & $\circ$ %AC8   
    & $\circ$ %AC9
    & $\bullet$ %AC10
    & $\halfcirc$ %AC11   
    & $\halfcirc$ %AC12
    & $\halfcirc$ %AC13
    & $\circ$ %AC14   
    & $\halfcirc$ %AC15
    & $\bullet$ %AC16
    & $\halfcirc$ %AC17
    & $\halfcirc$ %AC18
    & $\bullet$ %AC19
    & $\bullet$ %AC10
    \\
\scriptsize{Tier 5 (Organized criminals)}
    & $\bullet$ %AC1 
    & $\bullet$ %AC2   
    & $\bullet$ %AC3
    & $\bullet$ %AC4 
    & $\bullet$ %AC5   
    & $\bullet$ %AC6
    & $\bullet$ %AC7 
    & $\bullet$ %AC8   
    & $\bullet$ %AC9
    & $\halfcirc$ %AC10
    & $\bullet$ %AC11   
    & $\bullet$ %AC12
    & $\halfcirc$ %AC13
    & $\circ$ %AC14   
    & $\bullet$ %AC15
    & $\halfcirc$ %AC16
    & $\bullet$ %AC17
    & $\bullet$ %AC18
    & $\bullet$ %AC19
    & $\bullet$ %AC10
    \\
\scriptsize{Tier 6 (State actors)}
    & $\bullet$ %AC1 
    & $\bullet$ %AC2   
    & $\bullet$ %AC3
    & $\bullet$ %AC4 
    & $\bullet$ %AC5   
    & $\bullet$ %AC6
    & $\bullet$ %AC7 
    & $\bullet$ %AC8   
    & $\bullet$ %AC9
    & $\bullet$ %AC10
    & $\bullet$ %AC11   
    & $\bullet$ %AC12
    & $\bullet$ %AC13
    & $\bullet$ %AC14   
    & $\bullet$ %AC15
    & $\bullet$ %AC16
    & $\bullet$ %AC17
    & $\bullet$ %AC18
    & $\bullet$ %AC19
    & $\bullet$ %AC10
    \\
\hline
\end{tabular}
\end{center}
\end{table*}

The adversary model provides the capability context for interpreting the incident corpus analyzed in the next section.
\section{A Brief History of Satellite Attacks\label{sec:Satellite_Attacks}}

We now use the incident corpus to examine how the system exposures, vulnerabilities, and adversary capabilities described above appear in publicly reported satellite incidents. This section connects the architectural and threat-analysis discussion to empirical evidence and provides the basis for the lifecycle taxonomy developed in Section~\ref{sec:MITRE}.

\begin{figure*}[ht]
  \centering
  \includegraphics[width=\linewidth]{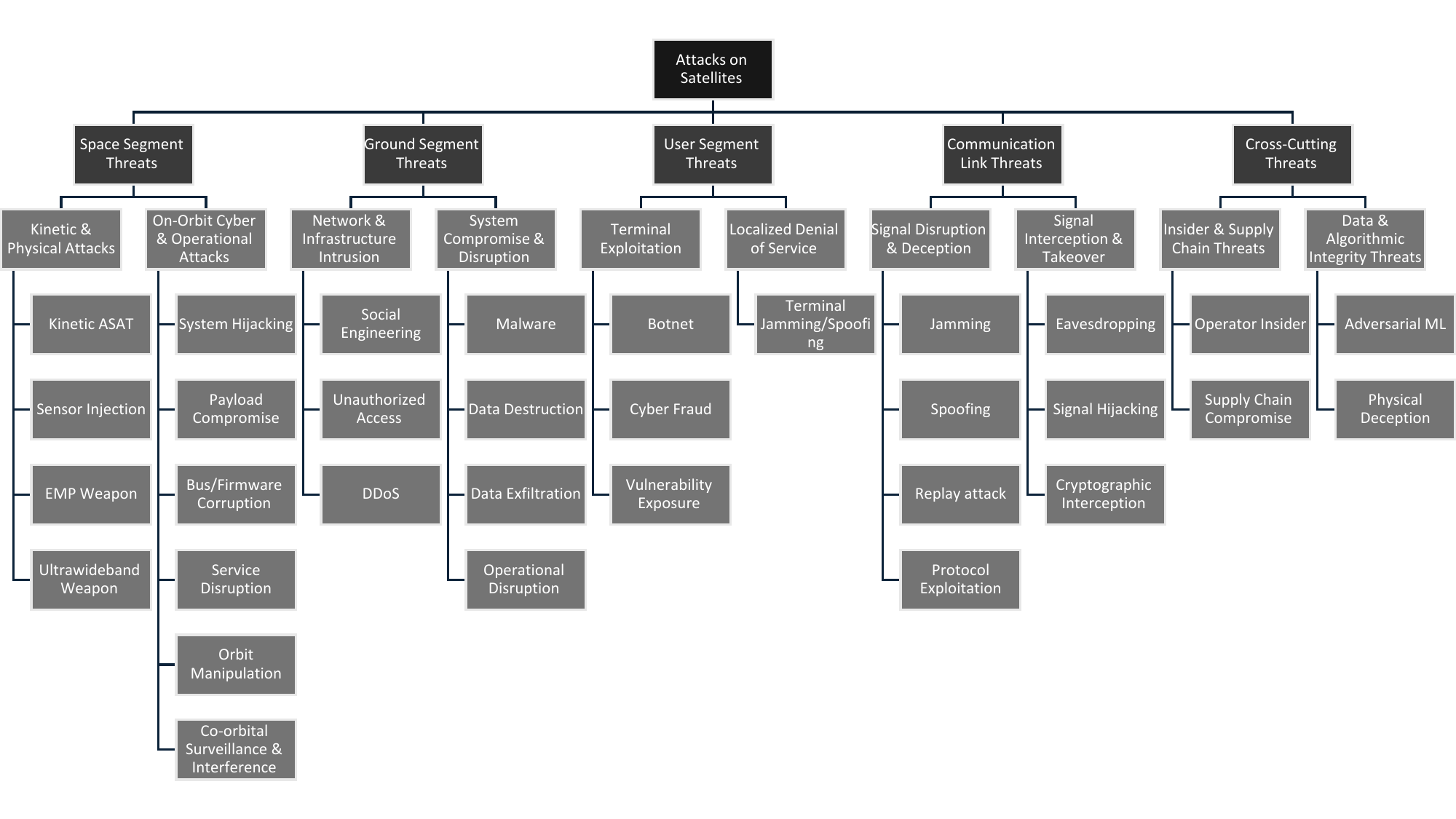}
  \caption{Segment-based taxonomy of satellite attack strategies, spanning space, ground, user, communication link, and cross-cutting threats.}
  \label{fig:atktax}
\end{figure*}

Prior work has documented more than one hundred satellite-related attacks since 1957~\cite{pavur2022building,spacesecurity2020}. Building on and extending these collections, we adopt a segment-based taxonomy (Figure~\ref{fig:atktax}) that reflects the satellite ecosystem across five domains: \textit{(i) Space Segment}, \textit{(ii) Ground Segment}, \textit{(iii) User Segment}, \textit{(iv) Communication Link}, and \textit{(v) Cross-Cutting Threats}. 
This framework extends earlier taxonomies focused primarily on RF interference by incorporating cyber intrusions, firmware manipulation, supply-chain compromise, and data-integrity attacks observed in modern space operations.

\subsection{Trends Over Time}

The trend analysis was derived from the coded incident fields in the corpus, including incident year or period, affected segment, attack category, threat domain, attacker type, victim type, and operational impact. Trends were therefore computed from the structured coding rather than from informal narrative interpretation.

In the 1960s and 1970s, only a small number of incidents were reported, primarily involving intentional jamming of broadcast links by nation-states. 
The 1980s introduced hijacking and unauthorized broadcast injection, while the 1990s saw increased signal interference, eavesdropping, and opportunistic access to satellite feeds. 
During the 2000s, attackers began experimenting with GPS spoofing and cryptographic exploitation. 
The 2010s marked a turning point, with state-sponsored actors targeting ground stations, mission control networks, and satellite operators through malware and credential theft. 
By the 2020s, attacks had evolved into coordinated, multi-vector operations affecting multiple segments simultaneously, including constellation-scale disruption and on-orbit manipulation.

To build a longitudinal perspective, we compiled a dataset of more than 200 publicly reported satellite incidents spanning 1962–2026, integrating prior historical surveys~\cite{pavur2022building,tedeschi2022satellite,willbold2023space} with more than eighty recent incidents from 2019 to 2026 that extend existing datasets. 
Each incident is mapped to the proposed taxonomy, enabling consistent analysis across decades despite variations in reporting detail.

\begin{figure}[ht]
  \centering
  \includegraphics[width=\linewidth]{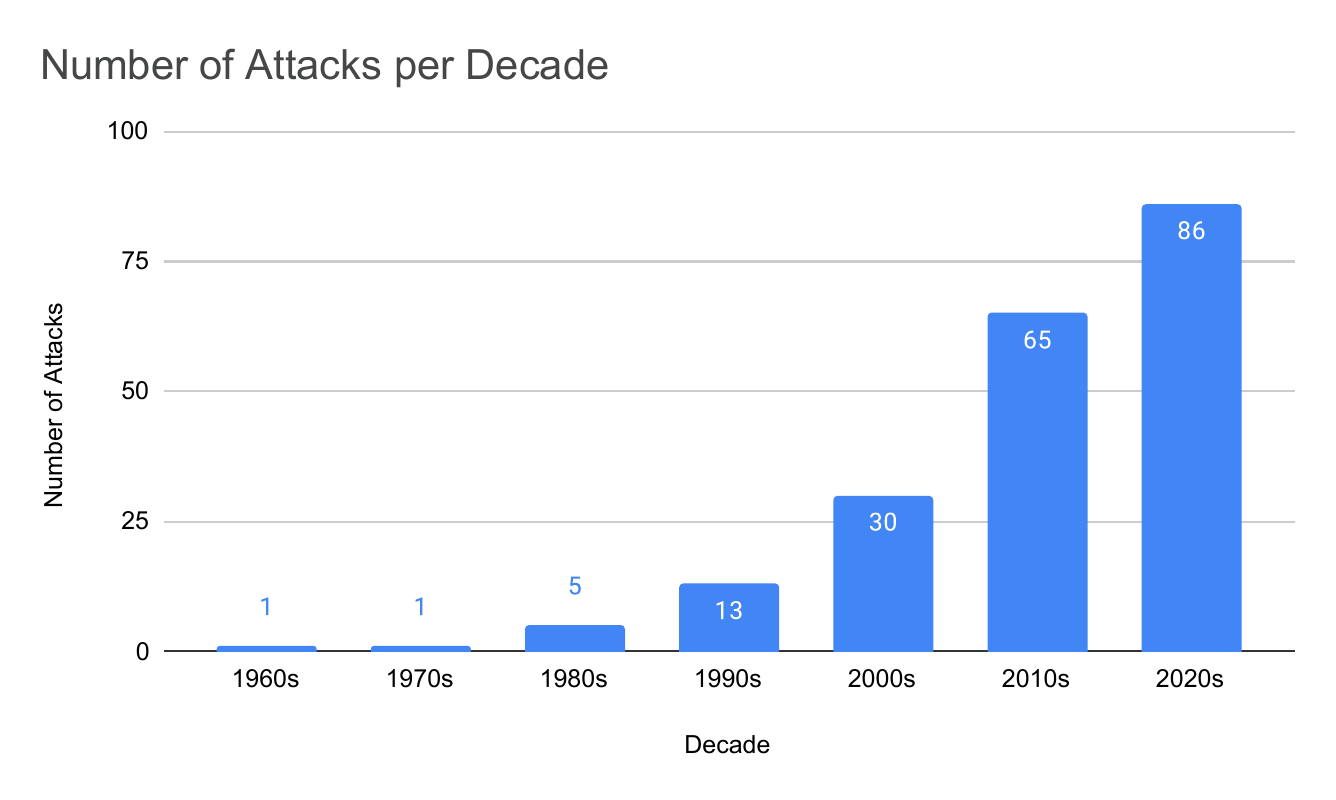}
  \caption{Documented satellite attacks by decade, including more than eighty incidents from 2019--2026 and more than two hundred incidents overall.}
  \label{fig:attk-decade}
\end{figure}

\begin{figure}[ht]
  \centering
  \includegraphics[width=\linewidth]{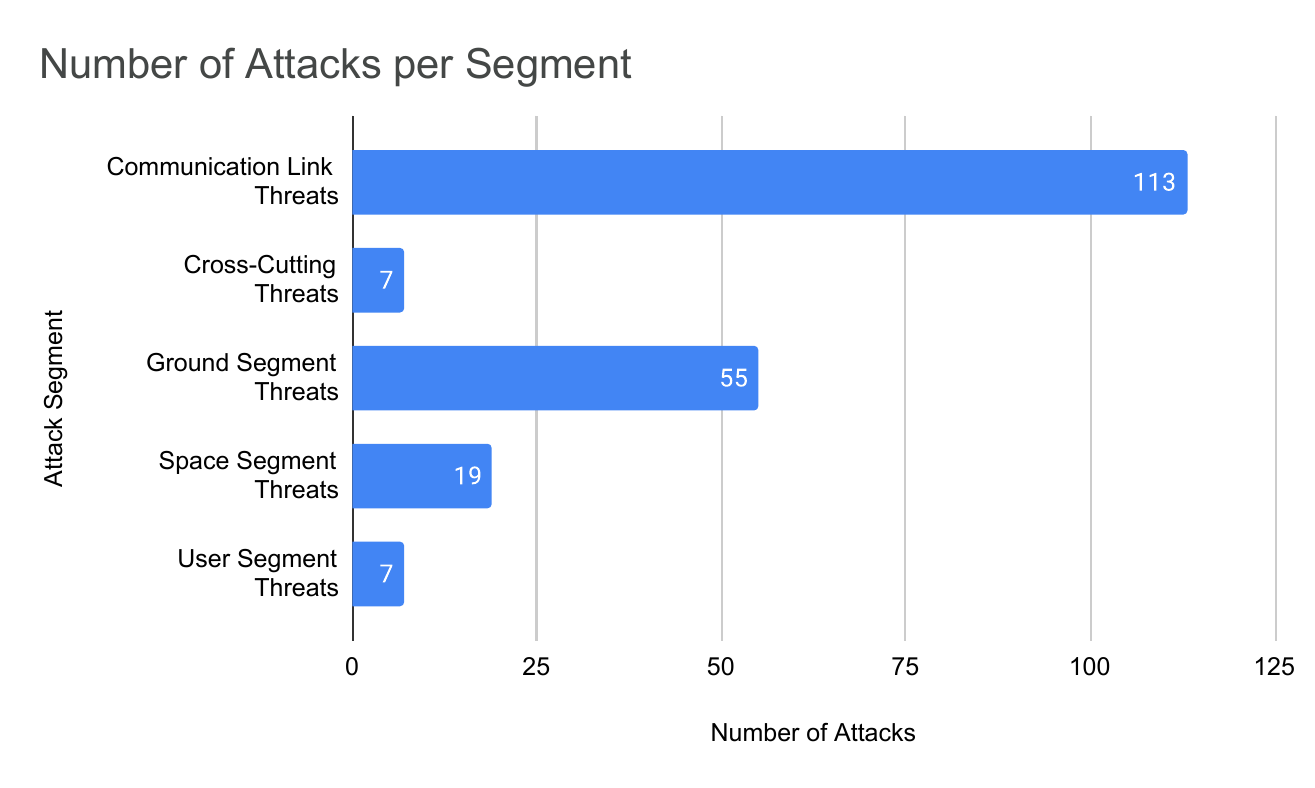}
  \caption{Attack classes by affected segment. Communication-link threats are the most frequent category, followed by ground-segment, space-segment, user-segment, and cross-cutting threats.}
  \label{fig:attk-type}
\end{figure}

As shown in Figure~\ref{fig:attk-decade}, reported attack frequency has increased across decades, with the partial 2020s already exceeding any previous decade in the corpus.
As shown in Figure~\ref{fig:attk-decade}, reported attack frequency has increased across decades, with the partial 2020s already exceeding any previous decade in the corpus.
Figure~\ref{fig:attk-type} shows that communication-link threats remain the most frequently reported category, reflecting the long history of RF interference, jamming, spoofing, interception, and signal hijacking.
At the same time, ground-segment threats have become increasingly important in recent decades, particularly through network intrusion, malware, credential abuse, and disruption of satellite-service infrastructure.

\begin{figure}[ht]
  \centering
  \includegraphics[width=\linewidth]{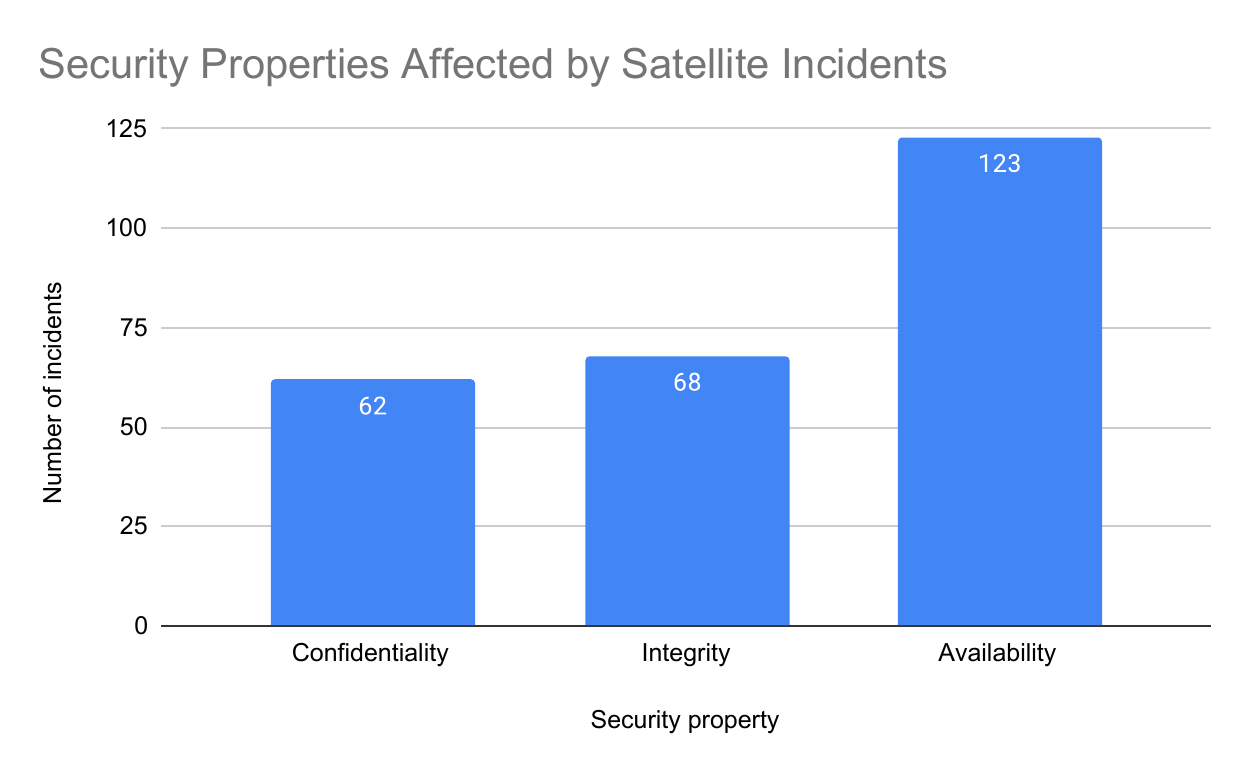}
  \caption{Security properties affected by incidents in the corpus. Availability is the most common effect, reflecting the prevalence of jamming, denial-of-service, and communication disruption, while integrity and confidentiality impacts appear through spoofing, signal hijacking, unauthorized access, and interception. Categories are multi-label and are therefore not mutually exclusive.}
  \label{fig:cia-impact}
\end{figure}

To further characterize the operational consequences of these incidents, we also analyzed the affected security properties and taxonomy-level impact categories.
As shown in Figure~\ref{fig:cia-impact}, availability is the most frequently affected security property, followed by integrity and confidentiality.
This pattern is consistent with the prevalence of jamming, denial-of-service attacks, service disruptions, spoofing, interception, and unauthorized access in public satellite incident reporting.
Because some incidents affect multiple security properties, these categories are not mutually exclusive.

\begin{figure}[ht]
  \centering
  \includegraphics[width=\linewidth]{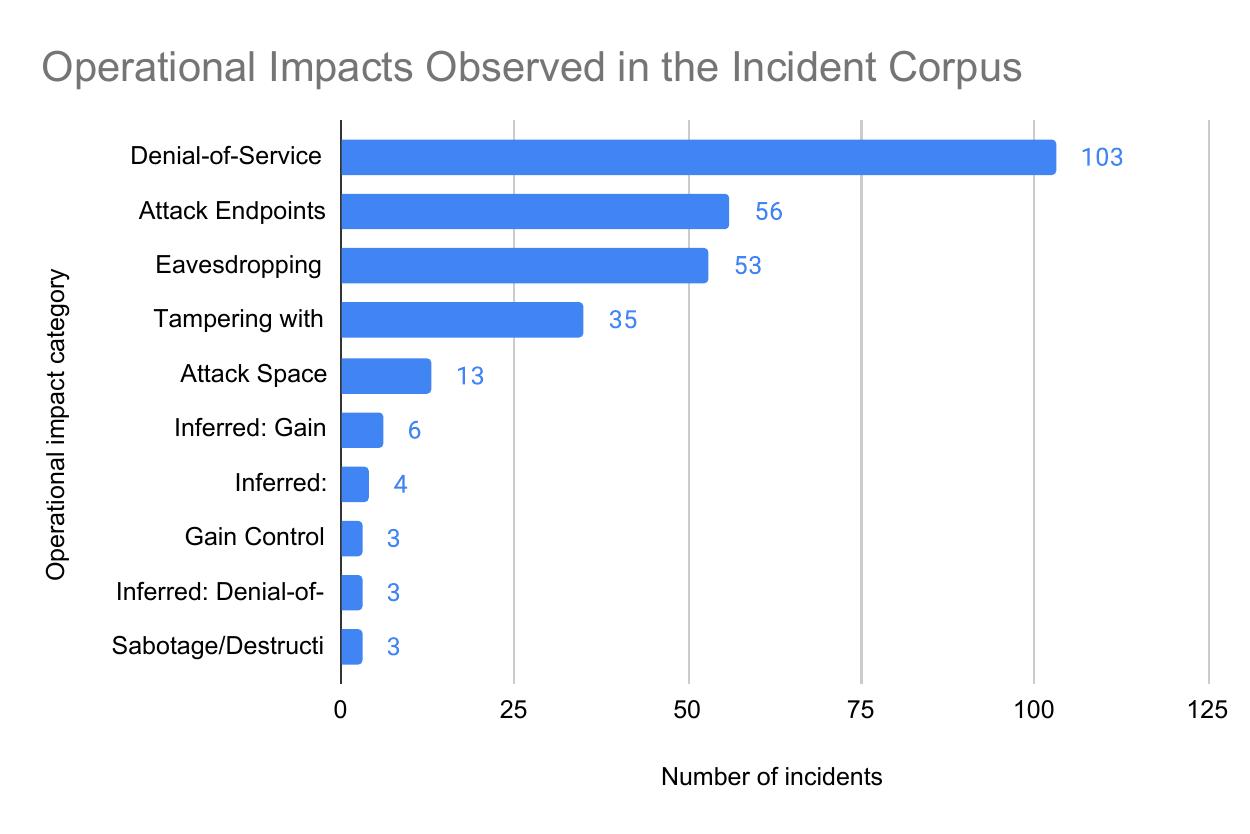}
  \caption{Operational impact categories observed in the incident corpus. Denial-of-service is the most common impact, followed by attacks on ground endpoints, eavesdropping, and tampering with network communications. Categories are multi-label and may overlap across incidents.}
  \label{fig:operational-impact}
\end{figure}

Figure~\ref{fig:operational-impact} summarizes the most common operational impacts observed in the corpus.
Denial-of-service is the dominant impact category, followed by attacks on ground endpoints, eavesdropping, and tampering with network communications.
This distribution underscores the need for a taxonomy that captures not only technical attack vectors but also the mission-level consequences of satellite cyber and electronic warfare activity.
As with the security-property analysis, operational-impact categories are multi-label and may overlap across incidents.

\textit{Dataset limitations.} This corpus reflects publicly available incident reporting, which may over-represent Western space operators and signal-layer attacks.
True adversary behavior may differ in unreported or classified incidents, and earlier historical events often lack the level of technical detail available in more recent reports.
\subsection{Space Segment Threats}

Space segment threats target satellites in orbit, including spacecraft buses, payloads, onboard firmware, and inter-satellite interfaces. 
These threats span both kinetic and cyber domains and may affect spacecraft availability, integrity, or operational control.

\subsubsection*{Kinetic \& Physical Attacks}
Kinetic anti-satellite (ASAT) capabilities have been demonstrated since the Cold War through destructive tests conducted by the United States and the Soviet Union~\cite{bateman2022mutually}, followed by China's 2007 and India's 2019 demonstrations. 
Such events highlighted both the direct risk of spacecraft loss and the long-term hazard posed by orbital debris. 
Non-kinetic physical threats have also been reported or hypothesized, including suspected laser illumination events in 2006~\cite{pavur2022building}, as well as electromagnetic pulse (EMP) and ultrawideband disruption scenarios that may degrade spacecraft subsystems without direct physical contact.

\subsubsection*{On-Orbit Cyber \& Operational Attacks}
On-orbit cyber threats involve unauthorized modification of spacecraft behavior or mission functions. 
These include system hijacking, payload compromise, firmware manipulation, service disruption, and orbit or trajectory interference. 
The growing reliance on software-defined subsystems, reprogrammable COTS components, and remote update mechanisms has expanded the attack surface and increased the feasibility of persistent on-orbit exploitation.

\subsection{Ground Segment Threats}

Ground segment threats encompass cyber operations targeting terrestrial infrastructure, including mission control centers, ground stations, GSaaS providers, data-processing environments, and operator enterprise networks.

\subsubsection*{Network \& Infrastructure Intrusion}
Attackers frequently exploit credential theft, phishing, and exposed remote-access services to gain initial footholds. 
Documented cases include the 2006 NASA compromise via spear-phishing~\cite{fritz2013satellite} and repeated intrusions attributed to APT groups targeting aerospace organizations~\cite{o2017insights,jensen2020fancy}. 
Such intrusions may enable lateral movement into mission systems and telemetry-processing infrastructure.

\subsubsection*{System Compromise \& Service Disruption}
Ground-segment compromises have repeatedly resulted in operational outages and degraded services. 
The 2022 Viasat KA-SAT incident~\cite{boschetti2022space} demonstrated how exploitation of remote-access infrastructure combined with destructive malware could disrupt tens of thousands of terminals. 
Other incidents include telemetry manipulation, data destruction, and denial of mission-control capabilities.

\subsection{User Segment Threats}

User segment threats target end-user infrastructure such as VSAT terminals, maritime and aviation antennas, and downstream service ecosystems.

\subsubsection*{Terminal Exploitation}
These threats include firmware exploitation, credential compromise, botnet recruitment, and subscription fraud. 
While early incidents focused on satellite-TV piracy, modern attacks increasingly affect operational users, including maritime, aviation, and emergency-response platforms.

\subsubsection*{Localized Denial of Service}
Localized denial-of-service attacks typically involve portable jamming or spoofing systems that disrupt individual receivers rather than space or ground assets directly. 
GNSS-dependent platforms are particularly vulnerable, and such disruptions can create immediate operational hazards.

\subsection{Communication Link Threats}

Communication-link threats target RF uplinks, downlinks, and inter-satellite links that connect space, ground, and user segments.

\subsubsection*{Signal Disruption \& Deception}
Signal disruption and deception represent persistent and widely observed threat classes.
Jamming has been documented since early satellite-broadcasting disruptions such as MED-TV in 1995~\cite{spacesecurity2020} and continues in modern GNSS conflict zones~\cite{8170760}. 
Spoofing operations, including the 2013 misdirection of the yacht \textit{White Rose of Drachs}~\cite{humphreys2013ut}, demonstrate deliberate manipulation of navigation signals to mislead receivers.

\subsubsection*{Signal Interception \& Takeover}
Signal interception includes passive eavesdropping and active takeover of satellite transmissions. 
Documented examples include the interception of NATO imagery in 2002~\cite{urban2002} and the interception of unencrypted UAV video feeds~\cite{guardian2009}. 
More advanced techniques include replay attacks and cryptographic interception using commercially available RF equipment.

\subsection{Cross-Cutting Threats}

Cross-cutting threats span multiple segments and include insider activity, supply-chain compromise, and manipulation of satellite-derived data.

\subsubsection*{Insider \& Supply-Chain Threats}
Insiders and compromised suppliers can enable privilege escalation, hardware backdoors, and the exfiltration of sensitive data~\cite {del2013towards,falco2021cubesat}. 
Supply-chain risks arise across design, manufacturing, integration, and maintenance phases.

\subsubsection*{Data \& Algorithmic Integrity Threats}
Emerging threats increasingly target downstream analytics rather than spacecraft directly. 
Physical adversarial patches~\cite{du2022physical} and deception tactics, such as painted decoys~\cite{businessinsider2024}, demonstrate that Earth-observation pipelines can be manipulated without compromising satellite infrastructure itself.

\subsection{Summary}

Satellite threats have evolved from early RF interference and broadcast hijacking toward complex, multi-segment cyber operations. 
Contemporary adversaries increasingly combine link-layer attacks, ground intrusions, and space-segment targeting within coordinated campaigns. 
This evolution underscores the value of structured threat frameworks, including the taxonomy developed in this paper and the MITRE-inspired satellite attack lifecycle taxonomy presented in Section~\ref{sec:MITRE}, for analyzing threats and supporting defense planning.

\subsection{Mitigation Framework}
\label{sec:mitigation-framework}

To complement the proposed taxonomy, Table~\ref{tab:mitigation} summarizes practical defenses mapped to the major threat classes and operational segments identified in our framework. 
The table follows a prevention-detection-restoration paradigm and highlights how mitigation strategies align with subsystem-level and cross-segment attack paths observed in real-world incidents~\cite{pavur2022building,tedeschi2022satellite,smailes2024sticky,weerackody2021satellite,jayaweera2018cognitive,seco2021detection,smailes2023watch,yue2023low,lin2023clextract,li2023active}.

\begin{table*}[ht]
\centering
\caption{Mitigation strategies mapped to the operational segments and threat classes identified in our taxonomy, organized under a prevention--detection--restoration framework.}
\label{tab:mitigation}
\begin{tabular}{|p{3.2cm}|p{4.2cm}|p{4.2cm}|p{4.2cm}|}
\hline
\textbf{Segment Targeted} & \textbf{Prevention} & \textbf{Detection} & \textbf{Restoration / Resilience} \\
\hline
\textbf{Space Segment} 
(Satellite Bus \& Payload) 
& Radiation-hardened components, command encryption, bus/payload isolation, authenticated boot, and redundant critical systems. 
& On-orbit anomaly detection, power-side anomaly monitoring (e.g., voltage/current checks), telemetry-integrity checks, and detection of unexpected mode changes. 
& Safe-mode entry and diagnostics, golden-image firmware reload from protected memory, and switchover to redundant units. \\
\hline
\textbf{Ground Segment} 
(Mission Control, GSaaS) 
& Zero-trust segmentation, multi-factor authentication (MFA), secure SDLC for ground software, and patching of remote-access appliances. 
& Endpoint monitoring (EDR), intrusion-detection systems (IDS), and regular supply-chain and software-composition audits. 
& Tested backup and restore procedures, rapid firmware redeployment, and geographically diverse ground infrastructure. \\
\hline
\textbf{User Segment} 
(Terminals, Receivers) 
& Secure boot for terminals, cryptographically signed firmware updates, and user training on social engineering and physical security. 
& Monitoring for abnormal terminal behavior (e.g., unusual data rates or geographic location) and fleet-wide anomaly detection. 
& Remote terminal disabling or "bricking" commands, rapid revocation of compromised credentials, and device-replacement programs. \\
\hline
\textbf{Communication Link} 
(RF Uplink/Downlink) 
& Spread spectrum (FHSS, DSSS), adaptive beamforming, and cryptographic signal authentication (e.g., OSNMA for GNSS). 
& PHY/MAC-layer anomaly checks, transmitter fingerprinting, and monitoring of command streams and signal characteristics. 
& Failover to redundant ground paths, dynamic rerouting, and rapid rekeying of encrypted channels. \\
\hline
\textbf{Cross-Cutting Threats} 
(Insider Access, Supply Chain, Data Integrity)
& Vendor security certification (ISO 27001/AS9100), strict supplier vetting and hardware-provenance checks, segregated manufacturing lines, secure firmware signing, least-privilege access, and personnel background screening. 
& Continuous configuration and checksum auditing, runtime attestation of onboard software, supply-chain telemetry verification, and AI-based anomaly detection for data and algorithmic integrity. 
& Rapid certificate and credential revocation, supplier recall and re-imaging workflows, digital-twin validation to restore trustworthy states, and diversity-based redundancy (e.g., cross-vendor component sourcing). \\
\hline
\end{tabular}
\end{table*}

These mitigations illustrate how the taxonomy can support defensive planning by linking adversarial techniques to concrete operational safeguards across satellite systems.
\section{\label{sec:MITRE}MITRE-Inspired Satellite Attack Lifecycle Taxonomy}

This section operationalizes the threat-analysis framework introduced in Section~\ref{app:threat_analysis}. 
Section~\ref{app:threat_analysis} defined the core relationships among adversaries, capabilities, vulnerabilities, target assets, tactics, techniques, procedures, and impacts. 
Here, we use those concepts to organize satellite-specific adversarial behavior into a lifecycle-oriented taxonomy grounded in the incident corpus and the satellite-system model developed in Sections~\ref{sec:space_env}--\ref{sec:satellite_vuln}.

We organize the satellite cybersecurity knowledge gathered in this study using a MITRE-inspired adversarial lifecycle taxonomy.

We use the following terminology throughout this section:

\begin{itemize}[leftmargin=*,noitemsep]
\item \textbf{Tactics} describe why a technique or sub-technique is used and represent the adversary's technical objective.
\item \textbf{Techniques} describe how the adversary attempts to achieve a tactical objective.
\item \textbf{Sub-techniques} provide more specific descriptions of adversarial behavior where sufficient technical detail is available.
\item \textbf{Procedures} describe concrete implementations of techniques or sub-techniques in real-world incidents, academic demonstrations, or industrial studies.
\end{itemize}

\begin{figure*}[ht]
\centering
\includegraphics[width=0.8\linewidth]{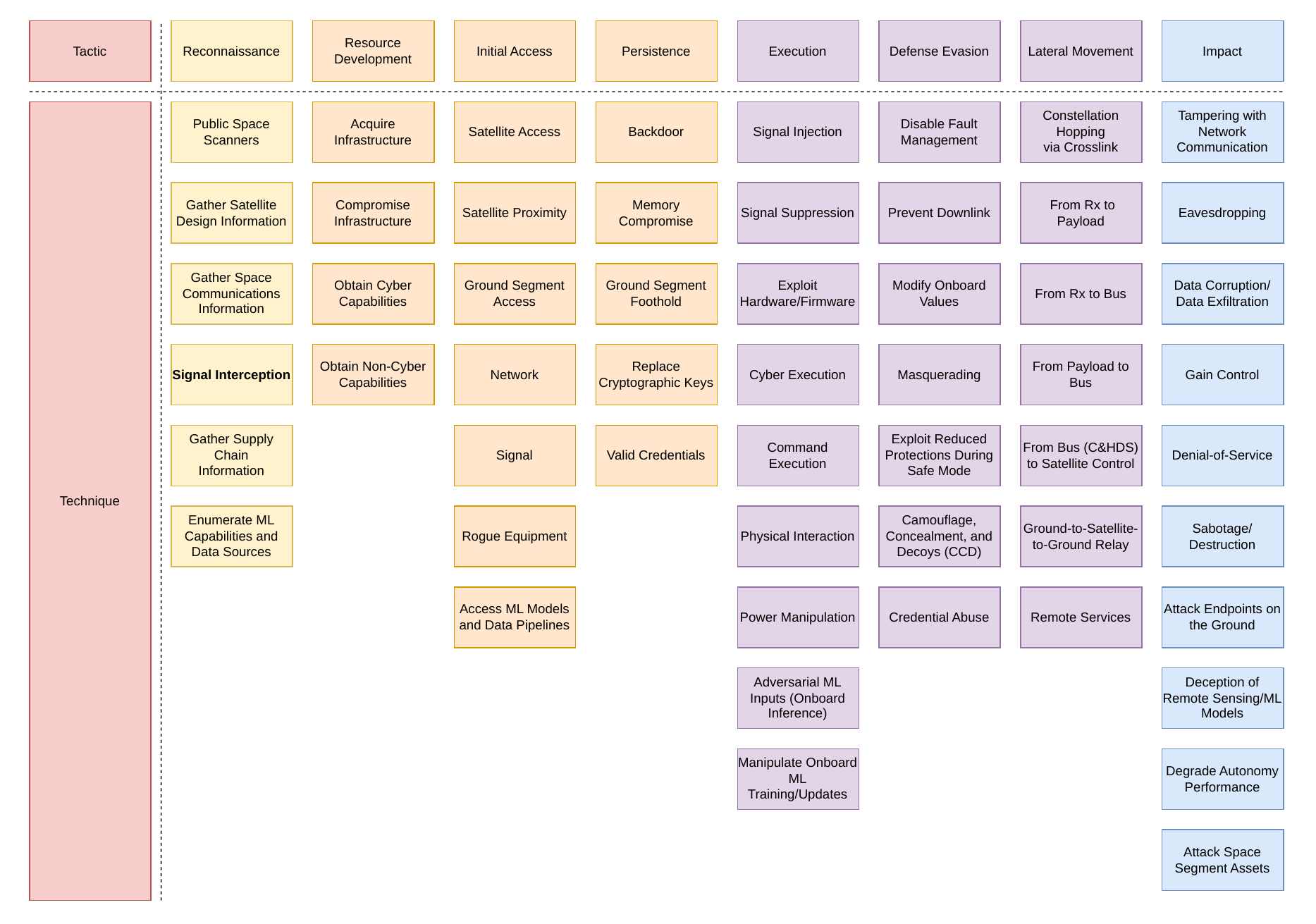}
\caption{MITRE-inspired satellite attack lifecycle taxonomy.}
\label{fig:MITRE}
\end{figure*}

The matrix in Figure~\ref{fig:MITRE} represents the proposed satellite attack lifecycle taxonomy.
Its columns correspond to broad adversarial phases, while its entries describe satellite-relevant techniques and procedures.
This representation supports structured analysis of satellite threats while preserving the attack-lifecycle perspective used in general adversarial frameworks.

\noindent\textbf{Subsystem-Level Exposure.}
To connect the lifecycle taxonomy to spacecraft operations, we associate representative attack behaviors with the satellite subsystems, payload functions, communication links, and mission interfaces most likely to be exposed.
This creates an operational layer between the subsystem model introduced in Section~\ref{sec:subsystem} and the taxonomy developed in this section.
The mapping is not treated as a mandatory incident-level field for every dataset row, because many public reports do not provide enough technical detail to identify the affected internal subsystem with confidence.
Instead, subsystem-level interpretation is applied only when supported by the incident description, cited sources, or the operational meaning of the attack.

Experimental satellite-security studies further motivate this subsystem-level view.
Willbold et al.~\cite{willbold2023space} show that onboard firmware threats often propagate through the COM--CDHS command path, dangerous telecommands, bus--payload interfaces, and insufficient access-control boundaries.
Similarly, Planta et al.~\cite{planta2026satbleed} show that COTS communication modules and their corresponding ground-station front-end processors (FEPs) can act as exposed TT\&C gateways, where weaknesses in authentication, routing, debug interfaces, and frame processing may affect both spacecraft and ground-side operations.
These studies support treating COMMS, TT\&C, C\&DHS, payload interfaces, and internal buses as connected operational exposure points when the evidence or attack model supports such interpretation.

\begin{table*}[t]
\centering
\footnotesize
\setlength{\tabcolsep}{4pt}
\renewcommand{\arraystretch}{0.95}
\caption{Connection between lifecycle attack behavior, subsystem exposure, and operational interpretation.}
\label{tab:mitre_subsystems}
\begin{tabular}{L{4.0cm} C{4.5cm} C{7.0cm}}
\toprule
\textbf{Attack behavior} & \textbf{Subsystems or interfaces exposed} & \textbf{Operational interpretation} \\
\midrule
RF interception, injection, replay, jamming, or frame manipulation & COMMS, TT\&C, FEP, Payload, ADCS, C\&DHS & Affects command, telemetry, payload data, navigation, synchronization, or service availability; may also exploit COM/FEP authentication, routing, or frame-processing flaws. \\

Ground-segment compromise or credential abuse & Ground interfaces, COMMS, C\&DHS, Payload & Enables unauthorized access to mission systems, command paths, or service infrastructure. \\

Command, firmware, or onboard software manipulation & C\&DHS, TT\&C, COMMS, Payload, bus--payload interfaces & May alter spacecraft behavior, command execution, COM configuration, payload operation, firmware state, or persistence through dangerous telecommands, insecure updates, or memory compromise. \\

Telemetry, sensor, or data manipulation & ADCS, EPS, Payload, C\&DHS & May mislead operators, control logic, anomaly detection, or autonomous decision-making. \\

Power or subsystem manipulation & EPS, C\&DHS, ADCS, Payload & May trigger resets, safe mode, degraded operation, or cascading subsystem effects. \\

Crosslink, proximity, or constellation-scale attacks & COMMS, PCS, ADCS, Payload, C\&DHS & May propagate effects across satellites or disrupt constellation-level services. \\

Adversarial ML inputs or model manipulation & Payload, ADCS, C\&DHS & May degrade perception, autonomy, detection, or mission decision-support systems. \\
\bottomrule
\end{tabular}
\end{table*}

This subsystem-level view links each attack pattern to three layers: the adversarial lifecycle behavior, the exposed satellite subsystem or interface, and the resulting operational impact.
For example, signal injection is not only an execution technique; its mission meaning depends on whether it affects TT\&C commands, ADCS inputs, payload data, or user-facing services.

\noindent\textbf{Taxonomy derivation.}
The techniques below were derived from three sources of evidence: recurring behaviors observed in the incident corpus, attack surfaces identified in the satellite ecosystem and subsystem analysis, and emerging attack patterns supported by recent satellite-security research.
Accordingly, not every technique appears in every historical incident.
Some techniques represent frequently observed incident classes, such as jamming, spoofing, signal hijacking, ground-segment compromise, and terminal exploitation, while others capture subsystem-level or emerging attack surfaces needed for threat modeling of modern satellite missions.

\subsection{Reconnaissance and Resource Development}

The adversary collects information, identifies targets, and prepares resources for later attack phases.
In satellite systems, reconnaissance may involve orbital tracking, RF observation, technical documentation collection, supply-chain analysis, and identification of ground or user-segment infrastructure.

\noindent\textbf{Public Space Scanners:}
Online services such as LEOLABS~\cite{leolabs} and N2YO~\cite{n2yo} enable adversaries to monitor satellite movement in real time.

\noindent\textbf{Gather Satellite Design Information:}
With COTS subsystems widely available~\cite{manulis2021cyber}, adversaries can obtain technical details from resources such as Theia~\cite{theia-esat} or CubeSatShop~\cite{cubeshop}.

\noindent\textbf{Gather Space Communications Information:}
Adversaries may collect frequencies, modulation parameters, command patterns, beacon data, or link behavior to prepare later exploitation or interference.

\noindent\textbf{Signal Interception:}
Signal interception refers to the passive acquisition of RF transmissions to gather operational or sensitive information.
This may include interception of telemetry, payload downlinks, navigation signals, or user communications using ground receivers or software-defined radio platforms.
Such activity enables adversaries to infer system behavior, extract data, or prepare subsequent attacks.

\noindent\textbf{Gather Supply-Chain Information:}
Adversaries may identify mission partners, suppliers, integration facilities, component dependencies, or known supply-chain vulnerabilities~\cite {falco2018vacuum,vanlyssel2025spychain}.

\noindent\textbf{Enumerate ML-Enabled Services:}
Adversaries may identify which satellites, ground services, or user applications rely on machine-learning models, for example, for anomaly detection, routing, autonomy, or image analysis.
This may be done by studying documentation, APIs, observable system behavior, or related software artifacts.

\noindent\textbf{Resource Development:}
Adversaries may prepare tools, infrastructure, malware, RF equipment, orbital resources, or access channels for later use.

\noindent\textbf{Acquire Infrastructure:}
This may include purchasing ground-station hardware, renting commercial ground-station-as-a-service resources, acquiring SDR equipment, or obtaining satellite-related components.

\noindent\textbf{Compromise Infrastructure:}
Adversaries may compromise or gain unauthorized access to ground-station systems, cloud-hosted mission infrastructure, contractor networks, mission-support networks, or third-party satellite-service providers to prepare for later attack stages.
This may include the use of stolen credentials, malware, exposed services, or compromised administrative access.

\noindent\textbf{Obtain Cyber Capabilities:}
Adversaries may acquire exploits, malware, stolen credentials, cryptographic material, or access to compromised infrastructure.

\noindent\textbf{Obtain Non-Cyber Capabilities:}
Adversaries may develop or acquire physical and RF capabilities, including lasers, high-power microwave systems, jammers, spoofers, or kinetic counterspace systems~\cite{way2019counterspace}.

\subsection{Initial Access}

Initial access refers to attempts to establish a foothold within a satellite mission, ground infrastructure, communication link, user segment, or associated operational environment.
In the satellite domain, entry may occur through physical, RF, cyber, supply-chain, or data-driven pathways depending on system exposure and mission architecture.

\noindent\textbf{Satellite Access:}
Satellite access refers to gaining direct access to spacecraft hardware during manufacturing, integration, testing, launch preparation, transport, or servicing.
This may include inserting malicious hardware, modifying onboard components, or tampering with subsystem interfaces before deployment.

\noindent\textbf{Satellite Proximity:}
Satellite proximity refers to achieving orbital proximity to a target spacecraft to enable physical, optical, RF, or cyber-enabled intrusion through rendezvous and proximity operations.

\noindent\textbf{Ground Segment Access:}
Ground segment access refers to entry into mission infrastructure through remote intrusion, credential theft, malware deployment, social engineering, exposed services, or physical presence at ground facilities and control centers.

\noindent\textbf{Network Access:}
Network access may involve exploiting exposed services, compromising user terminals, purchasing access to commercial SATCOM networks, or abusing networked mission interfaces.

\noindent\textbf{Signal Access:}
Signal access refers to exploiting RF links to establish operational influence through command injection, spoofing, replay, signal hijacking, unauthorized transmission, or crafted frame delivery.
In satellite TT\&C links, signal access may target not only the spacecraft receiver, but also the ground-station front-end processor or COM module that parses telemetry and telecommand traffic.
Recent experimental studies show that weaknesses in COM access control, replay protection, frame processing, and ground-side telemetry handling can allow adversaries to move from RF-layer access to command-path or cryptographic compromise~\cite{willbold2023space,planta2026satbleed}.

\noindent\textbf{Rogue Equipment:}
Rogue equipment includes unauthorized ground-station hardware, software-defined radios, custom transmitters, or spoofing infrastructure used to bypass operator-controlled communication paths.

\noindent\textbf{Access ML Models and Data Pipelines:}
Adversaries may gain access to model artifacts, training data, telemetry streams, storage systems, or messaging infrastructure used by ML-enabled satellite or ground applications.
Such access may support later adversarial ML, data poisoning, or inference-time manipulation.

\subsection{Persistence}\label{subsec:persistence}

Persistence refers to maintaining continued access or influence over satellite, ground, user, or supporting infrastructure after initial access has been obtained.
In satellite systems, persistence may be especially damaging because remediation can be difficult once a spacecraft is deployed.

\noindent\textbf{Backdoor:}
A backdoor may be inserted in hardware, firmware, SDR configurations, ground software, or supply-chain components to enable covert re-entry after detection or remediation.

\noindent\textbf{Memory Compromise:}
Memory compromise involves injecting or modifying code in volatile or non-volatile memory so that malicious logic survives resets, watchdog recovery, or safe-mode transitions.

\noindent\textbf{Ground Segment Foothold:}
A ground-segment foothold allows adversaries to maintain long-term access to mission control or supporting infrastructure and to repeatedly re-enter satellite operations.

\noindent\textbf{Replace Cryptographic Keys:}
Replacing or inserting cryptographic keys may allow adversaries to retain privileged access while excluding legitimate operators or bypassing authentication controls.

\noindent\textbf{Valid Credentials:}
Valid credentials may be reused after initial compromise to maintain authenticated access across resets, updates, or defensive remediation.
This technique is also relevant to defense evasion because legitimate credentials can help adversaries avoid detection.

\subsection{Execution}

Execution refers to adversary actions that actively manipulate or influence satellite systems, signals, data, or operational behavior through available attack surfaces.
In the satellite domain, execution may occur across RF, cyber, physical, and data-driven layers.

\noindent\textbf{Signal Injection:}
Signal injection refers to the transmission of crafted or replayed RF signals to manipulate satellite subsystems, user receivers, or dependent platforms.
This includes command replay, signal hijacking, GNSS spoofing, and injection of falsified telemetry designed to mislead control logic or navigation systems.

\noindent\textbf{Signal Suppression:}
Signal suppression refers to intentional disruption of satellite communication or navigation services through RF interference or jamming.
These attacks can deny signal availability to ground users, onboard receivers, or mission-control links and may degrade autonomy or mission continuity.

\noindent\textbf{Exploit Hardware or Firmware:}
Adversaries may exploit vulnerabilities in onboard software, firmware interfaces, embedded controllers, COM modules, or ground-station front-end processors to gain or extend access, escalate privileges, or alter system behavior~\cite{manulis2021cyber,willbold2023space,planta2026satbleed}.
Examples include vulnerable telecommand handlers, dangerous debug or maintenance commands, insecure update services, authentication bypasses, memory-corruption bugs, replay weaknesses, and frame-processing vulnerabilities in TT\&C communication paths.

\noindent\textbf{Cyber Execution:}
Cyber execution refers to the operational use of cyber capabilities through compromised systems, distributed endpoints, or attacker-controlled networks.
This may include malware execution, distributed denial-of-service, unauthorized command activity through authenticated channels, or coordinated traffic manipulation across networked assets.
Unlike exploitation, which focuses on gaining access through vulnerabilities, cyber execution describes the use of already available cyber capabilities to carry out operational effects.

\noindent\textbf{Command Execution:}
Command execution refers to issuing or relaying telecommands to spacecraft subsystems through legitimate, compromised, replayed, or spoofed channels.
This may trigger unintended behavior in onboard systems, payloads, or mission-control workflows.
As a concrete procedure, experimental work has shown that dangerous or overly permissive telecommands can allow adversaries to modify access-control state, trigger unsafe subsystem behavior, or interact with debug and configuration interfaces in ways that were intended for maintenance rather than adversarial use~\cite{willbold2023space,planta2026satbleed}.

\noindent\textbf{Firmware Injection:}
Firmware injection refers to installing modified firmware or altering update processes to overwrite satellite or ground-station functionality.
This may enable persistent control, stealthy manipulation of subsystem logic, or long-term mission degradation.

\noindent\textbf{Physical Interaction:}
Physical interaction refers to direct physical or energy-based interference with spacecraft systems.
Examples include kinetic ASAT actions, rendezvous and proximity operations, laser dazzling, or electromagnetic interference targeting spacecraft sensors or electronics.

\noindent\textbf{Power Manipulation:}
Power manipulation refers to modifying or disrupting the satellite's electrical power distribution or power-management behavior.
This may include overloading EPS power lines, forcing brownouts or resets in the onboard computer, manipulating battery charge states, or triggering automatic transitions into safe mode.
Such attacks exploit the tight coupling between power availability and subsystem autonomy and may indirectly affect ADCS, COMMS, payload, or C\&DHS operations.

\noindent\textbf{Adversarial ML Inputs:}
Adversarial ML inputs refer to crafted digital or physical perturbations to sensor data, imagery, navigation signals, or telemetry that cause onboard or ground-based ML models to produce incorrect outputs during inference~\cite{du2022physical,thummala2024adversarial}.

\noindent\textbf{Manipulate Onboard ML Training or Updates:}
Adversaries may modify training data, update schedules, model parameters, or retraining logic to degrade model performance, induce unstable behavior, or exhaust computational resources~\cite{thummala2024adversarial}.

\subsection{Defense Evasion}

Defense evasion refers to adversary behavior intended to avoid detection, conceal activity, or maintain covert operations.
In satellite systems, evasion may involve telemetry manipulation, suppression of fault management, masquerading as legitimate commands, or exploitation of operational modes.

\noindent\textbf{Disable Fault Management:}
Adversaries may suppress alerts, modify thresholds, or disable fault-handling logic to prevent anomaly-handling systems from responding.

\noindent\textbf{Prevent Downlink:}
Adversaries may block, delay, alter, or spoof telemetry streams to prevent ground operators from observing malicious activity.

\noindent\textbf{Modify Onboard Values:}
Adversaries may alter watchdog timers, counters, telemetry values, fault flags, mode indicators, or cryptographic states to bypass monitoring and detection mechanisms.

\noindent\textbf{Masquerading:}
Masquerading refers to impersonating authorized entities, commands, communication patterns, or operational states to evade authentication and detection systems.

\noindent\textbf{Exploit Reduced Protections During Safe Mode:}
Adversaries may exploit simplified operational logic or relaxed protections when satellites enter safe mode.

\noindent\textbf{Camouflage, Concealment, and Decoys:}
Camouflage, concealment, and decoys may physically or electronically obscure spacecraft behavior, signatures, position, or mission activity.

\noindent\textbf{Credential Abuse:}
Credential abuse refers to using compromised, weak, or legitimate credentials to avoid detection while maintaining covert access across systems or mission segments.

\subsection{Lateral Movement}

Lateral movement refers to the propagation of adversarial access across subsystems, payloads, satellites, ground systems, or mission segments after an initial foothold has been established.
Although public incident reports rarely confirm internal spacecraft lateral movement, experimental studies show that such paths are technically feasible in small-satellite architectures where COM modules, CDHS/OBC software, payload interfaces, and internal buses share trusted communication paths~\cite{willbold2023space,planta2026satbleed}.

\noindent\textbf{Constellation Hopping via Crosslink:}
Adversaries may attempt to use inter-satellite links, compatible transceivers, or compromised spacecraft communication modules to propagate effects, commands, or disruption across neighboring satellites in a constellation~\cite{lee2023vulnerabilities,planta2026satbleed}.

\noindent\textbf{From Receiver to Payload:}
Adversaries may exploit payload-facing user, service, or RF interfaces to influence payload behavior, payload data handling, or mission-data paths~\cite{willbold2023space}.

\noindent\textbf{From Receiver to Bus:}
Adversaries may attempt to move from an exposed receiver, COM module, or TT\&C communication path toward spacecraft bus interfaces through crafted RF traffic, replayed commands, vulnerable frame processing, or weak access control~\cite{willbold2023space,planta2026satbleed}.

\noindent\textbf{From Payload to Bus:}
Adversaries may abuse payload-to-bus interfaces, telecommand channels, or payload data-handling paths to affect platform-level functions or cross trust boundaries between mission payloads and bus subsystems~\cite{willbold2023space}.

\noindent\textbf{From Bus to Satellite Control:}
Adversaries may escalate from access to C\&DHS, OBC software, bus controllers, or onboard services toward broader spacecraft control, including command execution, configuration changes, or subsystem manipulation~\cite{willbold2023space}.

\noindent\textbf{Ground-to-Satellite-to-Ground Relay:}
A compromised satellite may be used as a relay, covert channel, or stepping stone to affect ground-segment networks, mission infrastructure, or other connected satellite-service components.

\noindent\textbf{Remote Services:}
Adversaries may leverage accessible onboard or ground-segment services, APIs, debug/configuration interfaces, maintenance functions, or front-end processor services to pivot between subsystems or mission elements after initial access has been established~\cite{planta2026satbleed}.

\subsection{Impact}

Impact refers to adversary actions intended to disrupt, degrade, compromise, deceive, or destroy satellite systems, services, or data.
In satellite missions, impact may affect spacecraft control, payload products, communication availability, navigation integrity, ground operations, user terminals, or downstream decision-making systems.

\noindent\textbf{Tampering with Network Communication:}
Adversaries may disrupt, reroute, manipulate, or degrade SATCOM channels and mission data flows.

\noindent\textbf{Eavesdropping:}
Adversaries may intercept and exploit user data, telemetry, payload products, or control-related communications~\cite{leyva2020leo}.

\noindent\textbf{Data Corruption:}
Data corruption refers to deliberate alteration of satellite, payload, ground-segment, or user-segment data stores.

\noindent\textbf{Gain Control:}
Adversaries may take full or partial control of a satellite, payload, terminal, or mission-support system.

\noindent\textbf{Denial of Service:}
Denial of service may involve overloading systems, blocking signals, disrupting user services, preventing command access, or forcing safe-mode transitions.

\noindent\textbf{Sabotage or Destruction:}
Sabotage or destruction may involve physically or digitally damaging mission assets, onboard components, ground systems, or supporting infrastructure.

\noindent\textbf{Attack Space Segment Assets:}
Adversaries may directly target spacecraft subsystems, payloads, sensors, onboard software, or orbital infrastructure through cyber, RF, physical, or kinetic means.

\noindent\textbf{Attack Endpoints on the Ground:}
Adversaries may target user terminals, service nodes, or other ground endpoints connected to satellite services~\cite{smailes2023dishing}.

\noindent\textbf{Deception of Remote Sensing or ML Models:}
Adversaries may manipulate observed scenes, sensor inputs, or data products so that satellite or ground-based analytics misidentify objects or events~\cite{businessinsider2024}.

\noindent\textbf{Degrade Autonomy Performance:}
Adversaries may reduce the stability, accuracy, or responsiveness of onboard autonomy systems or ML-enabled decision-support systems~\cite{thummala2024adversarial}.

\subsection{Relationship to SPARTA}

SPARTA, developed by The Aerospace Corporation, provides a valuable space-specific ATT\&CK-like knowledge base for organizing adversarial tactics, techniques, and procedures across the space-system attack lifecycle~\cite{sparta2023}.
It also supports defensive analysis through technique descriptions, countermeasure mappings, risk-oriented interpretation, and indicators of adversary behavior.
Our goal is not to replace SPARTA or redefine its adversarial techniques.
Rather, the proposed taxonomy complements SPARTA by adding an incident-grounded operational layer that connects adversarial behavior to satellite architecture, communication links, spacecraft subsystems, ground infrastructure, payload functions, and mission-level impacts.

This distinction is important because SPARTA already represents many core space-cyber behaviors, including jamming, spoofing, replay, ground-system compromise, cryptographic-key replacement, safe-mode abuse, AI/ML training-data poisoning, and constellation hopping via crosslink.
Therefore, the contribution of the proposed taxonomy is not that these themes are absent from SPARTA.
Instead, our taxonomy models them with greater emphasis on incident grounding, subsystem exposure, propagation pathways, communication-link context, and operational impact.

While SPARTA focuses primarily on organizing adversarial behavior as a structured TTP knowledge base, our taxonomy emphasizes how attacks manifest across satellite-system dependencies.
For example, satellite attacks may propagate through TT\&C command paths, COMMS links, ground-station front-end processors, payload-to-bus interfaces, EPS power distribution, ADCS pointing control, and C\&DHS command execution.
Recent experimental work on onboard firmware and COTS communication modules further motivates treating COM modules and ground-side front-end processors as exposed TT\&C gateways whose authentication, routing, debug, and frame-processing behavior can mediate progression from RF-layer access to command-path compromise~\cite{willbold2023space,planta2026satbleed}.

The proposed taxonomy contributes the following complementary perspectives:

\begin{itemize}[leftmargin=*,noitemsep]
\item \textbf{Subsystem-aware interpretation} --- linking techniques to spacecraft subsystems and mission functions, including EPS, ADCS, C\&DHS, payload, COMMS, TT\&C, TCS, and PCS, where supported by evidence or the attack model.
\item \textbf{Communication-link and gateway context} --- connecting adversarial behavior to link directionality, RF access, COM/FEP exposure, TT\&C pathways, payload links, crosslinks, and constellation-level dependencies.
\item \textbf{End-to-end operational context} --- capturing how attacks span ground, communication, space, and user segments, including cross-segment and constellation-scale effects.
\item \textbf{Cross-domain integration} --- combining cyber, RF, physical, deception, and adversarial machine learning threats while distinguishing confirmed historical incidents from emerging or research-supported attack surfaces.
\item \textbf{Empirical grounding} --- deriving and evaluating the taxonomy using a corpus of more than 200 publicly documented satellite incidents spanning 1962--2026.
\item \textbf{Network-centric attack modeling} --- supporting analysis of coordinated attacks such as distributed denial-of-service and constellation-scale disruption by linking traffic generation, routing constraints, communication links, and operational effects.
\end{itemize}

Table~\ref{tab:sparta_comparison} summarizes the relationship between SPARTA and the proposed taxonomy.
Both approaches adopt a lifecycle-based perspective, but SPARTA provides a defender-oriented TTP and countermeasure knowledge base, while this work focuses on connecting adversarial behavior to system architecture, empirical incidents, subsystem exposure, propagation paths, and mission-level impact.

\begin{table*}[ht]
\centering
\footnotesize
\setlength{\tabcolsep}{4pt}
\renewcommand{\arraystretch}{1.05}
\caption{Comparison between SPARTA and the proposed taxonomy.}
\label{tab:sparta_comparison}
\begin{tabular}{L{7.5cm} C{2.8cm} C{5.5cm}}
\toprule
\textbf{Feature} & \textbf{SPARTA} & \textbf{This Work} \\
\midrule
Lifecycle-based TTP modeling & \checkmark & \checkmark \\

Space-specific adversarial technique knowledge base & \checkmark & \checkmark \\

Defensive countermeasure and indicator mappings & \checkmark & Not primary focus \\

Empirical grounding in large-scale incident corpus & Not primary emphasis & \checkmark \\

Longitudinal analysis of attack trends & Not primary emphasis & \checkmark \\

Subsystem-aware operational interpretation & Partially represented & \checkmark \\

Communication-link and COM/FEP gateway interpretation & Partially represented & \checkmark \\

Explicit receiver--payload--bus--control propagation paths & Partially represented & \checkmark \\

Integration of RF, cyber, physical, deception, and AML threats & Partially represented & \checkmark \\

Distributed and constellation-scale network attack modeling & Partially represented & \checkmark \\

Support for incident-level trend analysis & Not primary design goal & \checkmark \\

Multi-label operational impact analysis & Partially represented & \checkmark \\

Standalone exfiltration tactic & \checkmark & Represented through interception, eavesdropping, and related impact categories \\
\bottomrule
\end{tabular}
\end{table*}

Accordingly, SPARTA and the proposed taxonomy serve complementary roles.
SPARTA provides a structured space-cyber knowledge base for organizing adversarial techniques and supporting defensive analysis.
The taxonomy developed in this work provides an empirically grounded and system-level perspective that supports analysis of operational impact, cross-segment interactions, subsystem exposure, propagation paths, longitudinal trends, and emerging threat patterns in modern satellite missions.
\section{\label{sec:demonstration} Demonstration: Using the Taxonomy to Model Attacks}

This section demonstrates the practical utility of the proposed taxonomy through two detailed case studies and six additional mapped incidents.
The case studies cover the 2022 Viasat KA-SAT cyberattack as a confirmed real-world incident and the ICARUS constellation-scale denial-of-service scenario as a simulation-based disruption model.
The additional examples in Table~\ref{tab:tax_map} show how the taxonomy supports consistent analysis across attack types, segments, adversary objectives, and operational impacts.

For each case study, we followed the same evidence-based mapping procedure used for the incident corpus.
Reported steps were mapped only when supported by public technical reports, incident analyses, or explicit simulation assumptions.
When a lifecycle step was necessary to explain the attack but was not directly described in public reporting, it was marked as a supported high-level inference rather than treated as confirmed evidence.
Unsupported intermediate steps were not reconstructed.
This differs from extrapolation-based approaches that generate multiple plausible attack chains from incomplete reports; our mappings represent only reported evidence, supported high-level inference, or explicitly modeled behavior.

\subsection{Case Study — Viasat Cyberattack}

The Viasat cyberattack occurred in the early hours of February 24, 2022, coinciding with the initial phase of Russia's invasion of Ukraine.
While the attack primarily targeted Ukrainian military communications~\cite{boschetti2022space}, it caused collateral disruption affecting civilian infrastructure, including German wind turbines~\cite{willuhn2022satellite} and thousands of Internet users across Europe~\cite{viasat2022,viasatt2022}.

\begin{figure}[!ht]
  \includegraphics[width=\linewidth]{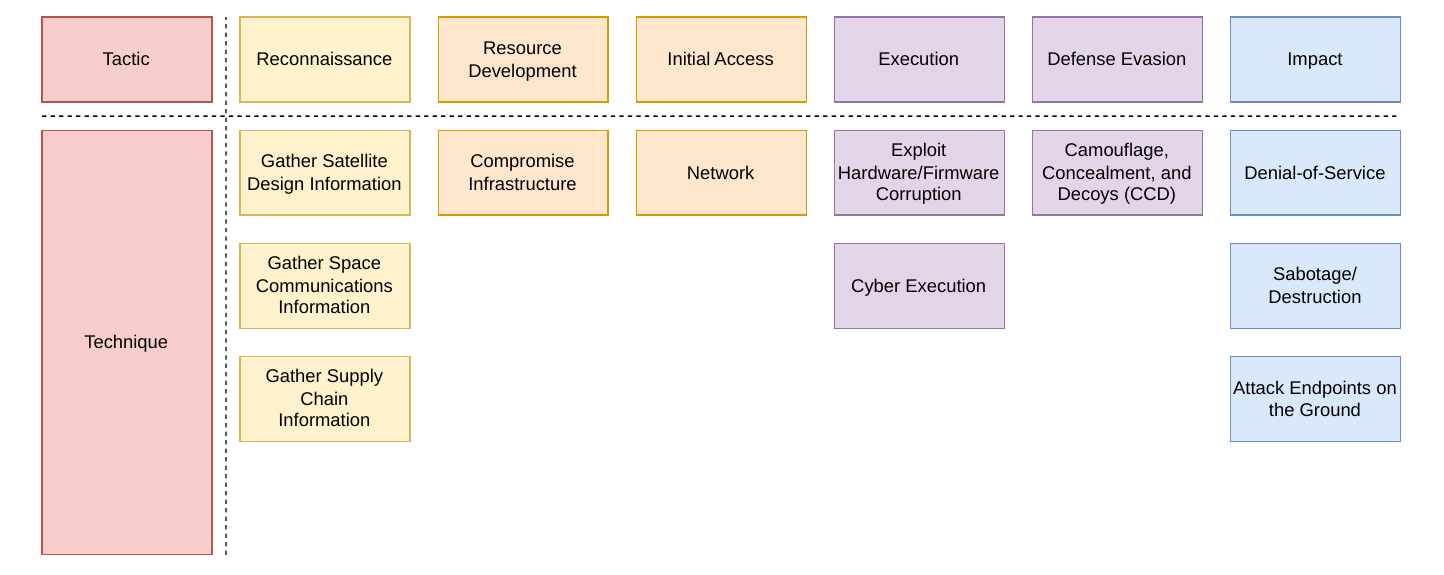}
  \caption{Case study — Viasat cyberattack mapped to the proposed taxonomy.}
  \label{fig:ViaSat}
\end{figure}

\noindent\textbf{Reconnaissance:}
Public reporting does not fully disclose the adversary's pre-compromise reconnaissance process.
We therefore treat reconnaissance of the KA-SAT ground-segment architecture and management environment as a supported high-level inference, rather than as a directly confirmed step.

\textbf{Resource Development:}
The preparation of attack infrastructure and destructive functionality is treated as a supported high-level inference.
Public reporting describes the later compromise and destructive effect, but does not provide enough detail to reconstruct the full tool-development or infrastructure-preparation process.

\textbf{Initial Access:}
Public reporting indicates that the adversaries gained access through the KA-SAT ground-segment management environment, including exploitation of a misconfigured VPN appliance~\cite{viasat2022}.
This step is mapped to ground-segment access because the attack path depended on compromise of terrestrial management infrastructure rather than direct RF access to the satellite.

\textbf{Execution:}
The adversaries deployed destructive functionality associated with the AcidRain wiper malware to overwrite modem flash storage remotely~\cite{sentinelone2022acidrain}.
This step is mapped to cyber execution because the attack used compromised operational pathways to trigger destructive behavior on satellite user terminals.

\textbf{Impact:}
The operation caused widespread disruption across the KA-SAT network.
Approximately 40,000 modems were disabled across Europe, affecting Ukrainian military communications and civilian users, including energy-sector infrastructure such as German wind turbines~\cite{viasatt2022,willuhn2022satellite}.
The attack is therefore mapped to denial-of-service and attack endpoints on the ground.

\subsection{Case Study — ICARUS}

ICARUS is treated as a simulation-based attack model rather than as a documented real-world incident.
Although some real-world campaigns, such as KillNet DDoS operations targeting Starlink-related infrastructure~\cite{killnet2025}, share partial characteristics, they primarily targeted terrestrial web services and did not exploit orbital dynamics, constrained routing, or inter-satellite links as assumed in the ICARUS scenario.

The ICARUS model describes a distributed denial-of-service attack against low Earth orbit satellite networks.
The scenario assumes an adversary that can generate traffic through compromised or attacker-controlled endpoints and use knowledge of constellation topology, orbital motion, and routing constraints to congest critical links~\cite{giuliari2021icarus}.
The mapping below therefore represents explicitly modeled behavior, not confirmed historical evidence.

\begin{figure}[!ht]
\includegraphics[width=\linewidth]{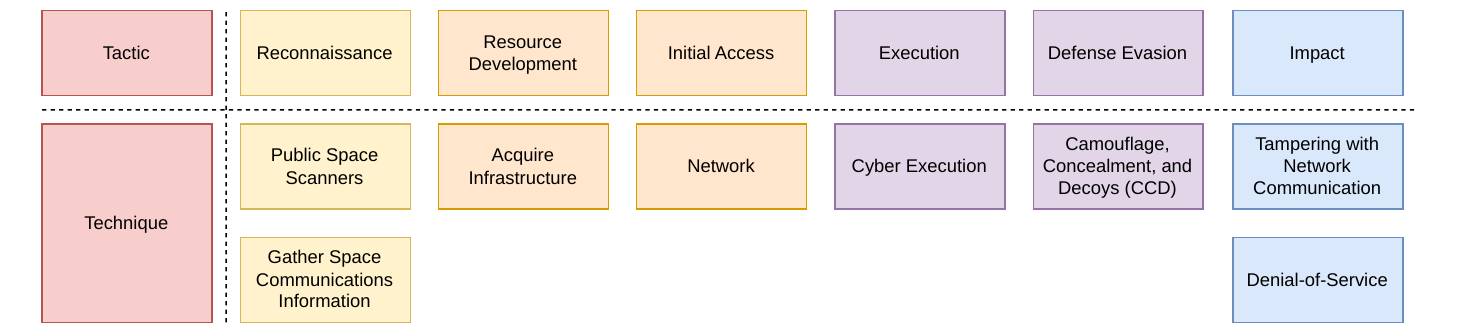}
\caption{Case study — ICARUS mapped to the proposed taxonomy.}
\label{fig:ICARUS}
\end{figure}

\noindent\textbf{Reconnaissance:}
The ICARUS scenario assumes adversary knowledge of constellation topology, satellite motion, and routing behavior.
This information is used to identify bottleneck links and timing windows for disruption.

\textbf{Resource Development:}
The scenario assumes a distributed traffic-generation capability, such as a botnet or attacker-controlled endpoints, capable of producing sustained traffic toward selected satellite-network paths.

\textbf{Initial Access:}
The adversary does not require direct compromise of spacecraft subsystems.
Instead, the modeled attack uses legitimate network entry points to inject traffic into the LEO satellite network.

\textbf{Execution:}
The modeled attack coordinates traffic flows to overload selected inter-satellite or satellite-to-ground links.
By exploiting limited routing diversity, the adversary may induce congestion across critical communication paths, resulting in degraded throughput and increased latency~\cite{giuliari2021icarus}.

\textbf{Defense Evasion:}
The scenario assumes that malicious traffic can resemble legitimate user traffic, making detection more difficult.
This behavior is mapped to camouflage, concealment, and decoys in the proposed taxonomy.

\textbf{Impact:}
The modeled impact is denial-of-service through sustained congestion, degraded communication performance, increased latency, and reduced availability across parts of the constellation.
In severe cases, congestion may propagate across interconnected links and amplify network-wide disruption.
\section{\label{sec:Conclusion}Conclusion and Future Work}

This paper presents a system-level analysis of the evolving cybersecurity landscape of modern satellite systems, with emphasis on LEO constellations and evidence from systems operating across LEO, MEO, and GEO.
The main finding of this work is that satellite cybersecurity cannot be analyzed effectively through isolated incident chronologies, generic cyber frameworks, or subsystem descriptions alone.
Instead, it requires an incident-grounded model that links adversarial behavior to satellite architecture, communication paths, spacecraft subsystems, adversary capabilities, and mission-level impact.

To address this gap, we constructed a reproducible incident-level corpus of publicly reported satellite cyber and electronic warfare incidents spanning 1962--2026, including more than 80 incidents from 2019--2026.
Unlike prior resources that primarily provide historical incident lists, risk/control mappings, or extrapolated attack chains, our corpus provides structured metadata, source references, uncertainty indicators, affected segments, communication links, and ATT\&CK-based lifecycle mappings.
This enables direct inspection, future validation, and consistent comparison across incidents.
For suspected or contested cases, we preserved source-level uncertainty rather than reconstructing unsupported attack-chain details as confirmed historical evidence.

Building on this corpus and the system analysis,we introduced a satellite-specific taxonomy inspired by MITRE ATT\&CK and selectively aligned with SPARTA where applicable by integrating cyber, radio-frequency (RF), physical, deception-oriented, and adversarial machine learning attack surfaces into a unified lifecycle-oriented framework.
The proposed taxonomy explicitly links adversarial behavior to satellite segments, communication link directionality, spacecraft subsystems, adversary capabilities, and operational consequences.
This provides a more operational view of how attacks can propagate across contemporary satellite ecosystems, particularly in distributed constellations and cross-segment architectures.

The longitudinal analysis revealed several important shifts in the satellite threat landscape.
Recent incidents show increasing operational use of GNSS jamming and spoofing, growing reliance on ground-segment compromise and user-terminal exploitation, and broader use of deception and service-disruption campaigns.
At the same time, adversarial machine learning remains primarily an emerging prospective threat rather than a publicly confirmed root cause of satellite incidents.
This distinction is important: the taxonomy includes AML-related attack surfaces to support future threat modeling, while the historical corpus separates confirmed incidents from suspected, contested, and research-motivated attack classes.

We demonstrated the practical utility of the taxonomy through two complementary case studies: the 2022 Viasat KA-SAT cyberattack as a confirmed real-world cyber incident, and the ICARUS constellation-scale denial-of-service scenario as a simulation-based disruption case.
Together with additional mapped incidents, these examples show how the framework can support threat analysis, detection design, mitigation planning, and consistent comparison across different attack types, segments, adversary objectives, and operational impacts.

The proposed dataset and taxonomy provide practical value for the design of resilient satellite architectures.
They can support subsystem-level intrusion detection, simulation-driven evaluation of defensive mechanisms, mission-level risk assessment, and more consistent threat-intelligence sharing across operators, agencies, and researchers.
However, securing satellite systems remains difficult due to constrained onboard resources, limited patching opportunities, long mission lifetimes, complex supply chains, and increasing reliance on cloud services, user terminals, and autonomous onboard functions.

Future research should address several open challenges.
First, constellation-scale security requires deeper study, particularly lateral movement across inter-satellite links, cascading failures, and coordinated attacks against distributed satellite services.
Second, AI-enabled payloads and onboard autonomy introduce new risks related to adversarial inputs, telemetry integrity, model-update security, and decision-support reliability.
Third, future work should improve methods for validating ambiguous public reports and distinguishing confirmed attacks from suspected or contested incidents.
Fourth, supply-chain dependencies and cross-organizational attack surfaces remain critical challenges for large commercial constellations.
Addressing these challenges will require closer collaboration among academia, industry, and government stakeholders to develop scalable, operationally grounded security frameworks for next-generation space systems.

\section*{Declaration of Generative AI Use}

The authors used generative AI tools to assist with literature exploration, identification of candidate publicly reported incidents, and organization and formatting of supplementary materials.
All incident entries, classifications, source mappings, and technical claims were independently reviewed and verified by the authors using cited sources.
The authors edited and validated the final manuscript and take full responsibility for its content.

% --------------------------------------------------
% Bibliography
% --------------------------------------------------

\bibliographystyle{elsarticle-num}
\bibliography{references}

@article{hassanien2020machine,
  title={Machine learning in telemetry data mining of space mission: basics, challenging and future directions},
  author={Hassanien, Aboul Ella and Darwish, Ashraf and Abdelghafar, Sara},
  journal={Artificial Intelligence Review},
  volume={53},
  pages={3201--3230},
  year={2020},
  publisher={Springer}
}

@incollection{nguyen2020communication,
  title={Communication subsystems for satellite design},
  author={Nguyen, Hung H and Nguyen, Peter S},
  booktitle={Satellite Systems-Design, Modeling, Simulation and Analysis},
  year={2020},
  publisher={IntechOpen}
}

@inproceedings{eger2008orion,
  title={Orion's Command and Data Handling Architecture},
  author={Eger, George},
  booktitle={AIAA Space 2008 Conference \& Exposition},
  pages={7743},
  year={2008}
}

@inproceedings{giuliari2021icarus,
  title={ICARUS: Attacking low Earth orbit satellite networks.},
  author={Giuliari, Giacomo and Ciussani, Tommaso and Perrig, Adrian and Singla, Ankit and Zurich, E},
  booktitle={USENIX Annual Technical Conference},
  pages={317--331},
  year={2021}
}

@article{pavur2022building,
  title={Building a launchpad for satellite cyber-security research: lessons from 60 years of spaceflight},
  author={Pavur, James and Martinovic, Ivan},
  journal={Journal of Cybersecurity},
  volume={8},
  number={1},
  pages={tyac008},
  year={2022},
  publisher={Oxford University Press}
}

@article{tedeschi2022satellite,
  title={Satellite-based communications security: A survey of threats, solutions, and research challenges},
  author={Tedeschi, Pietro and Sciancalepore, Savio and Di Pietro, Roberto},
  journal={Computer Networks},
  pages={109246},
  year={2022},
  publisher={Elsevier}
}

@article{falco2019cybersecurity,
  title={Cybersecurity principles for space systems},
  author={Falco, Gregory},
  journal={Journal of Aerospace Information Systems},
  volume={16},
  number={2},
  pages={61--70},
  year={2019},
  publisher={American Institute of Aeronautics and Astronautics}
}

@article{manulis2021cyber,
  title={Cyber security in new space: analysis of threats, key enabling technologies and challenges},
  author={Manulis, Mark and Bridges, Christopher P and Harrison, Richard and Sekar, Venkkatesh and Davis, Andy},
  journal={International Journal of Information Security},
  volume={20},
  pages={287--311},
  year={2021},
  publisher={Springer}
}

@article{hosel2014failure,
  title={Failure modes and fast repair procedures in high voltage organic solar cell installations},
  author={H{\"o}sel, Markus and S{\o}ndergaard, Roar R and J{\o}rgensen, Mikkel and Krebs, Frederik C},
  journal={Advanced Energy Materials},
  volume={4},
  number={7},
  pages={1301625},
  year={2014},
  publisher={Wiley Online Library}
}

@inproceedings{falco2021cubesat,
  title={Cubesat security attack tree analysis},
  author={Falco, Gregory and Viswanathan, Arun and Santangelo, Andrew},
  booktitle={2021 IEEE 8th International Conference on Space Mission Challenges for Information Technology (SMC-IT)},
  pages={68--76},
  year={2021},
  organization={IEEE}
}

@incollection{falco2020satellites,
  title={When satellites attack: Satellite-to-satellite cyber attack, defense and resilience},
  author={Falco, Gregory},
  booktitle={ASCEND 2020},
  pages={4014},
  year={2020}
}

@article{zhuo2021survey,
  title={Survey on security issues of routing and anomaly detection for space information networks},
  author={Zhuo, Ming and Liu, Leyuan and Zhou, Shijie and Tian, Zhiwen},
  journal={Scientific Reports},
  volume={11},
  number={1},
  pages={22261},
  year={2021},
  publisher={Nature Publishing Group UK London}
}

@article{fritz2013satellite,
  title={Satellite hacking: A guide for the perplexed},
  author={Fritz, Jason},
  journal={Culture Mandala},
  volume={10},
  number={1},
  pages={5906},
  year={2013},
  publisher={Bond University}
}

@article{humphreys2013ut,
  title={Ut austin researchers spoof superyacht at sea},
  author={Humphreys, T},
  journal={URL: https://cockrell. utexas. edu/news/archive/7649-superyacht-gps-spoofing},
  year={2013}
}

@article{o2017insights,
  title={Insights into Iranian cyber espionage: APT33 targets aerospace and energy sectors and has ties to destructive malware},
  author={O’Leary, Jacqueline and Kimble, Josiah and Vanderlee, Kelli and Fraser, Nalani},
  journal={Threat Research Blog},
  year={2017}
}

@inproceedings{del2013towards,
  title={Towards a cybersecurity policy for a sustainable, secure and safe space environment},
  author={del Monte, Luca},
  booktitle={Proceedings of the 64th International Astronautical Congress (IAC)},
  year={2013}
}

@article{zhang2022security,
  title={Security Performance Analysis of LEO Satellite Constellation Networks under DDoS Attack},
  author={Zhang, Yan and Wang, Yong and Hu, Yihua and Lin, Zhi and Zhai, Yadi and Wang, Lei and Zhao, Qingsong and Wen, Kang and Kang, Linshuang},
  journal={Sensors},
  volume={22},
  number={19},
  pages={7286},
  year={2022},
  publisher={MDPI}
}

@incollection{boschetti2022space,
  title={Space Cybersecurity Lessons Learned from The ViaSat Cyberattack},
  author={Boschetti, Nicol{\`o} and Gordon, Nathaniel G and Falco, Gregory},
  booktitle={ASCEND 2022},
  pages={4380},
  year={2022}
}

@incollection{guest2017telemetry,
  title={Telemetry, tracking, and command (TT\&C)},
  author={Guest, Arthur Norman},
  booktitle={Handbook of Satellite Applications},
  pages={1313--1324},
  year={2017},
  publisher={Springer}
}

@incollection{SOBOLEWSKI2003277,
title = {Data Transmission Media},
editor = {Robert A. Meyers},
booktitle = {Encyclopedia of Physical Science and Technology (Third Edition)},
publisher = {Academic Press},
edition = {Third Edition},
address = {New York},
pages = {277-303},
year = {2003},
isbn = {978-0-12-227410-7},
doi = {https://doi.org/10.1016/B0-12-227410-5/00165-4},
url = {https://www.sciencedirect.com/science/article/pii/B0122274105001654},
author = {John S. Sobolewski}
}

@article{lin20215g,
  title={5G from space: An overview of 3GPP non-terrestrial networks},
  author={Lin, Xingqin and Rommer, Stefan and Euler, Sebastian and Yavuz, Emre A and Karlsson, Robert S},
  journal={IEEE Communications Standards Magazine},
  volume={5},
  number={4},
  pages={147--153},
  year={2021},
  publisher={IEEE}
}

@incollection{strom2018mitre,
  title={Mitre att\&ck: Design and philosophy},
  author={Strom, Blake E and Applebaum, Andy and Miller, Doug P and Nickels, Kathryn C and Pennington, Adam G and Thomas, Cody B},
  booktitle={Technical report},
  year={2018},
  publisher={The MITRE Corporation}
}

@article{rawlins2022death,
  title={Death By A Thousand COTS: Disrupting Satellite Communications using Low Earth Orbit Constellations},
  author={Rawlins, Frederick and Baker, Richard and Martinovic, Ivan},
  journal={arXiv preprint arXiv:2204.13514},
  year={2022}
}

@techreport{lee2023vulnerabilities,
  title={The Vulnerabilities Less Exploited: Cyberattacks on End-of-Life Satellites},
  author={Lee, Frank and Falco, Gregory},
  year={2023},
  institution={EasyChair}
}

@article{way2019counterspace,
  title={Counterspace Weapons 101},
  author={Way, Tyler},
  journal={Aero Space Security (CSIS)},
  year={2019}
}

@inproceedings{falco2018vacuum,
  title={The vacuum of space cyber security},
  author={Falco, Gregory},
  booktitle={2018 AIAA SPACE and Astronautics Forum and Exposition},
  pages={5275},
  year={2018}
}

@incollection{falco2021security,
  title={A security risk taxonomy for commercial space missions},
  author={Falco, Gregory and Boschetti, Nicolo},
  booktitle={ASCEND 2021},
  pages={4241},
  year={2021}
}

@Inbook{You2021,
author="You, Rui
and Gao, Wenjun
and Wu, Chunbang
and Li, Hongbin",
title="Low-Frequency Antenna and Small Satellite Antenna",
bookTitle="Technologies for Spacecraft Antenna Engineering Design ",
year="2021",
publisher="Springer Singapore",
address="Singapore",
pages="231--247",
isbn="978-981-15-5833-7",
doi="10.1007/978-981-15-5833-7_8",
url="https://doi.org/10.1007/978-981-15-5833-7_8"
}

@article{bateman2022mutually,
  title={Mutually assured surveillance at risk: Anti-satellite weapons and cold war arms control},
  author={Bateman, Aaron},
  journal={Journal of Strategic Studies},
  volume={45},
  number={1},
  pages={119--142},
  year={2022},
  publisher={Taylor \& Francis}
}

@article{yahia2022optical,
  title={Optical satellite eavesdropping},
  author={Yahia, Olfa Ben and Erdogan, Eylem and Kurt, Gunes Karabulut and Altunbas, Ibrahim and Yanikomeroglu, Halim},
  journal={IEEE Transactions on Vehicular Technology},
  volume={71},
  number={9},
  pages={10126--10131},
  year={2022},
  publisher={IEEE}
}

@misc{kai2020handbook,
  title={Handbook of space security: policies, applications and programs},
  author={Kai, US and Peter, LH and Jana, R and others},
  year={2020},
  publisher={Springer New York}
}

@inproceedings{willbold2023space,
  title={Space Odyssey: An Experimental Software Security Analysis of Satellites},
  author={Willbold, Johannes and Schloegel, Moritz and V{\"o}gele, Manuel and Gerhardt, Maximilian and Holz, Thorsten and Abbasi, Ali},
  booktitle={IEEE Symposium on Security and Privacy},
  year={2023}
}

@INPROCEEDINGS{8170760,
  author={Wang, Qiwei and Nguyen, Thinh and Pham, Khanh and Kwon, Hyuck},
  booktitle={MILCOM 2017 - 2017 IEEE Military Communications Conference (MILCOM)}, 
  title={Satellite jamming: A game theoretic analysis}, 
  year={2017},
  volume={},
  number={},
  pages={141-146},
  doi={10.1109/MILCOM.2017.8170760}
}

@article{willuhn2022satellite,
  title={Satellite cyber attack paralyzes 11GW of German wind turbines},
  author={Willuhn, Marian},
  year={2022}
}

@article{leyva2020leo,
  title={LEO small-satellite constellations for 5G and beyond-5G communications},
  author={Leyva-Mayorga, Israel and Soret, Beatriz and R{\"o}per, Maik and W{\"u}bben, Dirk and Matthiesen, Bho and Dekorsy, Armin and Popovski, Petar},
  journal={Ieee Access},
  volume={8},
  pages={184955--184964},
  year={2020},
  publisher={IEEE}
}

@article{smailes2023dishing,
  title={Dishing out DoS: How to Disable and Secure the Starlink User Terminal},
  author={Smailes, Joshua and Salkield, Edd and Birnbach, Simon and Strohmeier, Martin and Martinovic, Ivan},
  journal={arXiv preprint arXiv:2303.00582},
  year={2023}
}

@article{perez2019signal,
  title={Signal processing for high-throughput satellites: Challenges in new interference-limited scenarios},
  author={Perez-Neira, Ana I and Vazquez, Miguel Angel and Shankar, MR Bhavani and Maleki, Sina and Chatzinotas, Symeon},
  journal={IEEE Signal Processing Magazine},
  volume={36},
  number={4},
  pages={112--131},
  year={2019},
  publisher={IEEE}
}

@article{darwish2022leo,
  title={LEO satellites in 5G and beyond networks: A review from a standardization perspective},
  author={Darwish, Tasneem and Kurt, Gunes Karabulut and Yanikomeroglu, Halim and Bellemare, Michel and Lamontagne, Guillaume},
  journal={IEEE Access},
  volume={10},
  pages={35040--35060},
  year={2022},
  publisher={IEEE}
}

@inproceedings{kodheli2017integration,
  title={Integration of Satellites in 5G through LEO Constellations},
  author={Kodheli, Oltjon and Guidotti, Alessandro and Vanelli-Coralli, Alessandro},
  booktitle={GLOBECOM 2017-2017 IEEE Global Communications Conference},
  pages={1--6},
  year={2017},
  organization={IEEE}
}

@article{di2019ultra,
  title={Ultra-dense LEO: Integration of satellite access networks into 5G and beyond},
  author={Di, Boya and Song, Lingyang and Li, Yonghui and Poor, H Vincent},
  journal={IEEE Wireless Communications},
  volume={26},
  number={2},
  pages={62--69},
  year={2019},
  publisher={IEEE}
}

@article{zhen2020energy,
  title={Energy-efficient random access for LEO satellite-assisted 6G internet of remote things},
  author={Zhen, Li and Bashir, Ali Kashif and Yu, Keping and Al-Otaibi, Yasser D and Foh, Chuan Heng and Xiao, Pei},
  journal={IEEE Internet of Things Journal},
  volume={8},
  number={7},
  pages={5114--5128},
  year={2020},
  publisher={IEEE}
}

@article{kim2020new,
  title={New era of air quality monitoring from space: Geostationary Environment Monitoring Spectrometer (GEMS)},
  author={Kim, Jhoon and Jeong, Ukkyo and Ahn, Myoung-Hwan and Kim, Jae H and Park, Rokjin J and Lee, Hanlim and Song, Chul Han and Choi, Yong-Sang and Lee, Kwon-Ho and Yoo, Jung-Moon and others},
  journal={Bulletin of the American Meteorological Society},
  volume={101},
  number={1},
  pages={E1--E22},
  year={2020}
}

@article{early2021spying,
  title={Spying from space: Reconnaissance satellites and interstate disputes},
  author={Early, Bryan R and Gartzke, Erik},
  journal={Journal of Conflict Resolution},
  volume={65},
  number={9},
  pages={1551--1575},
  year={2021},
  publisher={SAGE Publications Sage CA: Los Angeles, CA}
}

@article{shaofei2018analysis,
  title={Analysis of detection capabilities of LEO reconnaissance satellite constellation based on coverage performance},
  author={Shaofei, MENG and Jiansheng, SHU and Qi, YANG and Wei, XIA},
  journal={Journal of Systems Engineering and Electronics},
  volume={29},
  number={1},
  pages={98--104},
  year={2018},
  publisher={BIAI}
}

@article{de2019degree,
  title={The degree of the lack of regulation of space debris within the current space law regime and suggestions for a prospective legal framework and technological interventions},
  author={de Waal Alberts, Anton},
  journal={Space Security and Legal Aspects of Active Debris Removal},
  pages={93--106},
  year={2019},
  publisher={Springer}
}

@inproceedings{jones2018recent,
  title={The recent large reduction in space launch cost},
  author={Jones, Harry},
  year={2018},
  organization={48th International Conference on Environmental Systems}
}

@article{pavur2020sok,
  title={Sok: Building a launchpad for impactful satellite cyber-security research},
  author={Pavur, James and Martinovic, Ivan},
  journal={arXiv preprint arXiv:2010.10872},
  year={2020}
}

@book{li2014geostationary,
  title={Geostationary satellites collocation},
  author={Li, Hengnian},
  year={2014},
  publisher={Springer}
}

@inproceedings{hiriart2009comparative,
  title={Comparative reliability of GEO, LEO, and MEO satellites},
  author={Hiriart, Thomas and Castet, Jean-Francois and Lafleur, Jarret M and Saleh, Joseph H},
  booktitle={Proceedings of the international astronautical congress, IAC-09 D},
  volume={1},
  year={2009}
}

@article{vatalaro1995analysis,
  title={Analysis of LEO, MEO, and GEO global mobile satellite systems in the presence of interference and fading},
  author={Vatalaro, Francesco and Corazza, Giovanni Emanuele and Caini, Carlo and Ferrarelli, Carlo},
  journal={IEEE Journal on selected areas in communications},
  volume={13},
  number={2},
  pages={291--300},
  year={1995},
  publisher={IEEE}
}

@article{martin2012nasa,
  title={Nasa cybersecurity: An examination of the agency’s information security},
  author={Martin, Paul K and General, Inspector},
  journal={NASA, Testimony Before the Subcommittee on Investigations and Oversight, US House of Representatives, House Committee on Science, Space, and Technology},
  volume={29},
  year={2012}
}

@article{costin2023cybersecurity,
  title={Cybersecurity of COSPAS-SARSAT and EPIRB: threat and attacker models, exploits, future research},
  author={Costin, Andrei and Khandker, Syed and Turtiainen, Hannu and H{\"a}m{\"a}l{\"a}inen, Timo},
  journal={arXiv preprint arXiv:2302.08361},
  year={2023}
}

@article{falco2018job,
  title={Job one for space force: Space asset cybersecurity},
  author={Falco, Gregory},
  journal={Belfer Center, Harvard Kennedy School, Belfer Center for Science and International Affairs, Harvard Kennedy School},
  volume={79},
  year={2018}
}

@article{nussbaum2020cybersecurity,
  title={Cybersecurity implications of commercial off the shelf (COTS) equipment in space infrastructure},
  author={Nussbaum, Brian and Berg, George},
  journal={Space infrastructures: From risk to resilience governance},
  pages={91--99},
  year={2020}
}

@article{addaim2010design,
  title={Design of low-cost telecommunications CubeSat-class spacecraft},
  author={Addaim, Adnane and Kherras, Abdelhaq and Zantou, El Bachir},
  journal={Aerospace Technologies Advancements},
  pages={293},
  year={2010},
  publisher={IntechOpen}
}

@article{baker2019amazon,
  title={Amazon Web Services (AWS) Cloud Platform for Satellite Data Processing},
  author={Baker, Richard and MacHarrie, Peter and Phung, Hieu and Hansford, Jonathan and Reddy, Jakku and Causey, Stephen and Sobanski, John and Walsh, Steven and Niemann, Ronald and Beall, Daniel},
  year={2019}
}

@article{menzel1994introducing,
  title={Introducing GOES-I: The first of a new generation of geostationary operational environmental satellites},
  author={Menzel, W Paul and Purdom, James FW},
  journal={Bulletin of the American Meteorological Society},
  volume={75},
  number={5},
  pages={757--782},
  year={1994},
  publisher={American Meteorological Society}
}

@book{goodman2019goes,
  title={The GOES-R series: a new generation of geostationary environmental satellites},
  author={Goodman, Steven J and Schmit, Timothy J and Daniels, Jaime and Redmon, Robert J},
  year={2019},
  publisher={Elsevier}
}

@article{parkinson1995history,
  title={A history of satellite navigation},
  author={Parkinson, Bradford W and Stansell, Thomas and Beard, Ronald and Gromov, Konstantine},
  journal={Navigation},
  volume={42},
  number={1},
  pages={109--164},
  year={1995},
  publisher={Wiley Online Library}
}

@book{morton2021position,
  title={Position, navigation, and timing technologies in the 21st century: Integrated satellite navigation, sensor systems, and civil applications, volume 1},
  author={Morton, Y Jade and van Diggelen, Frank and Spilker Jr, James J and Parkinson, Bradford W and Lo, Sherman and Gao, Grace},
  year={2021},
  publisher={John Wiley \& Sons}
}

@article{morales2019survey,
  title={A survey on coping with intentional interference in satellite navigation for manned and unmanned aircraft},
  author={Morales-Ferre, Ruben and Richter, Philipp and Falletti, Emanuela and de la Fuente, Alberto and Lohan, Elena Simona},
  journal={IEEE Communications Surveys \& Tutorials},
  volume={22},
  number={1},
  pages={249--291},
  year={2019},
  publisher={IEEE}
}

@article{kodheli2020satellite,
  title={Satellite communications in the new space era: A survey and future challenges},
  author={Kodheli, Oltjon and Lagunas, Eva and Maturo, Nicola and Sharma, Shree Krishna and Shankar, Bhavani and Montoya, Jesus Fabian Mendoza and Duncan, Juan Carlos Merlano and Spano, Danilo and Chatzinotas, Symeon and Kisseleff, Steven and others},
  journal={IEEE Communications Surveys \& Tutorials},
  volume={23},
  number={1},
  pages={70--109},
  year={2020},
  publisher={IEEE}
}

@article{miraux2022environmental,
  title={Environmental limits to the space sector's growth},
  author={Miraux, Lo{\"\i}s},
  journal={Science of The Total Environment},
  volume={806},
  pages={150862},
  year={2022},
  publisher={Elsevier}
}

@article{karvinen2015using,
  title={Using hobby prototyping boards and commercial-off-the-shelf (COTS) components for developing low-cost, fast-delivery satellite subsystems},
  author={Karvinen, Kimmo and Tikka, Tuomas and Praks, Jaan},
  journal={Journal of Small Satellites},
  volume={4},
  number={1},
  pages={301--314},
  year={2015}
}

@article{strozzi2005survey,
  title={Survey and monitoring of landslide displacements by means of L-band satellite SAR interferometry},
  author={Strozzi, Tazio and Farina, Paolo and Corsini, Alessandro and Ambrosi, Christian and Th{\"u}ring, Manfred and Zilger, Johannes and Wiesmann, Andreas and Wegm{\"u}ller, Urs and Werner, Charles},
  journal={Landslides},
  volume={2},
  pages={193--201},
  year={2005},
  publisher={Springer}
}

@article{ilvcev2017users,
  title={Users Segment},
  author={Il{\v{c}}ev, Stoj{\v{c}}e Dimov and Il{\v{c}}ev, Stoj{\v{c}}e Dimov},
  journal={Global Mobile Satellite Communications Theory: For Maritime, Land and Aeronautical Applications},
  pages={511--580},
  year={2017},
  publisher={Springer}
}

@article{steinberger2008survey,
  title={A survey of satellite communications system vulnerabilities},
  author={Steinberger, Jessica A},
  year={2008}
}

@book{majumdar2010free,
  title={Free-space laser communications: principles and advances},
  author={Majumdar, Arun K and Ricklin, Jennifer C},
  volume={2},
  year={2010},
  publisher={Springer Science \& Business Media}
}

@book{hemmati2020near,
  title={Near-earth laser communications},
  author={Hemmati, Hamid},
  volume={1},
  year={2020},
  publisher={CRC press}
}

@misc{kopp1996electromagnetic,
  title={The electromagnetic bomb: a weapon of electrical mass destruction},
  author={Kopp, Carlo},
  year={1996},
  publisher={Monash University}
}

@inproceedings{wang2019power,
  title={Power grid resilience to electromagnetic pulse (EMP) disturbances: a literature review},
  author={Wang, Dingwei and Li, Yifu and Dehghanian, Payman and Wang, Shiyuan},
  booktitle={2019 North American Power Symposium (NAPS)},
  pages={1--6},
  year={2019},
  organization={IEEE}
}

@article{longmire1978electromagnetic,
  title={On the electromagnetic pulse produced by nuclear explosions},
  author={Longmire, Conrad L},
  journal={IEEE Transactions on Electromagnetic Compatibility},
  number={1},
  pages={3--13},
  year={1978},
  publisher={IEEE}
}

@article{schwelb1964nuclear,
  title={The nuclear Test Ban Treaty and international law},
  author={Schwelb, Egon},
  journal={American Journal of International Law},
  volume={58},
  number={3},
  pages={642--670},
  year={1964},
  publisher={Cambridge University Press}
}

@article{colgate1965phenomenology,
  title={The phenomenology of the mass motion of a high altitude nuclear explosion},
  author={Colgate, Stirling A},
  journal={Journal of Geophysical Research},
  volume={70},
  number={13},
  pages={3161--3173},
  year={1965},
  publisher={Wiley Online Library}
}

@incollection{jensen2020fancy,
  title={Fancy bears and digital trolls: Cyber strategy with a Russian twist},
  author={Jensen, Benjamin and Valeriano, Brandon and Maness, Ryan},
  booktitle={Military Strategy in the 21st Century},
  pages={58--80},
  year={2020},
  publisher={Routledge}
}

@inproceedings{mcgann2005analysis,
  title={An analysis of security threats and tools in SIP-based VoIP systems},
  author={McGann, Shawn and Sicker, Douglas C},
  booktitle={Second VoIP security workshop},
  year={2005}
}

@inproceedings{wu2018approach,
  title={An Approach of Security Protection for VSAT Network},
  author={Wu, Zhijun and Pan, Qingbo and Yue, Meng and Ma, Shaopu},
  booktitle={2018 17th IEEE International Conference On Trust, Security And Privacy In Computing And Communications/12th IEEE International Conference On Big Data Science And Engineering (TrustCom/BigDataSE)},
  pages={1511--1516},
  year={2018},
  organization={IEEE}
}

@misc{center2002cert,
  title={CERT Advisory CA-2002-03 Multiple Vulnerabilities in Many Implementations of the Simple Network Management Protocol (SNMP), 12 February},
  author={Center, CERT Coordination},
  year={2002}
}

@online{mario2008,
  author    = {Heiderich M},
  title     = {Total surveillance made easy with VoIP phones},
  year      = {2008},
  month     = {2},
  day       = {11},
  publisher = {Gnucitizen},
  url       = {https://www.gnucitizen.org/blog/total-surveillance-made-easy-with-voip-phones/}
}

@inproceedings{lenhart2021relay,
  title={Relay/replay attacks on GNSS signals},
  author={Lenhart, Malte and Spanghero, Marco and Papadimitratos, Panagiotis},
  booktitle={Proceedings of the 14th ACM Conference on Security and Privacy in Wireless and Mobile Networks},
  pages={380--382},
  year={2021}
}

@online{cyberscoop2023,
  author    = {CyberScoop},
  title     = {CISA researchers: Russia's Fancy Bear infiltrated US satellite network},
  year      = {2022},
  month     = {12},
  day       = {16},
  publisher = {CyberScoop},
  url       = {https://cyberscoop.com/apt28-fancy-bear-satellite/},
  note      = {Accessed: October 8, 2023}
}

@online{strom2019,
  author= {Strom A},
  title = {ATT\&CK Sub-Techniques Preview},
  year      = {2019},
  month     = {08},
  day       = {22},
  publisher = {MITRE ATT\&CK®},
  url       = {https://medium.com/mitre-attack/attack-sub-techniques-preview-b79ff0ba669a}
}

@online{rainbow2023,
  author    = {Rainbow J},
  title     = {Thuraya invests in Astrocast’s LEO constellation},
  year      = {2023},
  month     = {04},
  day       = {03},
  publisher = {spacenews},
  url       = {https://spacenews.com/thuraya-invests-in-astrocasts-leo-constellation/}
}

@online{jaffar2022ntn,
  author    = {Jaffar M, Chuberre N},
  title     = {NTN \& Satellite in Rel-17 \& 18},
  year      = {2022},
  month     = {07},
  day       = {01},
  publisher = {3GPP},
  url       = {https://www.3gpp.org/news-events/partner-news/ntn-rel17}
}

@online{nassa2022,
  author    = {NASA Ames Research Center},
  title     = {State-of-the-Art of Small Spacecraft Technology
},
  year      = {2022},
  month     = {07},
  day       = {28},
  publisher = {NASA},
  url       = {https://www.nasa.gov/smallsat-institute/sst-soa/soa-communications/}
}

@online{dhs2022,
  author    = {DHS},
  title     = {Electromagnetic Pulse (EMP) / Geomagnetic Disturbance (GMD)},
  year      = {2022},
  month     = {08},
  day       = {03},
  publisher = {DHS science and technology},
  url       = {https://www.dhs.gov/science-and-technology/electromagnetic-pulse-empgeomagnetic-disturbance}
}

@online{viasat2022,
  author = {cyber peace institute},
  title  = {Case Study Viasat},
  year   = {2022},
  month  = {06},
  publisher = {cyber peace institute},
  url ={https://cyberconflicts.cyberpeaceinstitute.org/law-and-policy/cases/viasat
}
}

@online{urban2002,
    author = {Urban M},
    title  = {Enthusiast watches Nato spy pictures},
    year   = {2002},
    month  = {06},
    day = {13},
    publisher = {BBC},
    url = {http://news.bbc.co.uk/2/hi/programmes/newsnight/2041754.stm}
}

@online{redteam2023,
  title     = {Satellite Hacking Demystified},
  year      = {2023},
  url       = {https://redteamrecipe.com/Satellite-Hacking-Demystified/},
  note      = {Accessed: October 28, 2023}
}

@article{mayo1963command,
  title={The command system malfunction of the Telstar satellite},
  author={Mayo, JS and Mann, H and Witt, FJ and Peck, DS and Gummel, HK and Brown, WL},
  journal={Bell System Technical Journal},
  volume={42},
  number={4},
  pages={1631--1657},
  year={1963},
  publisher={Wiley Online Library}
}

@inproceedings{kim2011satellite,
  title={Satellite electrical power subsystem: Statistical analysis of on-orbit anomalies and failures},
  author={Kim, So Young and Castet, Jean-Francois and Saleh, Joseph H},
  booktitle={2011 Aerospace Conference},
  pages={1--12},
  year={2011},
  organization={IEEE}
}

@inproceedings{fayyaz2014fault,
  title={Fault-tolerant distributed approach to satellite on-board computer design},
  author={Fayyaz, Muhammad and Vladimirova, Tanya},
  booktitle={2014 IEEE Aerospace Conference},
  pages={1--12},
  year={2014},
  organization={IEEE}
}

@online{antesky2023,
  title     = {L-band for Weather Satellite Data Receiving System},
  year      = {2023},
  url       = {https://www.antesky.com/l-band-for-weather-satellite-data-receiving-system/},
  note      = {Accessed: October 28, 2023}
}

@online{esa2023,
  title     = {Satellite frequency bands},
  year      = {2023},  
  publisher = {European Space Agency},
  url       = {https://www.esa.int/Applications/Connectivity_and_Secure_Communications/Satellite_frequency_bands},
  note      = {Accessed: October 28, 2023}
}

@online{spacesecurity2020,
  title     = {Space Attacks Open Database Project},
  year      = {2020},  
  publisher = {Space & Cybersecurity Info},
  url       = {https://www.spacesecurity.info/en/space-attacks-open-database/},
  note      = {Accessed: October 28, 2023}
}

@online{hewitson2020,
  author    = {Amy Hewitson},
  title     = {Satellite Ground Stations - Everything you ever wanted to know and more...},
  year      = {2020},
  month     = {10},
  day       = {22},
  publisher = {ESS Earth Sciences},
  url       = {https://www.essearth.com/satellite-ground-stations/},
  note      = {Accessed: October 28, 2023}
}

@online{werner2022,
  author    = {Debra Werner},
  title     = {Azure Orbital Space unveils software tools for space applications},
  year      = {2022},
  month     = {11},
  day       = {17},
  publisher = {SpaceNews},
  url       = {https://spacenews.com/azure-orbital-software-development-kit/},
  note      = {Accessed: October 28, 2023}
}

@online{andres2022,
  author    = {Carmen Reglero Andres and Nicholas Ansell and Eloy Salcedo and Jason Arora},
  title     = {Virtualizing satellite communication operations with AWS},
  year      = {2022},
  month     = {9},
  day       = {6},
  publisher = {AWS Public Sector Blog},
  url       = {https://aws.amazon.com/blogs/publicsector/virtualizing-satcom-operations-aws/},
  note      = {Accessed: October 28, 2023}
}

@online{dvb2023,
  title     = {Second Generation DVB Interactive Satellite System (DVB-RCS2); Part 2: Lower Layers for Satellite standard},
  year      = {2023},  
  publisher = {DVB},
  url       = {https://dvb.org/?standard=dvb-rcs2-lower-layer-satellite-specification},
  note      = {Accessed: October 28, 2023}
}

@online{mitre2023,
  title     = {CVE},
  year      = {2023},
  month     = {8},
  day       = {14},
  publisher = {The MITRE Corporation},
  url       = {https://cve.mitre.org/},
  note      = {Accessed: October 28, 2023}
}

@online{wainscott-sargent2022,
  author    = {Anne Wainscott-Sargent},
  title     = {Satellite Operators Respond to Cyber Threats in a Rapidly Changing Environment},
  year      = {2022},
  month     = {9},
  day       = {27},
  publisher = {Via Satellite},
  url       = {https://interactive.satellitetoday.com/via/october-2022/satellite-operators-respond-to-cyber-threats-in-a-rapidly-changing-environment/},
  note      = {Accessed: October 28, 2023}
}

@online{gedeon2023,
  author    = {Joseph Gedeon},
  title     = {For the first time, U.S. government lets hackers break into satellite in space},
  year      = {2023},
  month     = {8},
  day       = {11},
  publisher = {POLITICO},
  url       = {https://www.politico.com/news/2023/08/11/def-con-hackers-space-force-00110919},
  note      = {Accessed: October 28, 2023}
}

@online{guardian2009,
  title     = {Skygrabber Software Drones Hacked},
  year      = {2009},
  month     = {12},
  day       = {17},
  publisher = {The Guardian},
  url       = {https://www.theguardian.com/technology/2009/dec/17/skygrabber-software-drones-hacked},
  note      = {Accessed: October 28, 2023}
}

@article{curzi2020large,
  title={Large constellations of small satellites: A survey of near future challenges and missions},
  author={Curzi, Giacomo and Modenini, Dario and Tortora, Paolo},
  journal={Aerospace},
  volume={7},
  number={9},
  pages={133},
  year={2020},
  publisher={MDPI}
}

@online{viasatt2022,
  title     = {KA-SAT Network cyber attack overview},
  year      = {2022},
  month     = {3},
  day       = {30},
  publisher = {Viasat},
  url       = {https://news.viasat.com/blog/corporate/ka-sat-network-cyber-attack-overview},
  note      = {Accessed: October 28, 2023}
}

@misc{sparta-overview-2023,
  author       = {{The Aerospace Corporation}},
  title        = {SPARTA: Space Attack Research \& Tactic Analysis},
  howpublished = {\url{https://aerospace.org/sparta}},
  year         = {2023},
  note         = {Overview of the SPARTA matrix; tactics/techniques/countermeasures}
}

@techreport{ccsds-350-1-g3-2022,
  author       = {{Consultative Committee for Space Data Systems (CCSDS)}},
  title        = {Security Threats Against Space Missions},
  institution  = {CCSDS},
  number       = {CCSDS 350.1-G-3},
  year         = {2022},
  howpublished = {\url{https://public.ccsds.org/Pubs/350x1g3.pdf}}
}

@techreport{nist-ir-8270-2023,
  author       = {Scholl, Michael and Suloway, Thomas},
  title        = {Introduction to Cybersecurity for Commercial Satellite Operations},
  institution  = {NIST},
  number       = {NIST IR 8270},
  year         = {2023},
  howpublished = {\url{https://nvlpubs.nist.gov/nistpubs/ir/2023/NIST.IR.8270.pdf}}
}

@techreport{enisa-2025-space-threat,
  author       = {{ENISA}},
  title        = {Space Threat Landscape},
  year         = {2025},
  howpublished = {\url{https://www.enisa.europa.eu/sites/default/files/2025-03/Space_Threat_Landscape_Report_fin.pdf}}
}

@misc{leolabs,
  author       = {{LeoLabs}},
  title        = {LeoLabs LEO tracking service},
  howpublished = {\url{https://leolabs.space/seeleo/}},
  year         = {2025},
  note         = {Accessed: 24 Aug 2025}
}

@misc{n2yo,
  author       = {{N2YO}},
  title        = {N2YO Satellite Tracking},
  howpublished = {\url{https://www.n2yo.com/}},
  year         = {2025},
  note         = {Accessed: 24 Aug 2025}
}

@misc{theia-esat,
  author       = {{Universidad Politécnica de Madrid}},
  title        = {Theia ESAT Simulator},
  howpublished = {\url{https://www.theia.eusoc.upm.es/esat/}},
  year         = {2025},
  note         = {Accessed: 24 Aug 2025}
}

@misc{cubeshop,
  author       = {{CubeSatShop}},
  title        = {CubeSatShop Online Store},
  howpublished = {\url{https://www.cubesatshop.com/}},
  year         = {2025},
  note         = {Accessed: 24 Aug 2025}
}

@misc{ses-o3b,
  author       = {{SES S.A.}},
  title        = {O3b MEO Constellation Coverage},
  howpublished = {\url{https://www.ses.com/our-coverage/o3b-meo}},
  year         = {2025},
  note         = {Accessed: 24 Aug 2025}
}

@inproceedings{du2022physical,
  title     = {Physical Adversarial Attacks on an Aerial Imagery Object Detector},
  author    = {Du, Andrew and others},
  booktitle = {WACV},
  year      = {2022},
  url       = {https://openaccess.thecvf.com/content/WACV2022/papers/Du_Physical_Adversarial_Attacks_on_an_Aerial_Imagery_Object_Detector_WACV_2022_paper.pdf}
}

@online{businessinsider2024,
  title = {Satellite Images Suggest Russia Painted Fake Aircraft at Its Airfields to Trick Ukraine},
  author = {Business Insider},
  year = {2024},
  url = {https://www.businessinsider.com/satellite-images-russia-painted-fake-decoy-aircraft-fool-ukraine-weapons-2024-1}
}

@inproceedings{sparta2023,
  title={Applying the SPARTA Matrix to Develop Intelligent Security Controls for Space Systems},
  author={The Aerospace Corporation},
  booktitle={International Conference on Recent Advances in Space and Aerospace Science},
  year={2023},
  url={https://www.researchgate.net/publication/374833683_Applying_the_SPARTA_Matrix_to_develop_Intelligent_Security_Controls_for_Space_Systems}
}

@article{smailes2024sticky,
  title={Sticky fingers: resilience of satellite fingerprinting against jamming attacks},
  author={Smailes, Joshua and Salkield, Edd and K{\"o}hler, Sebastian and Birnbach, Simon and Strohmeier, Martin and Martinovic, Ivan},
  journal={arXiv preprint arXiv:2402.05042},
  year={2024}
}

@inproceedings{weerackody2021satellite,
  title={Satellite diversity to mitigate jamming in LEO satellite mega-constellations},
  author={Weerackody, Vijitha},
  booktitle={2021 IEEE International Conference on Communications Workshops (ICC Workshops)},
  pages={1--6},
  year={2021},
  organization={IEEE}
}

@techreport{jayaweera2018cognitive,
  title={Cognitive anti-jamming satellite-to-ground communications on NASA's SCaN testbed},
  author={Jayaweera, Sudharman K and Feng, Shuang and Mortensen, Dale and Holland, Abriel and Piasecki, Marie and Evans, Mike and Christodoulou, Christos},
  year={2018}
}

@article{seco2021detection,
  title={Detection of replay attacks to GNSS based on partial correlations and authentication data unpredictability},
  author={Seco-Granados, Gonzalo and G{\'o}mez-Casco, David and L{\'o}pez-Salcedo, Jos{\'e} A and Fern{\'a}ndez-Hern{\'a}ndez, Ignacio},
  journal={Gps Solutions},
  volume={25},
  number={2},
  pages={33},
  year={2021},
  publisher={Springer}
}

@inproceedings{smailes2023watch,
  title={Watch this space: Securing satellite communication through resilient transmitter fingerprinting},
  author={Smailes, Joshua and K{\"o}hler, Sebastian and Birnbach, Simon and Strohmeier, Martin and Martinovic, Ivan},
  booktitle={Proceedings of the 2023 ACM SIGSAC Conference on Computer and Communications Security},
  pages={608--621},
  year={2023}
}

@article{yue2023low,
  title={Low earth orbit satellite security and reliability: Issues, solutions, and the road ahead},
  author={Yue, Pingyue and An, Jianping and Zhang, Jiankang and Ye, Jia and Pan, Gaofeng and Wang, Shuai and Xiao, Pei and Hanzo, Lajos},
  journal={IEEE Communications Surveys \& Tutorials},
  volume={25},
  number={3},
  pages={1604--1652},
  year={2023},
  publisher={IEEE}
}

@article{lin2023clextract,
  title={CLExtract: Recovering Highly Corrupted DVB/GSE Satellite Stream with Contrastive Learning},
  author={Lin, Minghao and Cheng, Minghao and Luo, Dongsheng and Chen, Yueqi},
  journal={arXiv preprint arXiv:2310.08210},
  year={2023}
}

@article{li2023active,
  title={Active eavesdropping detection: a novel physical layer security in wireless IoT},
  author={Li, Mingfang and Dou, Zheng},
  journal={EURASIP Journal on Advances in Signal Processing},
  volume={2023},
  number={1},
  pages={119},
  year={2023},
  publisher={Springer}
}

@inproceedings{ear2023characterizing,
  title={Characterizing cyber attacks against space systems with missing data: Framework and case study},
  author={Ear, Ekzhin and Remy, Jose LC and Feffer, Antonia and Xu, Shouhuai},
  booktitle={2023 IEEE Conference on Communications and Network Security (CNS)},
  pages={1--9},
  year={2023},
  organization={IEEE}
}

@article{thummala2024adversarial,
  title={Adversarial machine learning threats to spacecraft},
  author={Thummala, Rajiv and Sharma, Shristi and Calabrese, Matteo and Falco, Gregory},
  journal={arXiv preprint arXiv:2405.08834},
  year={2024}
}

@article{tang2023adversarial,
  title={Adversarial patch attacks against aerial imagery object detectors},
  author={Tang, Guijian and Jiang, Tingsong and Zhou, Weien and Li, Chao and Yao, Wen and Zhao, Yong},
  journal={Neurocomputing},
  volume={537},
  pages={128--140},
  year={2023},
  publisher={Elsevier}
}

@inproceedings{czaja2018adversarial,
  title={Adversarial examples in remote sensing},
  author={Czaja, Wojciech and Fendley, Neil and Pekala, Michael and Ratto, Christopher and Wang, I-Jeng},
  booktitle={Proceedings of the 26th ACM SIGSPATIAL International Conference on Advances in Geographic Information Systems},
  pages={408--411},
  year={2018}
}

@article{morales2019jammer,
  title={Jammer classification in GNSS bands via machine learning algorithms},
  author={Morales Ferre, Ruben and De La Fuente, Alberto and Lohan, Elena Simona},
  journal={Sensors},
  volume={19},
  number={22},
  pages={4841},
  year={2019},
  publisher={MDPI}
}

@inproceedings{han2024deep,
  title={Deep Learning Based Satellite Communication Anti-Jamming System},
  author={Han, Weinan and Song, Xiaohan and Huang, Yixuan and Yan, Fangwei and Yin, Qingze and Zhang, Tongpo},
  booktitle={2024 IEEE/ACIS 24th International Conference on Computer and Information Science (ICIS)},
  pages={9--12},
  year={2024},
  organization={IEEE}
}

@article{ding2023few,
  title={Few-Shot Recognition and Classification Framework for Jamming Signal: A CGAN-Based Fusion CNN Approach},
  author={Ding, Xuhui and Zhang, Yue and Li, Gaoyang and Gao, Xiaozheng and Ye, Neng and Niyato, Dusit and Yang, Kai},
  journal={arXiv preprint arXiv:2311.05273},
  year={2023}
}

@article{han2020spatial,
  title={Spatial anti-jamming scheme for internet of satellites based on the deep reinforcement learning and stackelberg game},
  author={Han, Chen and Huo, Liangyu and Tong, Xinhai and Wang, Haichao and Liu, Xian},
  journal={IEEE Transactions on Vehicular Technology},
  volume={69},
  number={5},
  pages={5331--5342},
  year={2020},
  publisher={IEEE}
}

@article{li2025achieving,
  title={Achieving Hiding and Smart Anti-Jamming Communication: A Parallel DRL Approach against Moving Reactive Jammer},
  author={Li, Yangyang and Xu, Yuhua and Li, Wen and Li, Guoxin and Feng, Zhibing and Liu, Songyi and Du, Jiatao and Li, Xinran},
  journal={arXiv preprint arXiv:2502.02385},
  year={2025}
}

@article{xie2025multi,
  title={Multi-Satellite Beam Hopping and Power Allocation Using Deep Reinforcement Learning},
  author={Xie, Xia and Fan, Kexin and Deng, Wenfeng and Pappas, Nikolaos and Zhang, Qinyu},
  journal={arXiv preprint arXiv:2501.02309},
  year={32025}
}

@article{martinez5214437hybrid,
  title={Hybrid Moea with Problem-Specific Operators for Beam-Hopping Based Resource Allocation in Multi-Beam Leo Satellites},
  author={Mart{\'\i}nez Zamacola, Samuel and Luna Valero, Francisco and Mart{\'\i}nez Rogr{\'\i}guez-Osorio, Ram{\'o}n},
  journal={Available at SSRN 5214437},
  year = {2025}
}

@inproceedings{harris2023autonomous,
  title={Autonomous Command and Control for Earth-Observing Satellites using Deep Reinforcement Learning},
  author={Harris, Andrew and Naik, Kedar},
  booktitle={2023 IEEE Aerospace Conference},
  pages={1--12},
  year={2023},
  organization={IEEE}
}

@article{fourati2021artificial,
  title={Artificial intelligence for satellite communication: A review},
  author={Fourati, Fares and Alouini, Mohamed-Slim},
  journal={Intelligent and Converged Networks},
  volume={2},
  number={3},
  pages={213--243},
  year={2021},
  publisher={TUP}
}

@article{li2025mission,
  title={Mission Sequence Model and Deep Reinforcement Learning-Based Replanning Method for Multi-Satellite Observation},
  author={Li, Peiyan and Cui, Peixing and Wang, Huiquan},
  journal={Sensors},
  volume={25},
  number={6},
  pages={1707},
  year={2025},
  publisher={MDPI}
}

@inproceedings{hundman2018detecting,
  title={Detecting spacecraft anomalies using lstms and nonparametric dynamic thresholding},
  author={Hundman, Kyle and Constantinou, Valentino and Laporte, Christopher and Colwell, Ian and Soderstrom, Tom},
  booktitle={Proceedings of the 24th ACM SIGKDD international conference on knowledge discovery \& data mining},
  pages={387--395},
  year={2018}
}

@article{thangavel2024artificial,
  title={Artificial intelligence for trusted autonomous satellite operations},
  author={Thangavel, Kathiravan and Sabatini, Roberto and Gardi, Alessandro and Ranasinghe, Kavindu and Hilton, Samuel and Servidia, Pablo and Spiller, Dario},
  journal={Progress in Aerospace Sciences},
  volume={144},
  pages={100960},
  year={2024},
  publisher={Elsevier}
}

@article{herrmann2024unmasking,
  title={Unmasking overestimation: a re-evaluation of deep anomaly detection in spacecraft telemetry},
  author={Herrmann, Lars and Bieber, Marie and Verhagen, Wim JC and Cosson, Fabrice and Santos, Bruno F},
  journal={CEAS Space Journal},
  volume={16},
  number={2},
  pages={225--237},
  year={2024},
  publisher={Springer}
}

@article{rath2020security,
  title={Security approaches in machine learning for satellite communication},
  author={Rath, Mamata and Mishra, Sushruta},
  journal={Machine learning and data mining in aerospace technology},
  pages={189--204},
  year={2020},
  publisher={Springer}
}

@article{sitouah2022deep,
  title={Deep learning approach for interruption attacks detection in LEO satellite networks},
  author={Sitouah, Nacereddine and Merazka, Fatiha and Hedjazi, Abdenour},
  journal={arXiv preprint arXiv:2301.03998},
  year={2022}
}

@article{bourriez2023spacecraft,
  title={Spacecraft autonomous decision-planning for collision avoidance: A Reinforcement Learning approach},
  author={Bourriez, Nicolas and Loizeau, Adrien and Abdin, Adam F},
  journal={arXiv preprint arXiv:2310.18966},
  year={2023}
}

@article{pinto2020towards,
  title={Towards automated satellite conjunction management with bayesian deep learning},
  author={Pinto, Francesco and Acciarini, Giacomo and Metz, Sascha and Boufelja, Sarah and Kaczmarek, Sylvester and Merz, Klaus and Martinez-Heras, Jos{\'e} A and Letizia, Francesca and Bridges, Christopher and Baydin, At{\i}l{\i}m G{\"u}ne{\c{s}}},
  journal={arXiv preprint arXiv:2012.12450},
  year={2020}
}

@article{yao2025meta,
  title={Meta Reinforcement Learning Method for Dynamic Mission Scheduling of Earth Observation Satellites},
  author={Yao, Wei and Shen, Xin and Zhang, Guo and Lu, Zezhong and Wang, Jiaying and Gao, Gui},
  journal={Meta},
  volume={18},
  pages={5},
  year={2025}
}

@misc{3gpp_ntn_overview,
  author       = {{3rd Generation Partnership Project (3GPP)}},
  title        = {Non-Terrestrial Networks (NTN) Overview},
  year         = {2024},
  howpublished = {\url{https://www.3gpp.org/technologies/ntn-overview}},
  note         = {Accessed: 2025-09-10}
}

@article{feng2022improved,
  title={An improved X-means and isolation forest based methodology for network traffic anomaly detection},
  author={Feng, Yifan and Cai, Weihong and Yue, Haoyu and Xu, Jianlong and Lin, Yan and Chen, Jiaxin and Hu, Zijun},
  journal={Plos one},
  volume={17},
  number={1},
  pages={e0263423},
  year={2022},
  publisher={Public Library of Science San Francisco, CA USA}
}

@inproceedings{woo2022detecting,
  title={Detecting Satellite Laser Ranging Station Data and Operational Anomalies with Machine Learning Isolation Forests at NASA’s CDDIS},
  author={Woo, Justine and Belvins, Sandra and Michael, Benjamin Patrick and Yates, Taylor},
  booktitle={AGU Fall Meeting Abstracts, pp. IN12B--0269},
  year={2022}
}

@techreport{nasa_oig_ig_19_022,
  author       = {{NASA Office of Inspector General}},
  title        = {Cybersecurity Management and Oversight at the Jet Propulsion Laboratory},
  institution  = {National Aeronautics and Space Administration},
  number       = {IG-19-022},
  year         = {2019},
  month        = {June},
  url          = {https://oig.nasa.gov/wp-content/uploads/2024/02/IG-19-022.pdf},
  note         = {Accessed: 2025-05-21}
}

@article{vanlyssel2025spychain,
  title={SpyChain: Multi-Vector Supply Chain Attacks on Small Satellite Systems},
  author={Vanlyssel, Jack and Sobrados, Enrique and Anwar, Ramsha and Roman, Gruia-Catalin and Anwar, Afsah},
  journal={arXiv preprint arXiv:2510.06535},
  year={2025}
}

@article{shigolrisk,
  title={Risk Assessment for ML-Based Applications in Satellite Systems},
  author={Shigol, Simon and Peled, Roy and Shapira, Avishag and Elovici, Yuval and Shabtai, Asaf}
}

@online{s_isac,
  author = {{Space Information Sharing and Analysis Center (S-ISAC)}},
  title  = {S-ISAC Official Website},
  year   = {2024},
  url    = {https://s-isac.org/},
  note   = {Accessed: 2026-03-17}
}

@misc{satellitetoday2021_companies,
  author = {{Via Satellite / Satellite Today}},
  title  = {The 10 Hottest Satellite Companies in 2021},
  year   = {2021},
  howpublished = {\url{https://interactive.satellitetoday.com/via/july-2021/the-10-hottest-satellite-companies-in-2021/}},
  note   = {Accessed: 2026-03-17}
}

@misc{bbc2021_space_companies,
  author = {{BBC News}},
  title  = {The New Space Race: Small Satellite Companies Rising},
  year   = {2021},
  howpublished = {\url{https://www.bbc.com/news/business-55807150}},
  note   = {Accessed: 2026-03-17}
}

@misc{yale2025satellites,
  author = {{Yale Environment 360}},
  title = {Record Number of Objects Launched Into Space Last Year},
  year = {2026},
  howpublished = {\url{https://e360.yale.edu/digest/2025-satellite-launches}},
  note = {Accessed: April 9, 2026}
}

@article{ear2025characterizing,
  title={Characterizing Cyber Attacks against Space Infrastructures with Missing Data: Framework and Case Study},
  author={Ear, Ekzhin and Remy, Jose Luis Castanon and Chang, Caleb and Que, Qiren and Feffer, Antonia and Xu, Shouhuai},
  journal={arXiv preprint arXiv:2512.02414},
  year={2025}
}

@inproceedings{remy2025sok,
  title={SoK: Space infrastructures vulnerabilities, attacks and defenses},
  author={Remy, Jose Luis Castanon and Ear, Ekzhin and Chang, Caleb and Feffer, Antonia and Xu, Shouhuai},
  booktitle={2025 IEEE Symposium on Security and Privacy (SP)},
  pages={1028--1046},
  year={2025},
  organization={IEEE}
}

@article{kang2024survey,
  title={A survey on satellite communication system security},
  author={Kang, Minjae and Park, Sungbin and Lee, Yeonjoon},
  journal={Sensors},
  volume={24},
  number={9},
  pages={2897},
  year={2024},
  publisher={MDPI}
}

@techreport{harrison2020space,
  title        = {Space Threat Assessment 2020},
  author       = {Harrison, Todd and Johnson, Kaitlyn and Moye, Joe and Young, Makena},
  institution  = {Center for Strategic \& International Studies},
  year         = {2020}
}

@techreport{harrison2021space,
  title        = {Space Threat Assessment 2021},
  author       = {Harrison, Todd and Johnson, Kaitlyn and Young, Makena and Moye, Joe},
  institution  = {Center for Strategic \& International Studies},
  year         = {2021},
  month        = mar,
  note         = {Published March 31, 2021},
  url          = {https://www.csis.org/analysis/space-threat-assessment-2021}
}

@techreport{harrison2022space,
  title        = {Space Threat Assessment 2022},
  author       = {Harrison, Todd and Johnson, Kaitlyn and Young, Makena and Wood, Nicholas and Goessler, Alyssa},
  institution  = {Center for Strategic \& International Studies},
  year         = {2022},
  month        = apr,
  note         = {Published April 4, 2022},
  url          = {https://www.csis.org/analysis/space-threat-assessment-2022}
}

@techreport{bingen2023space,
  title        = {Space Threat Assessment 2023},
  author       = {Bingen, Kari A. and Johnson, Kaitlyn and Young, Makena and Raymond, John},
  institution  = {Center for Strategic \& International Studies},
  year         = {2023},
  month        = apr,
  note         = {Published April 14, 2023},
  url          = {https://www.csis.org/analysis/space-threat-assessment-2023}
}

@techreport{swope2024space,
  title        = {Space Threat Assessment 2024},
  author       = {Swope, Clayton and Bingen, Kari A. and Young, Makena and Chang, Madeleine and Songer, Stephanie and Tammelleo, Jeremy},
  institution  = {Center for Strategic \& International Studies},
  year         = {2024},
  month        = apr,
  note         = {Published April 17, 2024},
  url          = {https://www.csis.org/analysis/space-threat-assessment-2024}
}

@techreport{swope2025space,
  title        = {Space Threat Assessment 2025},
  author       = {Swope, Clayton and Bingen, Kari A. and Young, Makena and LaFave, Kendra},
  institution  = {Center for Strategic \& International Studies},
  year         = {2025},
  month        = apr,
  note         = {Published April 25, 2025},
  url          = {https://www.csis.org/analysis/space-threat-assessment-2025}
}

@incollection{Addaim10,
author = {Adnane Addaim and Abdelhaq Kherras and El Bachir Zantou},
title = {Design of Low-cost Telecommunications CubeSat-class Spacecraft},
booktitle = {Aerospace Technologies Advancements},
publisher = {IntechOpen},
address = {London},
year = {2010},
editor = {Thawar T. Arif},
chapter = {15},
doi = {10.5772/6925},
url = {https://doi.org/10.5772/6925}
}

@misc{guerra2018thermalanalysise,
      title={Thermal analysis of the electronics of a CubeSat mission}, 
      author={André G. C. Guerra and Diego Nodar-López and Ricardo Tubó-Pardavila},
      year={2018},
      eprint={1803.10468},
      archivePrefix={arXiv},
      primaryClass={physics.app-ph},
      url={https://arxiv.org/abs/1803.10468} 
}

@misc{delcastillo2024s,
      title={Unlocking the Potential of Small Satellites: TheMIS's Active Cooling Technology on the SpIRIT Mission}, 
      author={Miguel Ortiz del Castillo and Clint Therakam and Jack McRobbie and Andrew Woods and Robert Mearns and Simon Barraclough and Stephen Catsamas and Mika Ohkawa and Jonathan Morgan and Airlie Chapman and Michele Trenti},
      year={2024},
      eprint={2407.14031},
      archivePrefix={arXiv},
      primaryClass={astro-ph.IM},
      url={https://arxiv.org/abs/2407.14031} 
}

@article{Leomanni_2020,
   title={An adaptive groundtrack maintenance scheme for spacecraft with electric propulsion},
   volume={167},
   ISSN={0094-5765},
   url={http://dx.doi.org/10.1016/j.actaastro.2019.11.035},
   DOI={10.1016/j.actaastro.2019.11.035},
   journal={Acta Astronautica},
   publisher={Elsevier BV},
   author={Leomanni, Mirko and Garulli, Andrea and Giannitrapani, Antonio and Scortecci, Fabrizio},
   year={2020},
   month=Feb, pages={460–466} }

@techreport{enisa2025space,
  author      = {{European Union Agency for Cybersecurity (ENISA)}},
  title       = {Space Threat Landscape},
  institution = {ENISA},
  year        = {2025},
  month       = {March},
  doi         = {10.2824/8841206},
  url         = {https://europa.eu},
  note        = {Accessed: \text{July 2026}}
}

@misc{sentinelone2022acidrain,
  author       = {Guerrero-Saade, Juan Andr{\'e}s and van Amerongen, Max},
  title        = {{AcidRain: A Modem Wiper Rains Down on Europe}},
  howpublished = {SentinelLabs},
  year         = {2022},
  month        = mar,
  day          = {31},
  url          = {https://www.sentinelone.com/labs/acidrain-a-modem-wiper-rains-down-on-europe/},
  note         = {Accessed: 2026-07-07}
}

@misc{killnet2025,
  author       = {Dark Reading Staff},
  title        = {Killnet Gloats About DDoS Attacks on Starlink and Whitehouse.gov},
  year         = {2022},
  url          = {https://www.darkreading.com/threat-intelligence/kFillnet-gloats-ddos-attacks-starlink-whitehouse-gov},
  note         = {Accessed: 2025-01-20}
}

@inproceedings{planta2026satbleed,
  title={SATBLEED: Security of Commoditized Communication Modules in Satellites},
  author={Planta, Ulysse and Rederlechner, Julian and Strohmeier, Martin and Fischer, Mathias and Abbasi, Ali},
  booktitle={2026 IEEE Symposium on Security and Privacy (SP)},
  pages={2904--2921},
  year={2026},
  organization={IEEE}
}

@article{stan2017protecting,
  title={Protecting military avionics platforms from attacks on mil-std-1553 communication bus},
  author={Stan, Orly and Elovici, Yuval and Shabtai, Asaf and Shugol, Gaby and Tikochinski, Raz and Kur, Shachar},
  journal={arXiv preprint arXiv:1707.05032},
  year={2017}
}

@article{levy2024anomili,
  title={AnoMili: Spoofing Hardening and Explainable Anomaly Detection for the 1553 Military Avionic Bus},
  author={Levy, Efrat and Maman, Nadav and Shabtai, Asaf and Elovici, Yuval},
  journal={IEEE Transactions on Aerospace and Electronic Systems},
  volume={60},
  number={6},
  pages={7738--7753},
  year={2024},
  publisher={IEEE}
}

@article{batista2021using,
  author       = {Carlos Leandro Gomes Batista and
                  Andr{\'{e}} Corsetti and
                  F{\'{a}}tima Mattiello{-}Francisco},
  title        = {Using Fault Injection on the Nanosatellite Subsystems Integration
                  Testing},
  journal      = {CoRR},
  volume       = {abs/2102.11776},
  year         = {2021},
  url          = {https://arxiv.org/abs/2102.11776},
  eprinttype   = {arXiv},
  eprint       = {2102.11776},
  bibsource    = {dblp computer science bibliography, https://dblp.org}
}

% --------------------------------------------------
% Appendix
% --------------------------------------------------

\appendix
\clearpage
\onecolumn
\appendix
\section{Reported Attacks and Research Mapped to Our Taxonomy }
\begin{table*}[!ht]
\scriptsize
\centering
\caption{Concise descriptions of reported attacks used in Table~\ref{tab:tax_map}.}
\label{tab:tax_map_desc}
\begin{tabularx}{\textwidth}{@{}l c X c X@{}}
\toprule
\textbf{Ref} & \textbf{Year} & \textbf{One–two line description} & \textbf{Primary vector} & \textbf{Notable outcome} \\
\midrule
\cite{boschetti2022space} & 2022 & Infiltration of KA-SAT ground segment; AcidRain firmware wipe on user modems across Europe. & Ground segment compromise & Regional service outage; modem bricking \\
\midrule
\cite{giuliari2021icarus} & 2021 & ICARUS shows constellation-aware DDoS that blends with legitimate traffic in LEO networks. & Network flood (LEO-aware) & Link congestion; denial of service \\
\midrule
\cite{zhang2022security} & 2022 & DDoS strategies tailored for satellite networks; routing/queuing stress in moving topologies. & Network flood & Degradation; denial of service \\
\midrule
\cite{cyberscoop2023} & 2023 & Reporting on satellite-comm espionage ops using phishing, malware, and backdoors. & Phishing - malware & Long-lived access; eavesdropping \\
\midrule
\cite{lenhart2021relay} & 2021 & Demonstration of relay/replay signal manipulation against satellite/GNSS receivers. & Signal replay & Tampering; endpoint misuse \\
\midrule
\cite{smailes2023dishing} & 2023 & Reverse engineering of user terminals enables firmware-level modifications and traffic tampering. & Rogue/modified terminal & DoS or tamper with traffic \\
\bottomrule
\end{tabularx}
\end{table*}

\begin{table*}[!ht]
\scriptsize 
\centering
\caption{Reported attacks and research mapped to MITRE taxonomy presented in Section~\ref{sec:MITRE}.}
\label{tab:tax_map}
\begin{tabularx}{\textwidth}{@{}lX*{8}{X}@{}}
\toprule
\textbf{Ref} & \textbf{Violation Type} & \textbf{Reconnaissance} & \textbf{Resource Development} & \textbf{Initial Access} & \textbf{Execution} & \textbf{Persistence} & \textbf{Defense Evasion} & \textbf{Lateral Movement} & \textbf{Impact} \\
\midrule

\cite{boschetti2022space} 
& Availability 
& Gather space Comm Info 
& Compromise Infrastructure 
& Ground Segment Access, Network 
& Exploit Software/Hardware, Cyber Execution 
& - 
& - 
& - 
& Sabotage, Attack endpoints, Denial of service \\

\midrule

\cite{giuliari2021icarus} 
& Availability 
& Public space scanners, Gather space Comm Info 
& Acquire Infrastructure 
& Network 
& Cyber Execution 
& - 
& Camouflage, Concealment, and Decoys (CCD) 
& -
& Tampering with network communication, Denial of service \\

\midrule

\cite{zhang2022security} 
& Availability 
& Public space scanners, Gather space Comm Info 
& Acquire Infrastructure 
& Network 
& Cyber Execution 
& - 
& Camouflage, Concealment, and Decoys (CCD) 
& - 
& Tampering with network communication, Denial of service \\

\midrule

\cite{cyberscoop2023} 
& Integrity, Confidentiality 
& Gather space Comm Info 
& Acquire Infrastructure, Obtain Cyber Capabilities 
& Network, Valid Credentials 
& Cyber Execution 
& Backdoor 
& Credential Abuse 
& - 
& Eavesdropping \\

\midrule

\cite{lenhart2021relay} 
& Integrity 
& Gather space Comm Info 
& Acquire Infrastructure 
& Signal 
& Signal Injection / Replay 
& - 
& - 
& - 
& Tampering with network communication, Attack endpoints on the ground \\

\midrule

\cite{smailes2023dishing} 
& Availability 
& Gather space Comm Info 
& Acquire Infrastructure 
& Ground Segment Access 
& Cyber Execution 
& Backdoor / Firmware Injection 
& - 
& - 
& Tampering with network communication, Denial of service \\

\bottomrule
\end{tabularx}
\end{table*}

\end{document}